\documentclass[twocolumn]{aastex631}

\shorttitle{Catalog of the Galactic X-ray pulsars in HMXB systems}
\shortauthors{Vitaliy Kim, Ildana Izmailova \& Yerlan Aimuratov}

\graphicspath{{./}{figures/}}

\begin{document}

\title{Catalog of the Galactic population of X-ray pulsars in High-mass X-ray binary systems}

\correspondingauthor{Vitaliy Kim}
\email{kim@fai.kz, ursamajoris1987@gmail.com}

\author[0000-0003-1202-9751]{Vitaliy Kim}
\affiliation{Fesenkov Astrophysical Institute, Observatory 23, 050020, Almaty, Kazakhstan}
\author[0000-0001-9878-0989]{Ildana Izmailova}
\affiliation{Fesenkov Astrophysical Institute, Observatory 23, 050020, Almaty, Kazakhstan}
\author[0000-0001-5717-6523]{Yerlan Aimuratov}
\affiliation{Fesenkov Astrophysical Institute, Observatory 23, 050020, Almaty, Kazakhstan}

\begin{abstract}

A catalog of the Galactic population of X-ray pulsars in high-mass X-ray binary (HMXB) systems is presented. It contains information about 82 confirmed sources: 18 persistent and 64 transient pulsars. Their basic parameters include spin period, spin evolution with global and local spin-up/SPIN-down and duration, orbital period, X-ray luminosity, magnetic field strength measured by cyclotron line analysis, distance, spectral and luminosity class, observable parameters of massive companions, which are shown in the tables provided, with corresponding references. Candidates of the HMXB pulsars are also listed for further careful consideration.

\end{abstract}

 
\section{Introduction} \label{sec:intro}

High-mass X-ray binaries (HMXBs) are systems consisting of two components: the first is a massive star of the early spectral class, and the second can be presented as a degenerated object---a neutron star (NS), black hole, or white dwarf (in rare cases). A significant difference from other massive binaries is the presence of X-ray emission in these systems for most cases due to the accretion \citep{Liu...2006A&A...455.1165L, Fortin...2023A&A...671A.149F, Neumann...2023arXiv230316137N}. Observable parameters and characteristics of HMXBs give important information for modeling and understanding binary systems' accretion and stellar evolution processes. One of the largest HMXB subclasses is X-ray pulsars.  

HMXB X-ray pulsars are close pairs of stars, one of which is an NS, and its companion is a massive star of early (O--B or Be) spectral type. The massive components of these systems usually do not fill their Roche lobe, but experience a significant loss of matter in the form of a stellar wind with a rate of $10^{-6}$--$10^{-7} \, M_{\odot} \,\text{yr}^{-1}$ \citep{Liu...2006A&A...455.1165L}. As the NS orbits its companion, it picks up some of the matter from stellar wind and accretes it onto its surface. This mechanism of mass exchange is called accretion from the stellar wind or wind accretion \citep{Davidson...1973ApJ...179..585D}. The fundamental difference between HMXB X-ray pulsars and most other types of X-ray pulsars is that the mass exchange between the components of these systems occurs in the wind accretion mode. 

Periodic changes in the X-ray intensity of these pulsars are usually associated with their spin rotation, which modulates the pulsations, and with a sufficiently strong magnetic field, which affects the nature of the motion of matter near the NS and leads to an inhomogeneous temperature distribution of its surface \citep{Lyne...2012puas.book.....L}.

Until recently, the most complete catalog of HMXBs was published more than 15 yr ago and, to that date, collected information about 64 X-ray pulsars in binary systems \citep{Liu...2006A&A...455.1165L}. Since then, new sources have been discovered, a few candidates were confirmed to be not pulsars, and information about already known sources was clarified and supplemented. Recently, HMXB catalogs have been released by \citet{Fortin...2023A&A...671A.149F} and independently by \citet{Neumann...2023arXiv230316137N} with updated information about the Galactic population of HMXBs. Additionally, the progress in spaceborne missions resulted in various all-sky surveys, some of them focusing on the hard X-ray domain \citep{Krivonos...2022MNRAS.510.4796K}, which can also be used for an extensive search for pulsars in HMXBs. All of these catalogs give general information about HMXBs without focusing on X-ray pulsars. 

We conducted our independent crossmatching analysis of existing catalogs and databases, such as the Fortin (including Liu), Neumann, and Krivonos catalogs, as well as Compton Gamma Ray Observatory, Fermi, and SIMBAD databases, to search for X-ray pulsars identified with NSs in massive binary systems. Based on the analysis results, we have created a complete catalog (as of April 2023) of the Galactic population of X-ray pulsars in HMXB systems containing data from 82 sources. 

In contrast to the general HMXB catalogs, we have concentrated our efforts on finding and providing detailed information about the parameters of pulsars, such as spin and orbital periods, data on the evolution of pulsar spin periods, X-ray luminosity of the source, and magnetic fields measured from cyclotron lines. Moreover, the catalog contains, where available, detailed information about the parameters of the massive companions: proper name or identification number, spectral type, and luminosity class, photometric estimates of the brightness of the companion in the optical and infrared regions, color excess, as well as information about the coordinates of the objects and estimates of their distance.

For the sake of convenience, we split the catalog into two parts as persistent (Table~\ref{tab:chartable1}) and transient (Table~\ref{tab:chartable2}) pulsars in HMXBs. We provide discussions of notable confirmed HMXB pulsars and separately of candidates, highlighting the extreme or peculiar properties found during their observations. We attempt to provide comprehensive literature on the cataloged sources, although we do not claim to be complete.

The catalog can be useful for studying stellar evolution in binary systems in the late stages \citep{Chaty...2022abn..book.....C}, for studying the distribution of the NS population in the Galaxy \citep{Coleiro...2013ApJ...764..185C}, for studying and modeling a stellar wind and its parameters from hot stars \citep{Burgos...2023arXiv230500305D}, and for studying the massive components in HMXBs \citep{Kretschmar...2017IAUS..329..355K}, by analyzing the generated X-ray flux from the NS and its spin evolution, etc.

\section{Building the catalog}
\label{sec:buildingthecatlog}
We base our searches on catalogs and data from various space missions of the last decades as well as on an extensive review of the literature via the SAO/NASA Astrophysics Data System (ADS)\footnote{\url{https://ui.adsabs.harvard.edu}} and SIMBAD astronomical database \citep{Wenger...2000AAS..143....9W}. A coordinate crossmatching method of three catalogs was used to collect the core of our HMXB pulsars catalog. As a basic criterion, we took an NS to be the main X-ray active component of the binary, thus excluding black hole HMXBs from our consideration. A reduced sample has been searched for additional data to collect only confirmed HMXB pulsars, therefore, we excluded candidates, erroneous identifications, and misclassifications. To complete the catalog, we gathered information on massive companions, focusing on the main information, including spectral type, luminosity, and photometric brightness. Finally, we separately collected information on candidates of HMXB pulsars from elsewhere, for ease of further investigations.

\subsection{Reference catalogs}
\label{sec:referencecatalogs}
There are recently published catalogs on Galactic HMXBs. They came to update a commonly referenced work---the Fourth HMXB Catalog by \citet{Liu...2006A&A...455.1165L}, which in turn was the last in a sequence of works \citep{Bradt...1983ARAA...21...13B,vanParadijs...1995xrbi...nasa...536V,Liu...2000AAS...147...25L} dedicated to systematically collect information on HMXBs.

\citet{Fortin...2023A&A...671A.149F} provide a catalog of $152$~HMXBs in the Galaxy collected up to the end of 2022. Based on a catalog by \citet{Liu...2006A&A...455.1165L}, they formed a working sample via crossmatching with a catalog of hard X-ray sources by \citet{Bird...2016ApJS...223...15B}. A dataset has been further complemented by querying HMXBs from the SIMBAD database and enhanced by positional crossmatching with available counterparts from various space missions.

The catalog by \citet{Neumann...2023arXiv230316137N} presents information and characteristic properties of $169$~HMXBs in the Galaxy collected by 2022 October. The authors have also used Liu's catalog and cross-correlated it with INTEGRAL and Neil Gehrels Swift Observatory datasets obtained between 2002 and 2020 \citep{Kretschmar...2019NewAR...8601546K}. The catalog was enhanced by a systematic search of the literature and querying HMXBs from the SIMBAD and VizieR databases.  

The above catalogs aimed to provide updated basic information on Galactic HMXBs. Both have used the catalog by \citet{Liu...2006A&A...455.1165L} as a core and cross-correlated it with databases of space missions of different years. \citet{Fortin...2023A&A...671A.149F} used the INTEGRAL observations summarized in \citet{Bird...2016ApJS...223...15B}, where the data were collected between $2002$ and $2010$. \citet{Neumann...2023arXiv230316137N} used INTEGRAL data summarized in \citet{Kretschmar...2019NewAR...8601546K}. We aim to further utilize these reference catalogs and form a list of X-ray pulsars in HMXBs. To eliminate the discrepancies between Fortin's and Neumann's lists, since each of them used slightly different sources and methods for sampling, we support our coordinate crossmatch by including recently obtained all-sky survey data observed in hard X-ray within $17$~yr by the INTEGRAL/IBIS detector \citep{Krivonos...2022MNRAS.510.4796K}. Pulsar lists by the Fermi Gamma-Ray Burst Monitor \citep[GBM;][]{Fermi_pulsars} and CGRO-BATSE \citep{Batse} were also used to extract information on the collected objects. We gathered literature and extensively used the SIMBAD database to complete our catalog with the most comprehensive information on pulsars in HMXBs.

\subsection{Data collection and selection methods}
\label{sec:datacollection}
We conducted a coordinate crossmatch method for three catalogs: \citet{Krivonos...2022MNRAS.510.4796K}, \citet{Fortin...2023A&A...671A.149F}, and \citet{Neumann...2023arXiv230316137N}---cumulatively they contain $1000$~objects of various types. To search for the catalog data, the capabilities of the Virtual Observatory were used via TOPCAT \citep{Taylor...2005ASPC..347...29T}. The data were retrieved via the Table Access Protocol (TAP) query using the following algorithm. The keyword for finding a service containing the \citet{Fortin...2023A&A...671A.149F} catalog was \texttt{fortin}, which corresponded to TapVizier service.\footnote{\url{http://tapvizier.cds.unistra.fr/TAPVizieR/tap}} In the service, the request was made via \texttt{Description} and the \texttt{fortin} keyword. The required table was found through a manual search of the table descriptions. The keyword was necessary because when using the service, the user gets access to all data and tables, so a set of results correspond to a given query. In the future, for a faster search, the table name of the tab \texttt{J/A+A/665/A31/table1} could be used in \texttt{TapVizier}. The catalog by \citet{Krivonos...2022MNRAS.510.4796K} was retrieved from the website.\footnote{\url{https://integral.cosmos.ru}} The catalog data in \citet{Neumann...2023arXiv230316137N} were retrieved from the catalog website.\footnote{\url{http://astro.uni-tuebingen.de/\~xrbcat/HMXBcat.html}}

For the coordinate crossmatch of each pair of the three catalogs in TOPCAT, a search radius of $10'$ was used, since it is the maximal radius used in these catalogs that corresponds to the accuracy of the objects positioning of the objects by the INTEGRAL observatory. In addition, \texttt{Match Selection} with \texttt{``Best match, symmetrical''}, and \texttt{Join Type} with criteria \texttt{``1 and 2''} were used as \texttt{Output Rows} parameters. A pairwise comparison of the tables resulted, with the objects contained in each pair of directories (Table~\ref{tab:cross_match}). Further matching of any of the resulted pairs with the third catalog gives a list of objects contained in all of the three directories. 

\begin{figure*}[ht!]
    \centering
    \includegraphics[width=\textwidth]{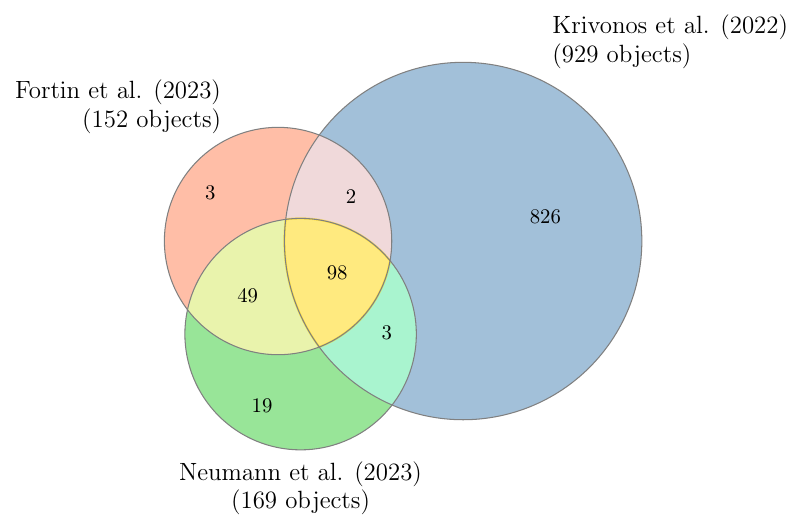}
    \caption{Venn diagram of the coordinate crossmatch of the three catalogs (not to the scale). Each colored sector indicates the number of unique objects present either in the single catalog (3, 19, and 826) or in two catalogs (2, 3, and 49), or in all three catalogs (98). \label{fig:venn_diag}}
\end{figure*}

A Venn diagram in Fig.~\ref{fig:venn_diag} schematically illustrates the crossmatching process. Each sector indicates the number of unique objects present either in the single catalog (3, 19, and 826, colored respectively in coral, light green, and steel blue) or in two catalogs (2, 3, and 49, colored respectively in light pink, cyan, and olive), or in all three catalogs (98, colored in gold), the latter being the working core of the catalog under construction. This golden sector and the remaining supplementary sectors are further investigated individually by applying selection criteria. 

From Neumann's and Fortin's data, we select sources that have regular pulsations in their X-ray emission, i.e., having the \texttt{P\_PULSE} parameter. However, not all objects with \texttt{P$\_$PULSE} are X-ray pulsars. We apply a criterion to exclude black holes, nonmagnetized (or weakly magnetized) NSs, and unidentified objects as active X-ray components. Among the remaining objects, we have X-ray pulsars (confirmed and candidates), pulsating white dwarfs, and objects with pulsations caused by other processes, i.e., not associated with spin rotation. In the resulting list, we examine each object in the SAO/NASA ADS and SIMBAD databases, thereby selecting the sources confirmed as X-ray pulsars. 

Another sampling criterion should be applied to Krivonos' dataset from the INTEGRAL all-sky survey, which is obtained mainly in the hard X-ray and gamma-ray ranges. First of all, we select only Galactic objects (\texttt{Category = Galactic}), and then we apply an additional selection for which the \texttt{Type} parameter meets the criterion of HMXBs (\texttt{Type = HMXB}).

The selection by the \texttt{P\_PULSE} parameter reduced the working core (golden sector) by $35$~discarded objects. The remaining $63$~sources in the golden core were complemented with $18$~confirmed HMXB pulsars found in either Fortin's or Neumann's lists and one object found in Krivonos' list. In the supplementary sectors, we were able to find that three objects (Swift~J1845.7–0037, IGR~J18179-1621, and IGR~J17200-3116) identified as X-ray pulsars are missing from Fortin's catalog, one object (Swift~J1845.7–0037) is missing from Neumann's catalog, and $17$~objects (SAX~J0635.2+0533, XTE~J1829-098, IGR~J18179-1621, Swift~J1626.6-5156, XTE~J1543-568, RX~J0812.4-3114, 1H~2138+579, XTE~J1906+09, 1H~1238-599, GRO~J2058+42, SGR~0755-2933, IGR~J21343+4738, IGR~06074+2205, MAXI~J1409-619, SAX~J2239.3+6116, SAX~J1324.4-6200, and AX~J1910.7+0917) are missing from Krivonos' catalog (see the last row in Table~\ref{tab:cross_match}). Thus, the complete catalog of confirmed pulsars in HMXB contains $82$~sources.

\begin{table*}[ht!]
\caption{Crossmatch of three catalogs and resulting HMXBs~Pulsar Catalog.}\label{tab:cross_match}%
\begin{tabular}{lccc}
\toprule
Catalog & Krivonos et al. (2022)  &  Fortin et al. (2023)  &    Neumann et al. (2023)   \\
\hline
Krivonos et al. (2022)  &  929      & 100       & 101 \\
Fortin et al. (2023)    &  100      & 152       & 147 \\
Neumann et al. (2023)   &  101      & 147       & 169 \\
\hline
$82$~HMXB~pulsars (this work)  & 65 & 79 & 81 \\
\botrule
\end{tabular}
\end{table*}

\section{Catalog objects} \label{sec:part1}
\subsection{HMXB Pulsars and Their Classification: Persistent Sources and Transients} \label{sec:part1:class}
\begin{figure*}[ht!]
    \includegraphics[width=\textwidth]{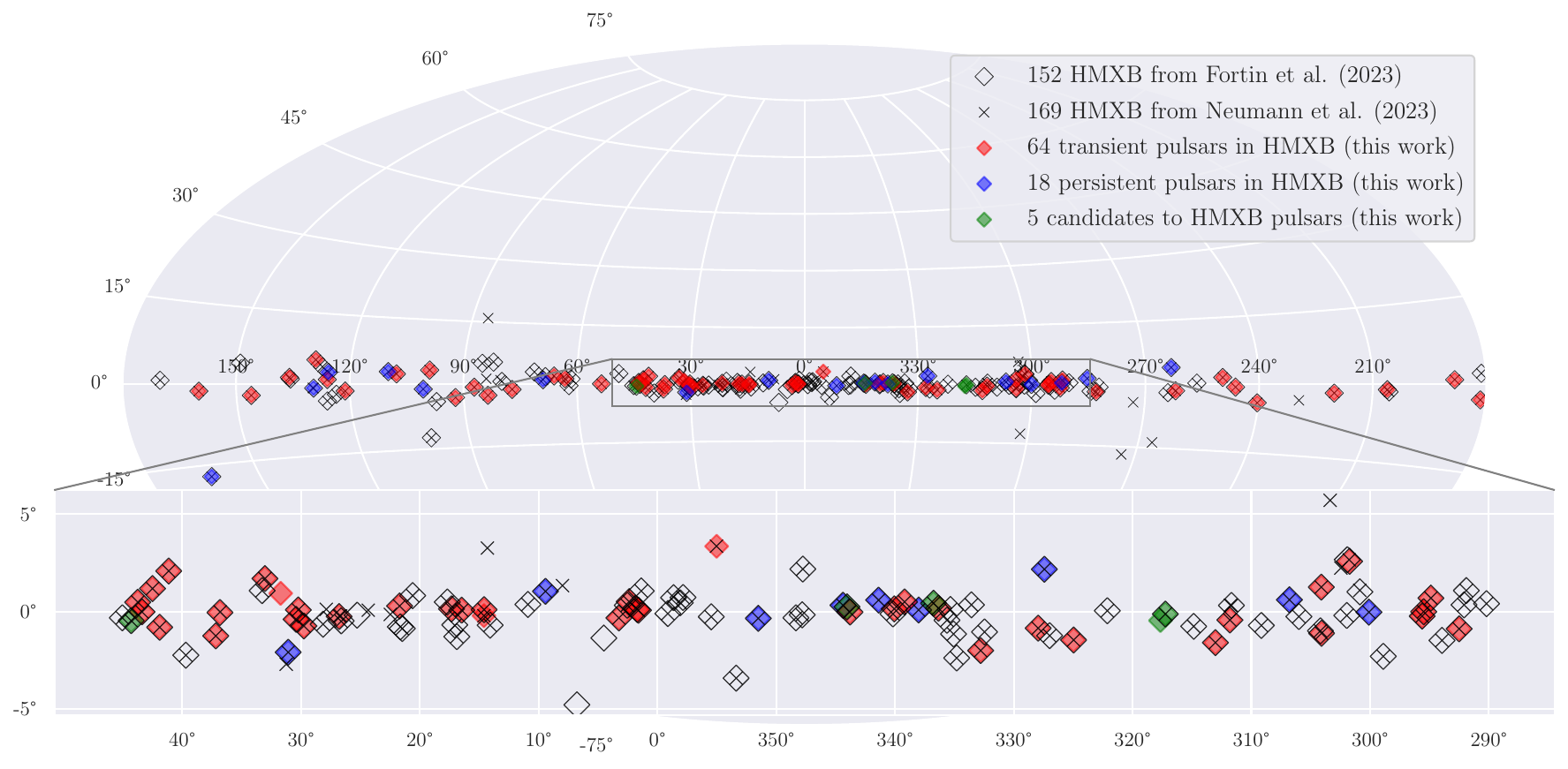}
    \caption{Distribution of the Galactic HMXB sources in Aitoff projection. The inset resolves the dense central region. Crosses indicate objects from the catalog by \citet{Neumann...2023arXiv230316137N}, while empty diamonds show the objects from the catalog by \citet{Fortin...2023A&A...671A.149F}. Transient (red) and persistent (blue) pulsars in HMXBs are indicated by the coloring of the corresponding crosses, diamonds, or both. Crosses and diamonds without coloring indicate HMXBs with a black hole as the main X-ray active component. Candidates of pulsars are colored green. All HMXB X-ray pulsars (except X Per) have absolute values of the galactic latitude not exceeding $4{\rlap{.}^\circ}1$.} \label{fig:galact}
\end{figure*}

X-ray pulsars in massive binary systems are relatively numerous compared to other types of X-ray pulsars. The current population of Galactic HMXB pulsars includes $82$ objects, and all of them (except X~Per) have absolute values of the galactic latitude not exceeding $4{\rlap{.}^\circ}1$ (see Fig.~\ref{fig:galact}). Most of them are located within  $\sim2$--$10$ kpc from the galactic plane \citep{Coleiro...2013ApJ...764..185C}.

\begin{figure*}[ht!]
    \includegraphics[width=\textwidth]{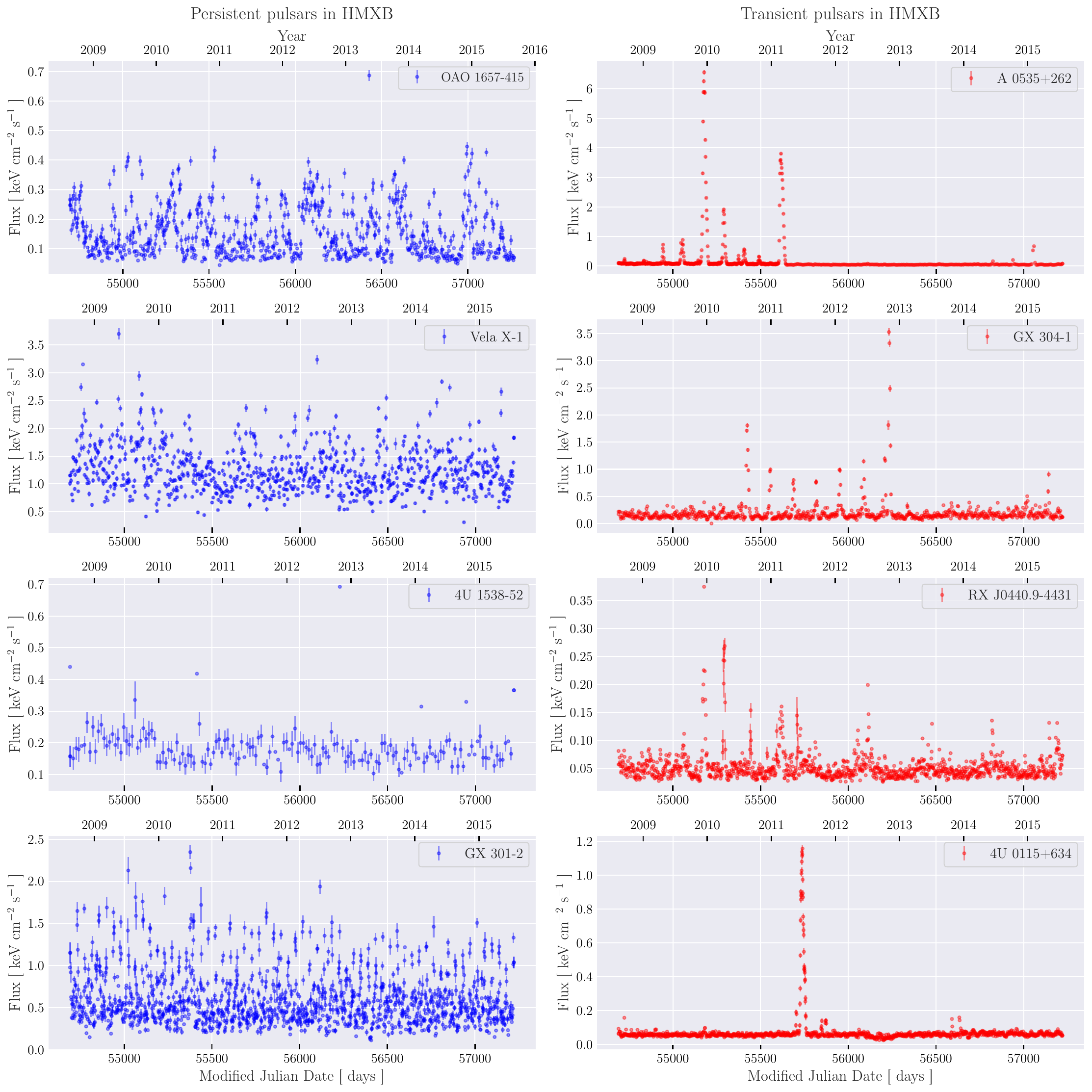}
    \caption{Observed flux (in keV cm$^{-2}$ s$^{-1}$) for a time in Modified Julian Date (MJD) by the Fermi-GBM \citep{Fermi_pulsars} from some persistent (left column) and transient (right column) X-ray pulsars in the range $12$--$50$ keV.  \label{fig:trans_lum}}
\end{figure*}

X-ray pulsars in HMXBs are usually divided into two main groups: persistent sources and transients. Persistent pulsars constitute a group of sources where the X-ray flux experiences only smooth variations without outburst activity over the entire period of observations \citep{Reig...1999MNRAS.306..100R}, see the left column in Fig.~\ref{fig:trans_lum}. This implies a relatively stable outflow of matter from their massive companions, with a stable accretion rate on the NS surface \citep{Reig...1999MNRAS.306..100R}.

The galactic population of persistent pulsars includes $18$ sources, the luminosity of which lies in the range $10^{34}$--$10^{37}$ erg s$^{-1}$. They have been observed in systems with relatively short orbital periods, for most cases $<20$~days, and small orbital eccentricity \citep{Tutukov...2020PhyU...63..209T}. Exceptions, however, are X Per, with an orbital period of $P_{\rm orb}\sim250$~days \citep{Delgado-Mart...2001ApJ...546..455D} and RX J0146.9+6121, with $P_{\rm orb}\sim 330$~days \citep{Sarty...2009MNRAS.392.1242S}. Most of the persistent X-ray pulsars are long-periodic, with $P_{\rm s} > 100$~s and reaching $9475$~s. Exceptions are Cen~X-3, OAO~1657, and IGR~J22534+6243; their spin periods are less than $100$~s. 

The galactic population of transient pulsars currently includes $64$~objects. On the contrary, they are characterized by rapid changes in the magnitude of the X-ray flux by several orders of magnitude; see the right column in Fig.~\ref{fig:trans_lum}. Transients, compared to persistent pulsars, have larger orbital periods and larger orbital eccentricity $\varepsilon > 0.2$ \citep{Tutukov...2020PhyU...63..209T, Fortin...2023A&A...671A.149F}, reaching $\varepsilon \simeq 0.88$ in GS~1843-02 \citep{Finger...1999ApJ...517..449F}.

The X-ray luminosities of persistent sources and the majority of transient pulsars in a quiescent state are found within $10^{34}$--$10^{37}$ erg s$^{-1}$. According to \citet{Negueruela...1998A&A...338..505N}, the outbursts of transient pulsars, in most cases, can be classified as follows:

\begin{enumerate}
    \item Short X-ray outbursts (or first-type outbursts) are characterized by a rapid increase in luminosity by an order of magnitude and have a durations of several days. For some sources, they occur near the periastron and have a periodicity comparable to the orbital period;
    \item Giant X-ray outbursts (or second-type outbursts) are characterized by a significant increase in X-ray luminosity, up to several orders of magnitude and several weeks in duration. This type of outburst, in most cases, does not correlate with the orbital period. 
\end{enumerate}

The spectra of X-ray pulsars within $1$--$10$~keV are predominantly of a blackbody origin. An exponential cutoff is generally observed in the hard part of the spectrum. However, power-law ``tails'' are observed occasionally \citep{Lipunov...1992ans..book.....L, Coburn...2002ApJ...580..394C}.

\subsection{Pulsar Spin and Its Evolution}
\label{subsec:spin}
The spin periods ($P_{\rm s}$) of X-ray pulsars in HMXBs are dispersed over a wide range of values. The periods of most pulsars whose radiation is generated during the accretion process are within the range from $2.76$ to $36,200$~s. The exceptions are the first object in Table~\ref{tab:chartable2} (SAX~J0635.2+0533), the transient pulsar whose period is less than a second. There is, however, a reason to believe that the radiation from these objects is not of an accretionary nature (see Section~\ref{sec:part3}). 

\begin{figure*}[ht!]
\centering
    \includegraphics[width=\textwidth]{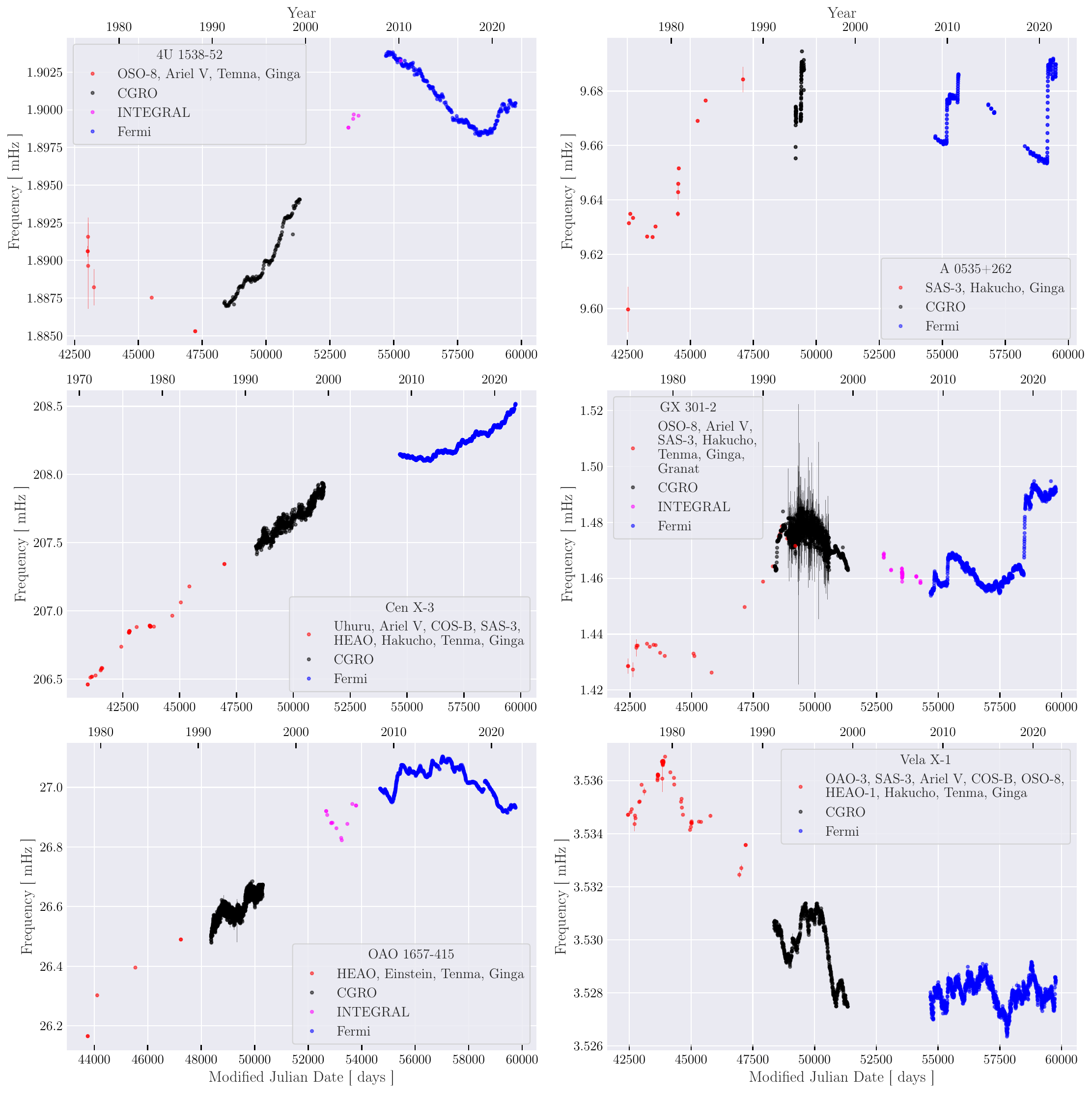}
    \caption{Evolution of the spin period (frequency) of some Galactic X-ray pulsars in HMXBs within the time range from the 1970s to 2022 received by HEAO-1, Tenma, Ginga, Einstein, OSO-8, Ariel~V, and Granat \citep[red;][]{Nagase...1989PasJ...396..P147}, the CGRO \citep[black;][]{Batse}, INTEGRAL \citep[magenta;][]{Integral_pulsars}, and Fermi-GBM \citep[blue;][]{Fermi_pulsars}. The error bars for later missions are within data points due to the increase in instrumental precision. \label{fig:long}}
\end{figure*}

It should be especially noted that the spin periods of pulsars are not constant and change over time (see Fig.~\ref{fig:long}), undergoing:

\begin{enumerate}
    \item Global trends of spin-up (or spin-down) taking place with rates usually $|\dot{\nu}| < 10^{-12}\, \text{Hz s$^{-1}$}$, where $\dot{\nu} = d \nu /dt$, and $\nu = 1 / P_{\rm s}$ is the spin frequency of a pulsar. These trends usually can last from several months to several years and, in some cases, up to decades.
    
    \item Episodic variations (local trends) of spin-up (or spin-down). These chaotic variations occur at an exceptionally high rate. They can be an order of magnitude or more superior to global trends and last from several days to several weeks. Episodic variations take place against the backdrop of global trends.
\end{enumerate}

Even in the first decades after the discovery of X-ray pulsars, it was noted that the vast majority of such objects experience a peculiar global spin-up \citep{Lipunov...1992ans..book.....L}. There are several hypotheses about the origin of this phenomenon:
\begin{enumerate}
    \item \citet{Siuniaev...1977SvAL....3..114S} proposed a hypothesis that during the evolution of the massive component of the system, the intensity of the outflow of its stellar wind can increase, which leads to the global spin-up of an NS in the system. However, this hypothesis raises serious doubts \citep{Lipunov...1992ans..book.....L}, since the evolution of stars is insignificant on a scale of several decades.
    \item Selection effects. According to this hypothesis, pulsar acceleration occurs at the accretion stage \citep{Ghosh...1979ApJ...232..259G, Ghosh...1979ApJ...234..296G}. The deceleration occurs when an NS does not manifest as a variable X-ray source at the propeller stage. This hypothesis also raises doubts, since accreting X-ray pulsars rotate near the equilibrium period, which can be much larger than the critical period for the transition to the propeller stage.
    \item The durations of the stages of spin-up and spin-down of the pulsar are asymmetric. The spin-up phase has a longer and gentler trend, while the spin-down phase occurs over a short time, but at a faster rate. Therefore, the probability of observing a pulsar in the spin-up stage is higher. Such a scenario can occur when the scalar potential is asymmetric on the side of the accretion flow \citep{Lipunov...1992ans..book.....L}.
    \item \citet{Kim...2017JPhCS.929a2005K} proposed a hypothesis about long-period variations in the stellar wind of massive companions using the pulsar OAO~1657-415 as an example. Local variations of the spin-up/spin-down of pulsars are approximately an order of magnitude higher than the value of its global acceleration trend, which speaks in favor of the equilibrium rotation of the pulsar, i.e., the spin rotation of an NS with an equilibrium period. The torque carried by the accretion flow to the NS essentially depends on the velocity of the stellar wind. The observed global trends in spin-up and spin-down may be due to the drift of the equilibrium period caused by apparently long-term changes in the wind velocity of massive components in HMXBs. Similar long-term velocity variations are also observed in the solar wind associated with the 11~yr activity cycle \citep{Li...2017MNRAS.472..289L}.
\end{enumerate}

In most cases, the observed changes in the spin periods of X-ray pulsars in HMXBs are associated with the angular momentum exchange between the NS and its accretion flow. In this case, the efficiency of the angular momentum exchange depends on the parameters of the HMXB system and the realized accretion scenario: quasi-spherical (QSp) accretion \citep{Davidson...1973ApJ...179..585D, Arons...1976ApJ...210..792A}, Keplerian disc \citep[Kd;][]{Pringle...1972A&A....21....1P, Shakura...1973SvA....16..756S}, MAD-disk \citep[or ML-disk;][]{Ikhsanov...2014ARep...58..376I}, etc.

\begin{figure*}[ht!]
    \includegraphics[width=\textwidth]{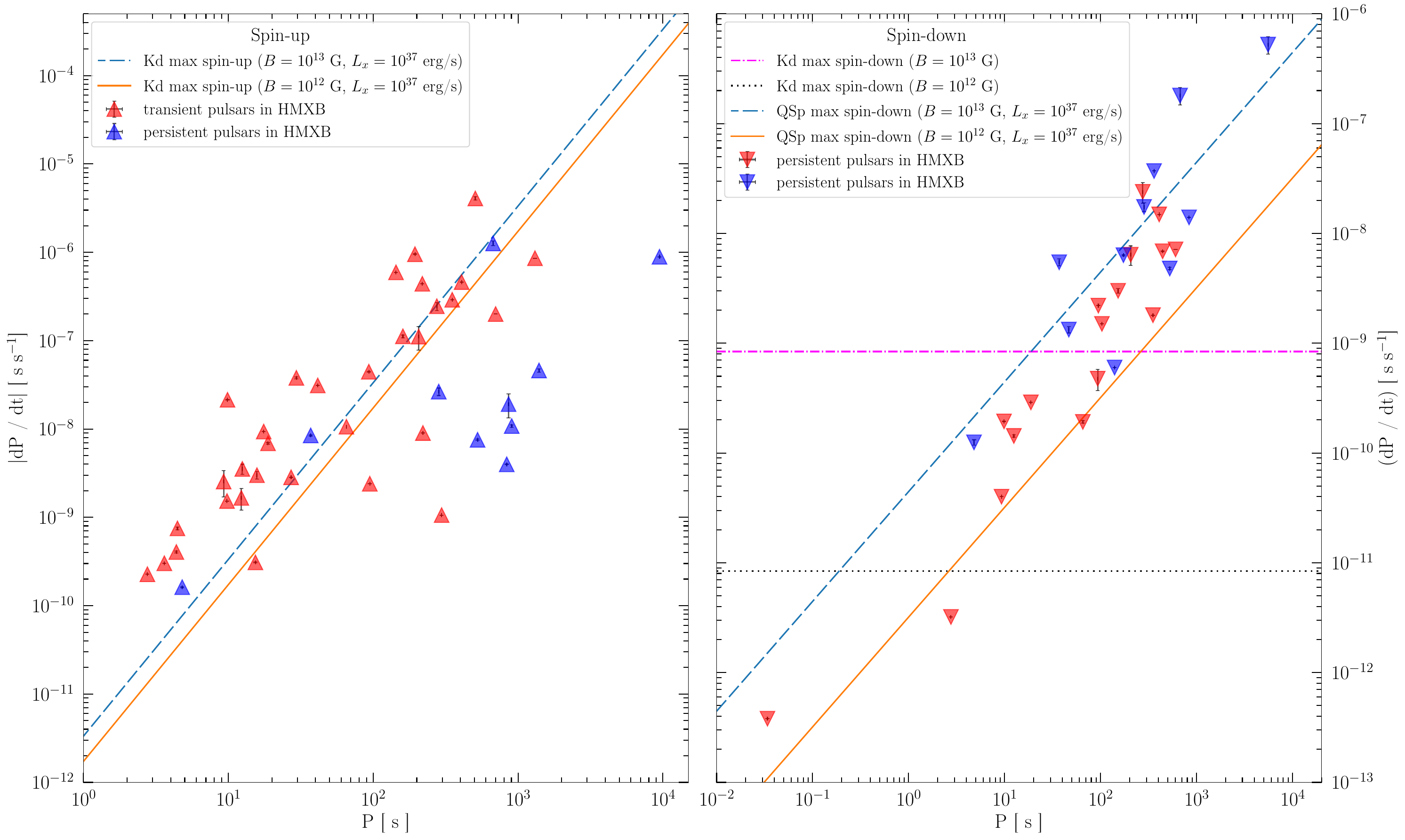}
    \caption{$P_{\rm s}$--$|\dot{P}|$ diagram for  Galactic X-ray pulsars in HMXBs. In the spin-up diagram (left panel), the lines correspond to the maximal possible spin-up for Kd scenario, with $B_{\rm ns} = 10^{13}\, \text{G}, \, L_{\rm x} = 10^{37}\, \text{erg s$^{-1}$}$ (dashed blue line) and with $B_{\rm ns} = 10^{12}\, \text{G}, \, L_{\rm x} = 10^{37}\, \text{erg s$^{-1}$}$ (solid orange line). In the spin-down diagram (right panel), the lines correspond to the maximal possible spin-down for the Kd scenario, with $B_{\rm ns} = 10^{13}\, \text{G}, \, L_{\rm x} = 10^{37}\, \text{erg s$^{-1}$}$ (horizontal dashed-dotted magenta line) and with $B_{\rm ns} = 10^{12}\, \text{G}, \, L_{\rm x} = 10^{37}\, \text{erg s$^{-1}$}$ (horizontal dotted line). The dashed blue line corresponds to the maximal possible spin-down for QSp, with $B_{\rm ns} = 10^{13}\, \text{G}, \, L_{\rm x} = 10^{37}\, \text{erg s}^{-1}$, and the solid orange line corresponds to QSp with $B_{\rm ns} = 10^{12}\, \text{G}, \, L_{\rm x} = 10^{37}\, \text{erg s$^{-1}$}$. Due to the high precision of the measurements, the error bars for some objects are within data points.}
    \label{spin-diag}
\end{figure*}

In Fig.~\ref{spin-diag} we show the $P_{\rm s}$--$|\dot{P}|$ diagram for Galactic X-ray pulsars in HMXBs with known spin evolution. As mentioned above, these observable processes are due to angular momentum exchange. In such a case, the torque applied to NS is 
\begin{equation}
  |K| = 2\pi\, I\,\dot{\nu} =2\pi\, I  \frac{\dot{P}}{P_{\rm s}^2},
  \label{eq1}
\end{equation}
where $I$ is a moment of inertia of the NS. The maximal possible value of the spin-up or spin-down torque $|K|$ depends on the type of accretion structure surrounding an NS. In the case of accretion from a Kd, these values for spin-down \citep{Lipunov...1982SvA....26..537L} are limited by $|K| \leq |K^{\rm _{Kd}}_{\rm sd}|$, where

\begin{equation}
|K^{\rm _{Kd}}_{\rm sd}| = k_{\rm t}\, \frac{\mu_{\rm ns}^2}{r_{\rm cor}^3 },
  \label{eq2}
\end{equation}
and the values for spin-up \citep{Pringle...1972A&A....21....1P} are limited by $|K| \leq |K^{\rm _{Kd}}_{\rm su}|$, where

\begin{equation}
   |K^{\rm _{Kd}}_{\rm su}| = \dot{M}\sqrt{GM_{\rm ns}\, r_{_A}}.
   \label{eq3}
\end{equation}
Here, $\mu_{\rm ns} = 0.5 B_{\rm ns}\, R_{\rm ns}^3$ is a dipole magnetic moment of the NS, $r_{\rm cor} = (GM_{\rm ns}/\omega_{s}^2)^{1/3}$ is a corotation radius, $\dot{M} = L_{\rm x}\, R_{\rm ns} / GM_{\rm ns}$ is an accretion rate onto the NS surface, $r_{_A} = [\mu_{\rm ns}^{2}/\dot{M}\,(2GM_{\rm ns})^{2}\,]^{2/7}$ is the Alfven radius, $R_{\rm ns}$ is a radius of the NS, and $k_{\rm t}$ is a dimensionless parameter of the order of unity. In the case of QSp, the spin-down torque is limited by $|K| \leq |K^{\rm _{QSp}}_{\rm sd}|$ \citep{Shakura...1975SvAL....1..223S}, where

\begin{equation}
|K^{\rm _{QSp}}_{\rm sd}| = k_{\rm t}\,\dot{M}\,\omega_{\rm s}\, r_{\rm _A}^2. 
  \label{eq4}
\end{equation}

On the diagrams, corresponding lines show the borders of maximal possible spin changes in frames of QSp and Kd accretion with the canonical parameters of an NS ($R_{\rm ns} = 10^6\,\text{cm}$, $M_{\rm ns} = 1.4\,M_{\odot}$, $I = 10^{45}\, \text{g cm}^2$) and a surface magnetic field $10^{12}$--$10^{13}\,\text{G}$ with upper luminosity $L_{\rm x} \sim 10^{37}\,\text{erg s}^{-1}$ in a quiescent state. Acceptable values of ($R_{\rm ns}, \,I\, , M_{\rm ns}$) for all NSs are in a narrow range close to the canonical parameters; their variations influence the evaluation insignificantly. As shown in Fig.~\ref{spin-diag}, some of the HMXB X-ray pulsars from the Galactic population demonstrate a spin evolution $\dot{P}$ exceeding the Kd and QSp theoretical upper limits. See the further discussion in Section~\ref{sec:part4}.

Since their discovery, and over the entire study period, X-ray pulsars' spin periods have changed insignificantly. The changes are usually a few seconds \citep{LaPalombara...2009A&A...505..947L}. In a very rare case, as for the pulsar 2S~0114+650, a period change value reaches several minutes, but is still $\sim 5 \%$ of its average spin period \citep[$P_{\rm s} \simeq 9475 \,\text{s}$;][]{Wang...2011MNRAS.413.1083W}. Period changes were discovered as a minimum for 52 (15 persistent and 37 transient sources) Galactic HMXB X-ray pulsars (shown in Fig.~\ref{spin-diag} and listed in the catalog). 

\subsection{Magnetic fields} 
There are two basic methods for estimating the surface magnetic field of an NS---namely, an X-ray pulsar. The first of them is based on the interaction (angular momentum exchange) between the accretion flow and magnetosphere of the rotating NS (see Section~\ref{subsec:spin} for more details and cited literature). If the spin period is known and the rate of its spin-down is observed, then from equations (\ref{eq1}), (\ref{eq2}, and \ref{eq3}), depending on the implemented accretion scenario, one can calculate the magnetic dipole moment $\mu_{\rm ns}$, and then the surface magnetic field strength $B_{\rm ns}$. This method essentially depends on the parameters of the models and may give contradictory results in the frameworks of various accretion scenarios \citep[see, for example,][]{Klus...2014MNRAS.437.3863K}; therefore, it is not a reliable way for the estimation of $B_{\rm ns}$.

The second method is based on spectral analysis. X-ray spectra of highly magnetized NSs may demonstrate the so-called cyclotron lines predicted by \citet{Gnedin...1974A&A....36..379G}. They are due to the resonant scattering of photons by electrons in a strong magnetic field of an NS. The analysis of these absorption lines allows the estimation of the surface magnetic field strength with high reliability. This method does not depend on accretion parameters and is the main one for the magnetic field estimation of X-ray pulsars \citep{Staubert...2019A&A...622A..61S}.

The spectra of $31$~objects ($6$~persistent and $25$~transient HMXB pulsars) demonstrate the presence of a cyclotron line. The measured surface magnetic field strength of these sources lies in a narrow range from $\sim 0.86\times10^{12}\,\text{G}$ to $\sim 7.8\times10^{12}\,\text{G}$. These estimations are in good agreement with the canonical value $B_{\rm ns} \sim 10^{12}\,\text{G}$ for NSs \citep{Lipunov...1992ans..book.....L, Lyne...2012puas.book.....L}.

\subsection{Massive companions} 
The massive companions of transient pulsars are usually represented by main-sequence stars. In contrast, the massive companions of persistent pulsars are primarily represented by hot stars of early luminosity classes (giants and supergiants). 

An interesting fact is that the massive components of the currently known HMXBs are in a relatively narrow range of spectral types, O6.5--B3, and a wide range of luminosity classes that extend from main-sequence stars to supergiants (see the catalog in the Appendix). One of the possible reasons for this situation is the relatively small number of stars whose spectral type is earlier than O6.5. On the other hand, the outflow rate of stars whose spectral type is later than B3 \citep{Puls...2008A&ARv..16..209P,Krticka...2014A&A...564A..70K} turns out to be too low to provide a sufficiently high luminosity of the accretion source (see Section~\ref{sec:part4}). 

\section{Some remarks} \label{sec:part3}
\subsection{Notable Galactic HMXB pulsars}
\textit{SAX~J0635.2+0533}. This transient source has the smallest spin period $P_{\rm s} \simeq 0.0338 \,\text{s}$ in the Galactic population of HMXB pulsars. It is comparable with those of radio pulsars (ejectors). But it is one of the unique cases in the population of X-ray pulsars in massive binary systems. Its X-ray emission is apparently due to rotation-powered processes \citep{Pacini...1970Natur.226..622P}. For the canonical parameters of an NS, in this case, a centrifugal barrier occurs, which prevents accretion or surface magnetic field from being less than $B_{\rm ns} \leq 10^{8}\, \text{G}$ \citep{Mereghetti...2009A&A...504..181M}. This value is four orders of magnitude less than was expected from a young accreting NS in an HMXB system, which is doubtful. 

\textit{Swift~J0243.6+6124}. The first discovered ultraluminous X-ray (ULX) pulsar in an HMXB in our Galaxy and, currently, the only known object from this class in the Galactic population. Swift~J0243.6+6124 was discovered in 2017 by Swift/Burst Alert Telescope (BAT) as a possible GRB, but further observations showed continuous X-ray flux from this object with 9.86~s periodicity. This fact indicated that the object was an accreting NS \citep{Kennea...2017ATel10809....1K}. Observations in the optical range allowed us to identify its normal companion---an O9.5Ve star at $\sim 5$ kpc distance \citep{Reig...2020A&A...640A..35R}. The distinguishing feature of this pulsar is its ultrahigh X-ray luminosity, reaching $> 10^{39}\, \text{erg s$^{-1}$}$ during outburst activity, exceeding the Eddington limit, but in the quiescent state, it dropped to $(0.3$--$6)\times 10^{34}\, \text{erg s$^{-1}$}$ \citep{Doroshenko...2020MNRAS.491.1857D}. To explain the ULX phenomenon, there are several possible hypotheses. (1) The presence of a multipolar magnetic field of the NS with superstrong $B_{\rm ns} \geq 10^{14}\,\text{G}$. In this case \citep{Israel...2017Sci...355..817I}, the high radiation flux will not block its accretion flow totally. (2) In \citet{Mushtukov...2015MNRAS.454.2539M}, ULX pulsars and super-Eddington accretion phenomenon are explained  by the specific geometry of the accretion column, where generated photons can escape not only in a radial direction (along the column), but also through the sides of the column. This approach also implies the presence of a superstrong $B_{\rm ns} \geq 10^{14}\,\text{G}$ magnetic field on the NS surface. However, at the moment, there is no evidence of the presence of a superstrong magnetic field in the Galactic population of HMXB pulsars (see the catalog in the Appendix).

\textit{X~Per (X~Persei, 4U~0352+309)}. This persistent pulsar has a large absolute galactic latitude $|b| \simeq 17{\rlap{.}^\circ}1$ (see Fig.~\ref{fig:galact}) as compared to other HMXB pulsars not exceeding $4{\rlap{.}^\circ}1$ from the galactic plane. The large galactic latitude of X Per can be explained either by the ``binary supernova mechanism'' or ``dynamical ejection'' \citep{Raddi...2021A&A...645A.108R}. The first one points to a mechanism of runaway through a supernova explosion in a binary system. The second possibility can occur in few-body interactions in clusters when binding energy transforms into kinetic energy \citep{Oh...2016A&A...590A.107O}. In both cases, there is a high probability of system decomposition \citep{Oh...2016A&A...590A.107O,Raddi...2021A&A...645A.108R}. Still, an NS and a normal component could form a binary system after gravitational capture has occurred after the escape from the Galactic plane.

\textit{SGR~0755-2933}. This object was discovered by the Swift Burst Alert Telescope in 2016 as a short ($< 128$~ms) and spectrally soft gamma-ray burst without detected optical afterglow. The source is located in the Galactic plane \citep[$b \sim - 0.6$ deg,][]{Barthelmy...2016GCN.19204....1B}. Initially, the source was assumed to be a soft gamma repeater (SGR)---an isolated magnetized NS with nonperiodic burst activity. Further observations by Swift/X-Ray Telescope (XRT), NuSTAR, and Chandra revealed the presence of modulations in the X-ray flux of SGR~0755-2933 associated with spin rotation and orbital motion in a massive binary system. Therefore the initial assumption about the isolated origin of this object was disproved \citep{Doroshenko...2021A&A...647A.165D}. However, the prefix ``SGR'' was saved for this HMXB pulsar.

\textit{2S~0114+650}. This persistent pulsar has the second-longest spin period \citep[$P_{\rm s} \simeq 9475\, \text{s}$;][]{Wang...2011MNRAS.413.1083W} among Galactic HMXB pulsars. For most of the known X-ray pulsars in HMXBs, especially persistent sources, their spin periods exceed $P_{\rm s} > 100\, \text{s}$, but are less than $1000$~s. Compared with radio pulsars, whose spin periods usually do not exceed $1$~s, HMXB X-ray pulsars are long-periodic. This phenomenon is related to the spin evolution of a magnetized NS from ejector to accretor with an intermediate propeller stage \citep{Lipunov...1992ans..book.....L}. Since the discovery of 2S~0114+650, its global spin-up trend with a high rate $\sim 10^{-6}\,\text{s s$^{-1}$}$ (see the catalog) has been observed \citep{Bonning...2005A&A...436L..31B}. The origin of its ultralong spin period and high-rate spin-up are still unknown.

\textit{AX~J1910.7+0917}. This transient pulsar has the longest spin period among all known pulsars, including other populations: low-mass X-ray binaries (LMXB), radio pulsars (RPs), anomalous X-ray pulsars (AXPs), soft gamma repeaters (SGRs), etc. The object was discovered as a Galactic faint X-ray source during the ASCA Galactic Plane Survey \citep{Sugizaki...2001ApJS..134...77S}, carried out from 1996 to 1999. Further photometric and spectral observations in the near-infrared range with the $3.58$ m Telescopio Nazioinale Galileo at Roque de los Muchachos Observatory (La Palma, Spain) made it possible to identify the optical counterpart as a massive B-class supergiant located at a distance $16.0\pm 0.5 \,\text{kpc}$. A timing analysis of its X-ray flux from observations performed in 2011 by the Chandra space observatory showed evident periodic modulations with $P_{\rm s} = 36200\pm110\, \text{s}$ \citep{Sidoli...2017MNRAS.469.3056S}, which was was identified as an NS spin period. According to \citet{Sidoli...2017MNRAS.469.3056S}, the anomalously long period of AX~J1910.7+0917 could be explained in the frame of QSp. Assuming a stellar wind velocity $\upsilon_{\rm w} = 1000\,\text{km s}^{-1}$, surface magnetic field $B_{\rm ns} = 10^{12}\,\text{G}$, and orbital period $P_{\rm orb} = 68\,\text{days}$,  a canonical NS accreting in QSp regime can spin down to $36,200 \, \text{s}$ on a $\tau \sim 10^{4}\,\text{yr}$ timescale. But the question remains: why do other HMXB pulsars with similar parameters did not reach identically long spin periods?

\subsection{Some candidates to HMXB X-ray pulsars}
We provide information on candidates of HMXB pulsars (the green diamonds in Fig.~\ref{fig:galact}). The list has been formed from the candidate objects of the studied catalogs \citep{Liu...2006A&A...455.1165L,Fortin...2023A&A...671A.149F,Krivonos...2022MNRAS.510.4796K,Neumann...2023arXiv230316137N} and extensively searched for additional information via SIMBAD and the literature. Candidates are of interest for further attentive investigation.

\subsubsection{Pulsar candidates with unclear spin period}
\textit{IGR~J14488-5942}. This object was discovered in 2009 by Swift/XRT as an X-ray point source \citep{Landi...2009ATel.2355....1L}. A period of $49.51\pm0.12$~days was found by analyzing the amplitude modulation in the IGR~J14488-594 light curve (covering the 2005--2009 interval in the $15$--$100$~keV range received from Swift/BAT). The period was expected to be the system's orbital period \citep{Corbet...2010ATel.2598....1C}. Based on optical and near-infrared spectral and photometric observations on a $3.58$ m telescope (ESO/New Technology Telescope, Chile), it was supposed that the system corresponded to an HMXB with an Oe/Be optical counterpart \citep{Coleiro...2013A&A...560A.108C}. According to \citet{Corbet...2017ApJ...846..161C}, there are some hints of possible pulsations with period $33.419 \pm 0.001$~s, but this needs further investigation.

\textit{IGR~J19140+0951}. This X-ray source was discovered in 2003 by INTEGRAL \citep{Swank...2003ATel..128....1S}. Near-infrared photometric observations on a $3.58$ m telescope (Roque de los Muchachos Observatory/TNG, La Palma, Spain) showed that the IGR~J19140+0951 optical counterpart could be a B0.5 supergiant (Iab luminosity class) located at a distance of $3.6\pm0.04$~kpc \citep{Torrejon...2010A&A...510A..61T}, confirming an HMXB status. According to \citet{Sidoli...2016MNRAS.460.3637S}, XMM–Newton observations of IGR~J19140+0951 performed in 2015 showed quasiperiodic oscillations of X-ray flux ($2$--$10$~keV range) with a $5937\pm219$~s period, which could be associated with the NS spin period, but this needs further investigation. 

\subsubsection{Pulsar Candidates with Unclear Status of Optical Counterpart}
\textit{SAX~J1452.8-5949, AX~J1700.1-4157 and IGR~J16358-4726}. Three Galactic X-ray pulsars in binary systems. SAX~J1452.8-5949 and AX~J1700.1-4157 were discovered in 1999 by BeppoSAX and ASCA, correspondingly. Initially, their origins were considered as X-ray pulsars in high-mass binary systems. But further detailed near-infrared and optical identification showed that their normal companions were late-type low-mass stars, thus SAX~J1452.8-5949 and  AX~J1700.1-4157 were classified as LMXBs \citep[LMXB,][]{Kaur...2009MNRAS..394..P1597, Kaur...2010MNRAS.402.2388K}, therefore the objects were not included in our catalog. However, both objects in the SIMBAD database and AX~J1700.1-4157 in Fortin's and Neumann's catalogs are marked as HMXBs. A similar history takes place with IGR~J16358-4726. This object was discovered in 2003 by Chandra as a point X-ray source with $P_{\rm s} \simeq 5860\,\text{s}$ periodicity \citep{Kouveliotou...2002cxo..prop.1360K}. Over the next decade, there was a discussion about its possible optical companion. \citet{Amico...2006AIPC..840...97D} based on near-infrared photometric observations on a $1.6$ m telescope (Laboratorio Nacional de Astrofisica, Brazil), it was supposed that the system corresponded to an HMXB, but \citet{Nespoli...2010A&A...516A..94N} based on infrared medium-resolution spectroscopy using a UT-1 telescope (ESO), showed that the optical companion belonged to late-type (K--M) giant, thereby classifying IGR~J16358-4726 as an LMXB. But the question about the spectral classification of IGR~J16358-4726's optical companion is still controversial: some articles provide arguments in favor of the HMXB hypothesis \citep[see, e.g.,][]{Coleiro...2013A&A...560A.108C}, but also there are arguments for LMXB origin \citep[see, e.g.,][]{Yungelson...2019MNRAS.485..851Y}. Therefore, IGR~J16358-4726 as an object with unclear status was not included in our catalog. 

\section{About the Catalog Tables}\label{sec:part2}
The catalog is split into two parts (tables): persistent and transient sources. Catalog objects are arranged in ascending order of their spin periods. The first column displays the serial number of the pulsar. In the second column is its name. The third double column are shown the equatorial coordinates, R.A., and decl. for the J2000 epoch from the SIMBAD Database \citep{Wenger...2000AAS..143....9W}, rounded to hundredths of arcseconds. The fourth column contains the values of the spin period ($P_{\rm s}$) in seconds. In parentheses, we indicate the year when the given $P_{\rm s}$ was measured. Also, in the third column, under the value of the spin period, there are given the changes in spin periods (rotational evolution): Loc.spin-up/down episodes (local trends) of the spin-up/spin-down of rotation and Glob.spin-up/down long-term (global) trends of the spin-up/spin-down of the pulsar's rotation. The fifth column ($\dot{P}$) gives their values in units ($\text{s s}^{-1}$), as well as the epoch of their observations in MJD. The sixth column shows the values of the orbital periods ($P_{\rm orb}$) in days. The seventh column gives estimates of the X-ray luminosity and the range of its variations ($L_{\rm x}$(erg s$^{-1}$)) in the indicated energy range (keV). The eighth column shows the value of the magnetic field strength on the surface of the NS $B_{\rm ns} (\times\,10^{12}\, \text{G})$ obtained from the analysis of cyclotron lines. The ninth column contains estimates of distances to objects (dist) in kiloparsecs. The tenth column lists information about massive companions: the first line is the star's name (or cataloged number), if available; the second line shows the spectral types and luminosity classes; the bottom lines contain data about photometric magnitudes in the optical (\textit{BVR}) and near-infrared (\textit{JHK}) passbands and also values of color excess $E(B-V)$, if available. The corresponding references support the values of all the parameters of the X-ray pulsars given in the catalogs. All catalog data are given with the same precision as in the cited literature. In the case of different values coexisting for the same object, we used the newest data as possible. The first object in Table~\ref{tab:chartable2} (SAX~J0635.2+0533) is a transient X-ray pulsar with a spin period of less than $1$~s. It seems to have a nonaccreting origin of their X-ray radiation \citep{Mereghetti...2009A&A...504..181M}; therefore, it is separated from the other objects in the table by a double line. 

\section{Discussion and Summary} \label{sec:part4}

The main issues of finding all pulsars exclusively by the coordinate crossmatching method are related to the mismatching of coordinates due to the measurement precisions of different instruments, as well as different namings of objects. While managing to minimize the divergence and make the data uniform, the reference catalogs still inherit some inconsistencies, making a batch comparison less efficient. This required us to search for objects manually after applying the basic selection criteria. The coordinates were different enough when the search radius increased during the crossmatching; false matches began to quickly appear in TOPCAT. These false match cases have to be examined individually, because the name search was complicated too. Some of the names in different directories are used differently and it is necessary to inspect them in SIMBAD, where multiple names can exist for the same object. We found $19$~additional HMXB~pulsars in the supplementary sectors of Fig.~\ref{fig:venn_diag} in two pairs of the three catalogs (Fortin--Neumann: $16$~objects; Krivonos--Neumann: one object) or two single catalogs (Krivonos: one object; Neumann: one object).

The number of known transients significantly exceeds those of persistent sources, $64$ versus $18$ ($\sim3.6$ times), in the Galactic population of HMXB pulsars. This is explained by transients mostly having orbits with a significant eccentricity, found in a wide range from $0.2$ and reaching $0.9$, while for most of the persistent pulsars, the eccentricity does not exceed $0.2$ \citep{Fortin...2023A&A...671A.149F}. According to the evolution of close massive binary systems, during HMXB formation, one of the massive companions normally ends its life with a core collapse followed by a supernova explosion. It significantly changes the orbital parameters of the system, including an increase of eccentricity \citep{Heuvel...2017hsn..book.1527V, Tutukov...2020PhyU...63..209T}. Thus, an NS born in a high-mass binary system is more likely to have an orbit with a relatively high value of eccentricity. This, in turn, leads to transient X-ray activity, due to the dynamic increase of the accretion rate during its orbital motion from apoaster to periaster \citep{Negueruela...1998A&A...338..505N}.

The massive components of the currently known Galactic HMXB pulsars are in a relatively narrow range of spectral types O6.5--B3; see the last column of the catalog tables. One of the possible reasons for this situation is the relatively small number of stars whose spectral type is earlier than O6.5. The number of O-class stars in the Galaxy (including all subclasses) lies within $\sim (1.8$--$5)\times 10^{4}$, and it does not exceed one-ten-thousandth of a percent of the Galactic star population \citep{Apellaniz...2013msao.confE.198M}.  On the other hand, the outflow rate of stars whose spectral type is later than B3 turns out to be too low to provide a sufficiently high luminosity of the accretion source. As mentioned in \citet{Krticka...2014A&A...564A..70K} and \citet{Puls...2008A&ARv..16..209P}, the mass-loss rate of B-class stars rapidly decreases as subclass goes from B0 to B9; moreover, for the B5 subclass and later, the stellar wind is not homogeneous.
    
As shown in Fig.~\ref{spin-diag} the spin changes (spin-up and spin-down) for some of the Galactic X-ray pulsars in HMXBs significantly exceed the maximal possible values predicted within classical accretion scenarios, QSp and accretion from Kd. Isolated cases of this phenomenon were known already in the early era of X-ray astronomy \citep{Shakura...1975SvAL....1..223S}. The number of such cases even increased over time, as new sources were discovered and data accumulated (see Fig.~\ref{spin-diag}). One possible explanation for the rapid spin evolution can be the formation of non-Keplerian accretion structures surrounding an NS in an HMXB. As was shown in \citet{Ikhsanov...2012ApJ...753....1I} and \citet{Ikhsanov...2015MNRAS.454.3760I}, the presence of magnetized stellar wind from a massive companion can influence the structure of the accretion with the formation non-Keplerian magnetically arrested disk (MAD-disk or ML-disk). In this case, the angular momentum exchange between an accretion flow and a NS can be more efficient than the QSp and Kd scenarios. 

In summary, we have carried out work on searching and identifying Galactic X-ray pulsars in HMXB systems by crossmatching various catalogs and databases. We created a catalog of X-ray pulsars with 82 sources, consisting of $18$~persistent pulsars and $64$~transients. The catalog contains individual information for each object: its period and its evolution, orbital period, X-ray luminosity in the indicated energy range, magnetic field strength, distance, and the characteristics of a massive component. We provide Table~\ref{tab:chartable1} with $18$~persistent pulsars in HMXBs, and Table~\ref{tab:chartable2} with $64$~transient pulsars in HMXBs, in a machine-readable format.

\begin{acknowledgments}
This research has been funded by the Science Committee of the Ministry of Education and Science of the Republic of Kazakhstan (grant~No.~AP09258811). We thank Dr.~Francis Fortin for the useful discussion on methods. I.I.~acknowledges the efficient training on VO tools and services given at the Second ESCAPE-VO School, funded by the European Union's Horizon 2020---grant No.~824064 (ESCAPE Project). The authors acknowledge the editor and anonymous referees for comments and suggestions that improved the presentation of the results.

This research has made use of data and/or software provided by the High Energy Astrophysics Science Archive Research Center (HEASARC), which is a service of the Astrophysics Science Division at NASA/GSFC. This research has made use of the SIMBAD database, operated at CDS, Strasbourg, France \citep{Wenger...2000AAS..143....9W}. This research has made use of NASA's Astrophysics Data System.
\end{acknowledgments}

\software{NumPy \citep{Harris...Nature2020...585...7825H}, pandas \citep{Mckinney...PPSC2010...56M}, Astropy \citep{astropy:2013, astropy:2018, astropy:2022},  TOPCAT \citep{Taylor...2005ASPC..347...29T}, Matplotlib \citep{Hunter...CSE2007...9...90}.}

\appendix

\begin{longrotatetable}
\begin{deluxetable*}{llccccccccc}
\tablecaption{Catalog of Galactic Population of Persistent Pulsars in HMXB Systems. Table is available in a machine-readable format (persistent\_table.mrt).}\label{tab:chartable1}
\tablewidth{700pt}
\setlength{\tabcolsep}{0.3em}
\tabletypesize{\tiny}
\tablehead{
\colhead{$\sharp$} & \colhead{Name} & \multicolumn{2}{c}{Eq. coord. J2000} &
\colhead{$P_{\rm s}$ (s)} & \colhead{$\dot{P} \, \text{(s/s)}$} & 
\colhead{$P_{\rm orb}$ (d)} & \colhead{$L_{\rm x}$ \,\text{erg/s}} & 
\colhead{$B_{\rm ns} (\times\,10^{12}\, \text{G})$} & \colhead{dist(kpc)} & 
\colhead{Comp. name} \\ 
\colhead{} & \colhead{} & \colhead{RA} & \colhead{DEC} & \colhead{Spin and its evol.} & \colhead{MJD} & 
\colhead{} & \colhead{range(keV)} & \colhead{} &
\colhead{} & \colhead{Spec. class and photom.}} 
\startdata
1 & 4U 1119-603 & 11:21:15.09 & -60:37:25.63 &  4.80188$\pm$85E-6  &   & 2.033$\pm$0.029    &  (0.058--1.4)E37 & 2.4--3  & 5.7$\pm$1.5   &  V779 Cen \\
 & (Cen X-3) & & &  (2016) [1] &  & [1] & 1--10 [2] & [3] & [4] & O6.5 II--III [5] \\
 &  & & & Loc.spin-down $\rightarrow$ & (1.25$\pm$0.07)E-10 [7] &  &  &  &  & B 14.37, V 13.30 [6]\\
 & & & & MJD $\rightarrow$ & 56834--56857 & &  & & &  J 10.736, H 10.311\\
 & & & & Loc.spin-up $\rightarrow$ & (-1.62$\pm$0.02)E-10 [7] & & & & & K 10.093 [6]\\
 & & & & MJD $\rightarrow$ & 56263--56315 & & & & & E(B-V) 1.40 [6]  \\
 & & & & Glob.spin-up $\rightarrow$ & (-2.26$\pm$0.01)E-11 [7--9]& & & & & \\
 & & & & MJD $\rightarrow$ & 40960--57220 & & & & & \\
\hline
2& OAO 1657-415 & 	17:00:48.88 & -41:39:21.46 &  37.024578$\pm$5E-6  & & 10.44729$\pm$21E-4 & (0.1--2)E37  & 3.29$\pm$0.23   & 6.4$\pm$1.5  & -- \\
 & & & & (2019) [10] & & [11] & 1--10 [2] & [2] & [12] & Ofpe/WNL [13] \\
 & & &  & Loc.spin-down $\rightarrow$ & (5.46$\pm$0.39)E-9 [14] & &  & & & J 14.09$\pm$0.05, H 11.68$\pm$0.04\\
 & & & & MJD $\rightarrow$ & 56188--56283 & & & & & K 10.38$\pm$0.04 [13a] \\
 & & & & Loc.spin-up $\rightarrow$ & (-8.46$\pm$0.13)E-9 [14] & & & & & \\
 & & & & MJD $\rightarrow$ & 55192--55365 & & & & & \\
 & & & & Glob.spin-up $\rightarrow$ & (-1.08$\pm$0.01)E-9 & & & & & \\
 & & & & & [8],[9],[14],[15] & & & & & \\
 & & & & MJD $\rightarrow$ & 43756--57250 & & & & & \\
\hline
3 &  IGR J22534+6243 & 	22:53:55.13 & 62:43:36.79 &  46.6784$\pm$37E-3 & & $>22$ [16] & $\sim$3E34   & -- & 9.7$\pm$1.7  &  -- \\
  & & & & (2009) [16] & &   & 17--60 [16] & & [16]  & B0--1 III--Ve [16] \\
 & & & & Loc.spin-down $\rightarrow$ & (1.33$\pm$0.09)E-9 [16]  & &  & & & B 17.38$\pm$0.1, V 15.78$\pm$0.09 [16]\\
 & & & & MJD $\rightarrow$ & 54937--54959 & & & & &  R 13.3, I 13.29 [13a]  \\
 & & & & Glob.spin-down $\rightarrow$ & $\sim$5.3E-10 [16]  & & & & & J 11.64$\pm$0.02, H 10.96$\pm$0.02 \\
 & & & & MJD $\rightarrow$ & 48988--55561 & & & & & K 10.46$\pm$0.03 [16]\\
\hline
4 & IGR J18027-2016 & 18:02:39.90 & -20:17:13.00  & 139.866$\pm$0.001 &  & 4.5696$\pm$9E-4 & (0.27--1.3)E37 & $\sim$3  & 12.4$\pm$0.1  & -- \\
& & & & (2015) [17]  & & [19] & 1--10 [2] & [17] & [18] & B1 Ib [18]\\
 & & & & Glob.spin-down $\rightarrow$ & $\sim$6E-10 [19]& &  & & & J 12.79$\pm$0.06, H 11.96$\pm$0.08 \\
 & & & & MJD $\rightarrow$ & 52640--57387 & & & & & K 11.48$\pm$0.05  [18]\\
& & & & & & & & & & E(B-V) 3.04$\pm$0.02 [18]\\
\hline
5 & SAX J1324.4-6200 & 	13:24:26.70 & -62:01:19.50 &  172.86$\pm$0.02 & & 1.1250$\pm$ 0.0417 & $\sim$1.1E34  & -- & $>$3.4 & --\\
  & & & & (2008) [20] & & [20a]  & 1--10 [21]  & & [21] & Be [20] \\
 & & & & Glob.spin-down $\rightarrow$ & (6.34$\pm$0.08)E-9 [20] & & & & & \\
 & & & & MJD $\rightarrow$ & 49353--54831 & & & & & \\
\hline
6 & 4U 0900-40  & 09:02:06.86 & -40:33:16.90 &  283.4290$\pm$6E-4 & &  8.964368$\pm$4E-5 & (0.0061--1)E37 & $\sim$2.6  & 1.9$\pm$0.2 & HD 77581 \\
  & (Vela X-1) & & & (2013) [25] & & [23] & 1--10 [2]  & [25] & [26] & B0.5 Ia [27] \\
  & & & & Loc.spin-down $\rightarrow$ & (1.74$\pm$0.16)E-8 [28]&  &  & & &  B 7.37, V 6.87 \\
  & & & & MJD $\rightarrow$ & 54770--54803 & & & & & R 6.31, I 6.05 [28a] \\
  & & & & Loc.spin-up $\rightarrow$ & (-2.65$\pm$0.26)E-8 [28]& & & & & J 5.833$\pm$0.020, H 5.705$\pm$0.034\\
  & & & & MJD $\rightarrow$ & 54806--54844 & & & & & K 5.596$\pm$0.024 [13a]\\
  & & & & Glob.spin-down $\rightarrow$ & (8.55$\pm$0.19)E-10 [8],[9] & & & & & E(B-V) 0.689$\pm$0.018 [28b] \\
  & & & & MJD $\rightarrow$ & 42449--51343 & & & & & \\
  & & & & Glob.spin-up $\rightarrow$ & (-3.34$\pm$0.13)E-10 [28]& & & & & \\
  & & & & MJD $\rightarrow$ & 54690--57217 & & & & & \\
\hline
7 & XTE J1855-026  & 18:55:30.41 & -02:36:16.74 &  360.741$\pm$0.002 & -- &  6.0724$\pm$9E-4  & (0.21--1.9)E37 & -- & $\sim$10  & --\\
  & & & & (2006) [30a] & & [30] & 1--10 [2] & & [29] & B0 Iaep [29a]  \\
 & & & & Loc.spin-down $\rightarrow$ & $\sim$3.7E-8 [30] & & & & & J 10.564$\pm$0.026, H 10.089$\pm$0.024 \\
 & & & & MJD $\rightarrow$ & ? & & & & & K 9.799$\pm$0.021 [13a] \\
\hline
8 & EXO 1722–363  & 17:25:11.39 & -36:16:57.53 &   413.7$\pm$0.3  & -- & 9.7403$\pm$4E-4 & (0.47--9.2)E36 & -- & 7.1--7.9 & --\\
 & & & & (2004) [31] & & [32] & 20--60 [33] & & [33] & B0--1 Ia [33] \\
 & & & & & & & & & & V 14.3$\pm$0.1 [33a]\\
  & & & & & & & & & & J 14.218$\pm$0.035, H 11.811$\pm$0.028\\
& & & & & & & & & & K 10.672$\pm$0.026 [13a]\\
\hline
9 & 4U 1538-52 & 15:42:23.36 & -52:23:09.58 &  526.41$\pm$0.07 &  &  3.72831$\pm$2E-5 & (0.42--4.3)E36 & 2.1--2.3  & 5$\pm$0.5  & QV Nor \\
  & & & & (2016) [33a] & & [33a] & 1--10 [2] & [34] & [2] & B0.2 Ia [36] \\
 & & & & Glob.spin-up $\rightarrow$ & (-7.54$\pm$0.16)E-9 [8],[9] & &  & & & B 16.3, V 14.5 [35] \\
 & & & & MJD $\rightarrow$ & 47223--51315 & & & & & J 10.358$\pm$0.025, H 9.910$\pm$0.023\\
 & & & & Glob.spin-down $\rightarrow$ & (4.79$\pm$0.12)E-9 [37]& & & & & K 9.677$\pm$0.022 [13a]\\
 & & & & MJD $\rightarrow$ & 54690--57220 & & & & & E(B-V) 2.10 [6]\\
\hline
10 & 4U 1223-624  & 12:26:37.56 & -62:46:13.26 & 672.51$\pm$0.05  & & 41.508$\pm$0.007  &   (0.023-3)E37  & 3.4--4.2  & 3.5$\pm$0.5  &  BP Cru \\
 & (GX 301-2) & & & (2019) [38] & & [39] & 1--10 [2] & [41] & [2] & B1.5 Ia [40] \\
 & & & & Loc.spin-down $\rightarrow$ & (1.80$\pm$0.33)E-7 [42]& & & & &  B $12.70 \pm 0.36$, V $10.66 \pm 0.07$  \\
 & & & & MJD $\rightarrow$ & 55094--55122 & & & & & [42a]\\
 & & & & Loc.spin-up $\rightarrow$ & (-1.26$\pm$0.08)E-6 [42]& & & & & J 6.717$\pm$0.021, H 6.077$\pm$0.029\\
 & & & & MJD $\rightarrow$ & 55354--55402 & & & & & K 5.672$\pm$0.017 [13a] \\
 & & & & Glob.spin-down $\rightarrow$ & (1.30$\pm$0.01)E-8  [9],[42]& & & & & E(B-V) 1.84 [6]\\
 & & & & MJD $\rightarrow$ & 49324--57222 & & & & & \\
 & & & & Glob.spin-up $\rightarrow$ & (-3.67$\pm$0.17)E-8 [8],[9] & & & & & \\
 & & & & MJD $\rightarrow$ & 42417--49200 & & & & & \\
\hline
11 & 4U 0352+309  & 03:55:23.08 & 31:02:45.04 &  835.29$\pm$0.29  &  & 250.3$\pm$0.6& (1.2-6.3)E34 & $\sim$5.6  & 0.801$\pm$0.138 &  HD 24534\\
& (X Per) & & & (2012) [43] & &  [44] & 1--10 [2] & [46] & [44a] & B0 Ve [45] \\
 &  & & & Loc.spin-down $\rightarrow$ & $\sim$1.4E-8 [47]& &  & & & B 6.840$\pm$0.007, V 6.720$\pm$0.009  \\
 & & & & MJD $\rightarrow$ & 43413--43532 & & & & &  [47a] \\
 & & & & Glob.spin-up $\rightarrow$ & (-3.98$\pm$0.03)E-9 [44] & & & & & J 6.149$\pm$0.018, H 6.073$\pm$0.024 \\
 & & & & MJD $\rightarrow$ & 42778--44969 & & & & &  K 5.920$\pm$0.016 [13a] \\
 & & & & Glob.spin-down $\rightarrow$ & $\sim$3.45E-9 [46] & & & & & E(B-V) 0.356$\pm$0.003 [47b]\\
 & & & & MJD $\rightarrow$ & 44000--50000 & & & & & \\
\hline
12 & 4U 1036-56 & 10:37:35.31 & -56:47:55.87 &  860$\pm$2  &  & 61$\pm$0.2 &  (2--4.5)E35 & -- & $\sim$5 &  LS 1698 \\
& & & & (1998) [48] & & [49]  & 3--30 [48] & &  [50] & B0 V--IIIe [50] \\
 & & & & Glob.spin-up $\rightarrow$ &  (-1.91$\pm$0.58)E-8 [51] & &  & & &  B 12.00$\pm$0.03, V 11.48$\pm$0.03 \\
 & & & & MJD $\rightarrow$ & 51179--55196 & & & & &  R 11.38$\pm$0.04 [51a] \\
& & & & & & & & & & J 10.122$\pm$0.024, H 9.954$\pm$0.022\\
& & & & & & & & & &  K 9.891$\pm$0.023 [13a]\\
& & & & & & & & & &  E(B-V) 0.75$\pm$0.25 [51b]\\
\hline
13 & Swift J2000.6+3210& 20:00:21.86  & 32:11:23.13 &  887.6$\pm$2.8  &  & --  & (2--4)E35  & -- & $\sim$ 8  &  -- \\
 & & & & (2006) [51c] & & & 0.3--10 [51c] &  & [51d] &  B V--III [51d]\\
 & & & & & & & &  & &  V 13.20$\pm$0.09 [51e]\\
 & & & & & & & & & &  R 16.1 [51d]\\
 & & & & & & & & & &  J 11.953$\pm$0.021, H 11.290$\pm$0.019 \\
 & & & & & & & & & &  K 10.847$\pm$0.018 [13a]\\
\hline
14 & IGR J16393-4643 & 16:39:05.50 & -46:42:14.00 &  904.0$\pm$0.1  &  & $\sim$4.2378472  & (5.2--9.9)E35  & 2.5$\pm$0.1 & $\sim 10$  &  -- \\
 & & & & (2014) [54] & & [54a] & 1--10 [2] & [54]  & [52] & Be V [55] \\
 & & & & Glob.spin-up $\rightarrow$ & (-1.08$\pm$0.03)E-8 [53] & & & & & \\
 & & & & MJD $\rightarrow$ & 53215--53815 & & & & & \\
\hline
15 & IGR J16493-4348 & 16:49:26.95 & -43:49:09.00 &  1093.1036$\pm$4E-4  & -- & 6.7828$\pm$4E-4 & (0.9--2.4)E36  & 3.7$\pm$0.4  &  16.1$\pm$1.5  & --\\
 & & & & (2019) [56] & & [56] & 1--10 [57] & [58] & [56] & B0.5 Ia [56] \\
 & & & & & & & & & & J 14.595$\pm$0.054, H 12.859$\pm$0.056 \\
 & & & & & & & & & & K 11.935$\pm$0.038 [13a]\\
\hline
16 & RX J0146.9+6121 & 01:47:00.21 & 61:21:23.66 & 1396.14$\pm$0.25  &  & $\sim$ 330 & (1--1.1)E35 &  -- &   2.3$\pm$0.5  & V831 Cas \\
 & & & & (2006) [60] & & [61]  & 1--10 [2] & & [62] &  B1 Ve [62a] \\
 & & & & Glob.spin-up $\rightarrow$ & (-4.6$\pm$0.2)E-8 [60] & & & & & B 12.09, V 11.42 \\
 & & & & MJD $\rightarrow$ & 45700--53370  & & & & & R 11.00, I 10.52 [62a] \\
  & & & & &  & & & & & J 9.899$\pm$0.027, H 9.700$\pm$0.036  \\
 & & & & &  & & & & & K 9.486$\pm$0.020 [13a]\\
  & & & & &  & & & & & E(B-V) 0.88$\pm$0.03 [62a]\\
\hline
17 & 4U 2206+54 & 22:07:56.24 & 54:31:06.41 & 5554$\pm$9  &  & 19.25$\pm$0.8  & (1.3--6.7)E35 & $\sim$ 3.3  & 3.0$\pm$0.7  &  BD+53 2790 \\
  & & & & (2010) [63] & & [64] & 1--10 [2] & [65] & [62a] &  O9.5Ve [62a] \\
 & & & & Glob.spin-down $\rightarrow$ & (5.24$\pm$0.93)E-7 [63] & & & & & B 10.11, V 9.84\\
 & & & & MJD $\rightarrow$ & 51141--54237   & & & & & R 9.64, I 9.43 [62a]\\
 & & & & & & & & & & J 9.218$\pm$0.025, H 9.116$\pm$0.027\\
 & & & & & & & & & &   K 9.038$\pm$0.022 [13a] \\
 & & & & & & & & & &  E(B-V) 0.51$\pm$0.03 [62a] \\
 & & & & & & & & & &   \\
\hline
18 & 2S 0114+650 & 01:18:02.70 & 65:17:29.83 & 9475$\pm$25  &  &   11.591$\pm$0.003  &  (0.067--1.2)E37  & $\sim$ 2.5  & 5.9$\pm$1.4  & V662 Cas\\
 & & & & (2008) [67] & & [68] & 1--10 [2]  & [69] & [70] & B1 Ia [70]\\
 & & & & Glob.spin-up $\rightarrow$ & $\sim$-8.9E-7 [69] & &  & & & B 12.17, V 11.03\\
 & & & & MJD $\rightarrow$ & 50451--53348   & & & & & R 10.33, I 9.58 [62a]\\
 & & & & & & & & & & J 8.597$\pm$0.026, H 8.296$\pm$0.034\\
 & & & & & & & & & & K 8.107$\pm$0.024 [13a] \\
 & & & & & & & & & & E(B-V) 1.33$\pm$0.04 [62a] \\
 & & & & & & & & & & \\
\enddata 
\end{deluxetable*}
\end{longrotatetable}

References for Table~\ref{tab:chartable1}: [1] - \citet{Shirke...2021JApA...42...58S}, [2] - \citet{Sidoli...2018MNRAS.481.2779S}, [3] -\citet{Santangelo...1998AandAP..340..L55}, [4] - \citet{Thompson...2009ApJ...691..P1744}, [5] - \citet{Hutchings...1979ApJ...229..P1079}, [6] - \citet{Liu...2006A&A...455.1165L}, [7] - \citet{Fermi_GBM_Cen_X_3}, [8] - \citet{Nagase...1989PasJ...396..P147}, [9] - \citet{Batse}, [10] - \citet{Falanga...2015AandAP..577..P1}, [11] - \citet{Jenke...2012ApJ...759..P124}, [12] - \citet{Chakrabarty...2002ApJ...573..P789}, [13] - \citet{Mason...2009A&A...505..281M}, [13a] - \citet{Cutri...2003yCat.2246....0C}, [14] - \citet{Fermi_GBM_OAO-1657}, [15] - \citet{Barnstedt...2008AandAP..486..P293}, [16] - \citet{Esposito...2013MNRAS..433..P2028}, [16a] - \citet{Fortin...2023A&A...671A.149F} [17] - \citet{Lutovinov...2017MNRAS.466..593L}, [18] - \citet{Torrejon...2010AandAP..510..P61}, [19] - \citet{Hill...2005A&A...439..255H}, [20] - \citet{Kaur...2009MNRAS..394..P1597}, [20a] - \citet{Lin...2002MNRAS.337.1245L}, [21] - \citet{Angelini...1998AandAP..339..L41}, [23] - \citet{Quaintrell...2003A&A...401..313Q}, [25] - \citet{Furst...2014ApJ...780..133F}, [26] - \citet{Sadakane...1985ApJ...288..284S}, [27] - \citet{Houk...1978mcts.book.....H}, [28] - \citet{Fermi_GBM_Vela_X_1}, [28a] - \citet{Ducati...2002yCat.2237....0D}, [28b] - \citet{Kretschmar...2021A&A...652A..95K}, [29] - \citet{Corbet...1999ApJ...517..956C}, [29a] - \citet{Negueruela...2008ATel.1876....1N}, [30] - \citet{Corbet...2002ApJ...577..923C}, [30a] - \citet{Bodaghee...2007A&A...467..585B}, [31] - \citet{Zurita...2006A&A...448..261Z}, [32] - \citet{Thompson...2007ApJ...661..447T}, [33] - \citet{Mason...2010A&A...509A..79M}, [33a] - \citet{Hemphill...2019ApJ...873...62H}, [34] - \citet{Hemphill...2013ApJ...777...61H}, [35] - \citet{Maccarone...2014MNRAS.440.1626M}, [36] - \citet{Parkes...1978MNRAS.184P..73P}, [37] - \citet{Fermi_GBM_4U_1538_52}, [38] - \citet{Nabizadeh...2019A&A...629A.101N}, [39] - \citet{Sato...1986ApJ...304..241S}, [40] - \citet{Kaper...1995A&A...300..446K}, [41] - \citet{Kreykenbohm...2004A&A...427..975K}, [42] - \citet{Fermi_GBM_GX_301_2}, [42a] - \citet{Hog...2000A&A...355L..27H},[43] - \citet{Maitra...2017MNRAS.470..713M}, [44] - \citet{Delgado-Mart...2001ApJ...546..455D}, [44a] - \citet{Bodaghee...2012ApJ...744..108B}, [45] - \citet{Lyubimkov...1997MNRAS.286..549L}, [46] - \citet{DiSalvo...1998ApJ...509..897D}, [47] - \citet{Ikhsanov...2014ARep...58..376I}, [47a] - \citet{Oja...1991A&AS...89..415O}, [47b] - \citet{Nikolov...2017BlgAJ..27...10N}, [48] - \citet{Reig...1999MNRAS.306..100R}, [49] - \citet{Cusumano...2013MNRAS.436L..74C}, [50] - \citet{Motch...1997A&A...323..853M}, [51] - \citet{LaPalombara...2009A&A...505..947L}, [51a] - \citet{Zacharias...2012yCat.1322....0Z}, [51b] - \citet{Sarty...2011RAA....11..947S}, [51c] - \citet{Pradhan...2013MNRAS.436..945P}, [51d] - \citet{Masetti...2008A&A...482..113M}, [51e] - \citet{Alfonso...2012A&A...548A..79A}, [52] - \citet{Bodaghee...2006A&A...447.1027B}, [53] - \citet{Thompson...2006ApJ...649..373T}, [54] - \citet{Bodaghee...2016ApJ...823..146B}, [54a] - \citet{Nazma...2015MNRAS.446.4148I}, [55] - \citet{Chaty...2008ChJAS...8..197C}, [56] - \citet{Pearlman...2019ApJ...873...86P}, [57] - \citet{Coley...2019ApJ...879...34C}, [58] - \citet{DAi...2011A&A...532A..73D}, [60] - \citet{LaPalombara...2006A&A...455..283L}, [61] - \citet{Sarty...2009MNRAS.392.1242S}, [62] - \citet{Reig...1997A&A...322..183R}, [62a] - \citet{Reig...2015A&A...574A..33R}, [63] - \citet{Finger...2010ApJ...709.1249F}, [64] - \citet{Corbet...2007ApJ...655..458C}, [65] - \citet{Torrejon...2004A&A...423..301T}, [67] - \citet{Wang...2011MNRAS.413.1083W}, [68] - \citet{Corbet...2013ApJ...778...45C}, [69] - \citet{Bonning...2005A&A...436L..31B}, [70] - \citet{Reig...1996rftu.proc..181R}

\begin{longrotatetable}
\begin{deluxetable*}{llccccccccc}
\tablecaption{Catalog of Galactic population of \textbf{transient} pulsars in High-mass X-ray binary systems. Table is available in a machine-readable format (transient\_table.mrt).}\label{tab:chartable2}
\tablewidth{700pt}
\setlength{\tabcolsep}{0.3em}
\tabletypesize{\tiny}
\tablehead{
\colhead{$\sharp$} & \colhead{Name} & \multicolumn{2}{c}{Eq. coord. J2000} &
\colhead{$P_{\rm s}$ (s)} & \colhead{$\dot{P} \, \text{(s/s)}$} & 
\colhead{$P_{\rm orb}$ (d)} & \colhead{$L_{\rm x}$ \,\text{erg/s}} & 
\colhead{$B_{\rm ns} (\times\,10^{12}\, \text{G})$} & \colhead{dist(kpc)} & 
\colhead{Comp. name} \\ 
\colhead{} & \colhead{} & \colhead{RA} & \colhead{DEC} & \colhead{Spin and its evol.} & \colhead{MJD} & 
\colhead{} & \colhead{range(keV)} & \colhead{} &
\colhead{} & \colhead{Spec. class and photom.}} 
\startdata
\hline
1 & SAX J0635.2+0533  & 06:35:17.40 & 05:33:21.00 &  0.0338565$\pm$1E-7   &  & 11.2$\pm$0.5 & (3--27.6)E32 & -- & 2.5--5  & -- \\
 & & & & (1997) [71] & & [72] & 0.2--12 [74] & & [73] & B1--2 IIIe--Ve [75]\\
 & & & & Loc.spin-down $\rightarrow$ & $\sim$3.8E-13
[72] & &  & & & B 13.81, V 12.83 \\
 & & & & MJD $\rightarrow$ & 51417--51420 & & & & & R 11.98 [73]\\
  & & & & & & & & & & E(B-V) 1.2$\pm$0.2 [73]\\
\hline 
\hline
2 & 4U 1901+03  & 19:03:39.39 & 03:12:15.76 & 2.761$\pm$0.001  & & 22.5827$\pm$2E-4
& (0.27--1.9)E37 & -- & $>$10  & --\\
 & & & & (2019) [76] & & [77] & 1--10 [2] & & [77] & Be [79] \\
 & & & & Loc.spin-up $\rightarrow$ & (-2.272$\pm$3E-4)E-10 [77] & &  & & & J 14.138$\pm$0.028, H 13.220$\pm$0.033\\
 & & & & MJD $\rightarrow$ & 52680--52837 & &  & & & K 12.581$\pm$0.028 [13a]\\
 & & & & Glob.spin-down $\rightarrow$ & $\sim$3.21E-12 [78] & & & & & \\
 & & & & MJD $\rightarrow$ & 55914--58522 & &  & & & \\
\hline
3 & 4U 0115+634 & 	01:18:31.97 & 63:44:33.07 & 3.61398$\pm$2E-5 & &  24.31643$\pm$7E-5 &   (0.099--3.7)E37  & $\sim$1  & 6$\pm$1.5 & V635 Cas \\
 & & & & (2017) [80] & & [81] & 1--10 [2]  & [82] & [62a] & B0.2 Ve [83] \\
 & & & & Glob. spin-up $\rightarrow$ & (-8.36$\pm$0.04)E-11 [84] & &  & & & B 16.92, V 11.03 \\
 & & & & MJD $\rightarrow$ & 44605--54831 & &  & & & R 14.34, I 13.22 [62a]\\
 & & & & Loc. spin-up $\rightarrow$ & (-3.02$\pm$0.01)E-10 [84] &  & & & & E(B-V) 1.71$\pm$0.05 [62a]\\
 & & & & MJD $\rightarrow$ & 54555--54566 & &  & & & \\
\hline
4 & V 0332+53 & 03:31:14.87 & 53:00:24.20 &   4.3748$\pm$9E-5  & &  34.67$\pm$0.38  &  (0.094--6.4)E37 & $\sim$2.5   &  6$\pm1.5$  &  BQ Cam\\
 & & & & (2005) [86] & & [86] & 1--10 [2] & [87] & [62a] & O8.5 Ve [88]  \\
 & & & & Loc. spin-up $\rightarrow$ & (-4.05$\pm$0.11)E-10  [86] & &  & & & B 17.16, V 15.42\\
 & & & & MJD $\rightarrow$ & 53367--53392 & &  & & & R 14.26, I 13.04 [62a]\\
 & & & & & & &  & & & J 11.817$\pm$0.023, H 11.21$\pm$0.03\\
 & & & & & & &  & & & K 10.744$\pm$0.025 [13a]\\
 & & & & & & &  & & & E(B-V) 1.94$\pm$0.03 [62a]\\
\hline
5 &  GRO J1750-27 & 17:49:12.97 & -26:38:38.93 & 4.451271$\pm$2E-6  &  & 29.817$\pm$0.009 & (0.78--8.5)E37  & $\sim$3.7 & $>$12 & -- \\
 & & & & (2022) [90] & & [91] & 1--10 [2] & [90] & [91a] & Be [91] \\
 & & & & Loc. spin-up $\rightarrow$ &  (-7.5$\pm$0.3)E-10 [92] & &  & & & \\
 & & & & MJD $\rightarrow$ & 54507--54576 & &  & & & \\
\hline
6 & AX J1841.0-0536 & 18:41:00.43 & -05:35:46.50 & 4.7394$\pm$8E-4 & -- & $\sim$25 & (0.023--1.1)E36  & -- & 3.2$\pm$2  &  --\\
 & & & & (1999) [93] & & [93] & 2--10 [93] & & [94] & B1 Ib [94]\\
  & & & & & & & & & & K 8.9 [94] \\
\hline
7 & XTE J1829-098 & 18:29:44.01 & -09:51:23.00 & 7.8448$\pm$2E-5 & -- & 244.2$\pm$0.2 & 3E32--3E36   & $\sim$1.7 & 4.5--18  &  -- \\
 & & & & (2018) [95]& & [95a] & 2--10 [96] & [95] & [97] & B0 V--B0.5 Ib  [97] \\
 & & & & & & & & & & J 16.147$\pm$0.009, H 14.064$\pm$0.003 \\
 & & & & & & & & & & K 12.591$\pm$0.002  [96a]\\
\hline
8 & 2S 1553-542 & 15:57:49.00 & -54:24:54.00 & 9.27949$\pm$8E-6 & & 31.303$\pm$0.027 & 1.1E34--3.2E37 & $\sim$3 & $>$15  &  -- \\
 & & & & (2021) [97a] & & [98] & 0.5--10 [97b] & [98] & [97b] & B1--2 V [97b] \\
 & & & & Loc. spin-up $\rightarrow$  & (-2.54$\pm$0.83)E-9 [97c] & &  & & & J 15.78$\pm$0.08, H 14.46$\pm$0.14 \\
 & & & & MJD $\rightarrow$ & 54480--54540 & & & & & K 13.45$\pm$0.10 [97b]\\
 & & & & Glob. spin-down $\rightarrow$  & $\sim$4E-11 [98] & &  & & & \\
 & & & & MJD $\rightarrow$ & 54101--57387 & & & & & \\
\hline
9 & XTE J1859+083 & 18:59:01.57 & 08:14:44.20 & 9.79156$\pm$1E-5 & & 60.65$\pm$0.08  & -- & -- & 6.1--8.7 &  --\\
 & & & & (2021) [98a] & & [98b] &  & &  [98a] & Be [98a]\\
 & & & & Loc. spin-up $\rightarrow$  & $\sim$-1.53E-9 [78] & & & & & \\
 & & & & MJD $\rightarrow$ & 57048--57149 & & & & & \\
\hline
10 & Swift J0243.6+6124 & 02:43:40.43 & 61:26:03.76 & 9.8661$\pm$3E-4  & & 28.3$\pm$0.2  & 6E33--2E39 & -- & 5.5$\pm$0.4  &  -- \\
  & & & & (2017) [98c] & & [98d] & 0.1-10 [98e, 98f] & & [98g] & O9.5Ve [98h] \\
 & & & & Loc. spin-up $\rightarrow$ & $\sim$-2.14E-8 [98d]& &  & & & B 13.83$\pm$0.03, V 12.86$\pm$0.01\\
  & & & & MJD $\rightarrow$ & 58027--58084 & & & & & R 12.18$\pm$0.01, I  11.45$\pm$0.02 [98h]\\
   & & & & Glob. spin-down $\rightarrow$ & $\sim$1.94E-10 [78]& &  & & & J 10.589$\pm$0.023, H 10.208$\pm$0.021 \\
  & & & & MJD $\rightarrow$ & 58161--58468 & & & & & K 9.961$\pm$0.018 [13a]\\
  & & & & & & & & & & E(B-V) 1.24$\pm$0.02 [98h]\\
\hline
11 & IGR J17544-2619 & 17:54:25.27 & -26:19:52.58 & 11.58$\pm$0.03 &  & 4.926$\pm$0.001  & (0.2--5.3)E36 & 1.45$\pm$0.03  & 3$\pm$0.2  &  -- \\
 & & & & (2015) [99]  & & [100] & 1--10 [2] & [101] & [2] & O9 Ib [102] \\
  & & & &  & & & & & & B  14.62$\pm$0.05,  V 12.89$\pm$0.05\\
  & & & &  & & & & & & R  11.76$\pm$0.05, I  10.39$\pm$0.05 [102a]\\
 & & & &  & & & & & & J 8.791$\pm$0.021, H 8.310$\pm$0.031\\
  & & & &  & & & & & & K 8.018$\pm$0.026 [13a]\\
\hline
12 & IGR J18179-1621 & 18:17:52.18 & -16:21:31.68 & 11.8268$\pm$1E-4   &  & -- & $<$1E37 & $\sim$1.7  & $<$ 8  & -- \\
  & & & & (2020) [102b] & & & 1.5--50 [102c] & [103] &  [102c] & O--B  [103] \\
    & & & & & & & & & & H 16$\pm$0.1, K 13.14$\pm$0.04 [103]\\
\hline
13 & GS 0834-430 & 08:35:54.00 & -43:11:17.50 & 12.294$\pm$0.034  &  & 105.8$\pm$0.4 & (0.1--5.5)E35  & -- & 5 &   \\
 & & & & (2012) [104] & & [105] & 0.1--2.4 [106] & & [106] & B0--2 III--Ve \\
 & & & & Loc. spin-up $\rightarrow$ & (-1.66$\pm$0.45)E-9 [104]& &  & & & [106] \\
 & & & & MJD $\rightarrow$ & 56106--56110 & & & & & \\
\hline
14 &  IGR J19294+1816 & 19:29:55.91 & 18:18:38.25 & 12.485065$\pm$15E-6  &  & 116.93$\pm$0.05  & (0.067--3.4)E36 & $\sim$5  & 11$\pm$1 & -- \\
 & & & & (2019) [107] & & [108] & 3--79 [109] & [109] & [109a] &  B1 Ve [109a] \\
 & & & & Loc. spin-up $\rightarrow$ & (-3.53$\pm$0.46)E-9 [109b] & &  & & & J 14.56$\pm$0.03, H 12.99$\pm$0.03\\
 & & & & MJD $\rightarrow$ & 58767--58793 & & & & & K 12.115$\pm$0.023 [109a]\\
 & & & & Glob. spin-down $\rightarrow$  & (1.43$\pm$0.04)E-10 [109b] & &  & & & \\
 & & & & MJD $\rightarrow$ & 56377--59349 & & & & & \\
\hline
15 & Swift J1626.6-5156 & 16:26:36.52 & -51:56:30.60 & 15.3714$\pm$3E-4  & & 132.89$\pm$0.03  & $\sim$ 3.4E35 & $\sim$ 0.86  & 10.7$\pm$3.5 & -- \\
 & & & & (2005) [110] & & [113] &  2--60 [112] & [112] & [113] & B0 Ve [113] \\
 & & & & Glob. spin-up $\rightarrow$ & (-3.09$\pm$0.05)E-10 [111] & & & & & B 16.81, V  15.54\\
 & & & & MJD $\rightarrow$ & 53724--54424 & &  & & & R 15.81, I 14.27 [113]\\
  & & & & & & &  & & & J 13.436$\pm$0.024, H 12.948$\pm$0.025\\
  & & & & & & &  & & & K 12.537$\pm$0.023 [13a]\\
\hline
16 & XTE J1946+274  & 19:45:39.36 & 27:21:55.50 &  15.757119$\pm$86E-6  &  & 169.2$\pm$0.9  & (0.5--5)E37  & 3.1$\pm$0.1  & 7$\pm$2 &  -- \\
 & & & & (2021) [114] & & [115] & 3--60 [117] & [117] & [62a] & B0--B1 IV--Ve [115] \\
 & & & & Loc. spin-up $\rightarrow$ & (-3.0$\pm$0.3)E-9 [116] & &  & & &  B 18.76, V 16.92\\
 & & & & MJD $\rightarrow$ & 55367--55377 & & & & & R 15.62, I 14.38 [62a]\\
  & & & & & & & & & & J 12.539$\pm$0.036, H 11.831$\pm$0.035\\
  & & & & & & & & & & K 11.333$\pm$0.024 [13a]\\
    & & & & & & & & & & E(B-V) 1.18$\pm$0.04 [62a]\\
\hline
17 & 2S 1417-624 & 14:21:12.80 & -62:41:54.00 &  17.544701$\pm$8E-6 & & 42.12$\pm$0.03 &  1.4E34--1.3E36 & -- & 7.4$\pm$3  &   -- \\
 & & & & (1999) [118] & & [119] & 3--20 [120] &  & [16a] & B1 Ve [121] \\
 & & & & Glob. spin-up $\rightarrow$ & $\sim$-9.35E-9 [119] & & & & & B 18.64, V 16.92 [121a]\\
 & & & & MJD $\rightarrow$ & 49600--49900 & &  & & & \\
\hline
18 & KS 1947+300 & 19:49:35.48 & 30:12:31.78 &  18.78896$\pm$7E-5  &  & 40.415$\pm$0.01  & (0.17--1.7)E36 & $\sim$1.1  &  8$\pm$2 & --\\
 & & & & (2013) [122] & & [123] & 1--10 [2] &  [124]& [62a] & B0 Ve [125] \\
 & & & & Glob. spin-down $\rightarrow$ & $\sim$2.89E-10 [123] & &  & & & B 15.16, V 14.22\\
 & & & & MJD $\rightarrow$ & 49353--51909  & &  & & & R 13.53, I 12.88 [62a]\\
 & & & & Loc. spin-up $\rightarrow$ &  (-6.88$\pm$0.13)E-9 [126] & &  & & & J 11.831$\pm$0.021, H 11.434$\pm$0.024\\
 & & & & MJD $\rightarrow$ & 56567--56671 & &  & & & K 11.054$\pm$0.020 [13a]\\
 & & & & & & &  & & & E(B-V) 2.01$\pm$0.05 [62a]\\
\hline
19 & IGR J18483-0311  & 18:48:17.21 & -03:10:16.87 &  21.0526$\pm$5E-4  & -- & 18.52$\pm$0.01  & 1.6E34--7.8E36 & -- & 3--4 & -- \\
 & & & & (2007) [127] & & [128] & 20--100[127] & & [128] & B0.5a (SG) [128] \\
 & & & & & & & & & & B $>$25.162, V 21.884$\pm$0.313\\
  & & & & & & & & & & R 17.888$\pm$0.041, I  14.382$\pm$0.061 \\
  & & & & & & & & & & J 10.840$\pm$0.028, H 9.376$\pm$0.018 \\
  & & & & & & & & & & K 8.472$\pm$0.023 [128]\\
\hline
20  & XTE J1543-568  & 15:44:01.90 & -56:42:43.00 &  27.12156$\pm$59E-4 & & 75.56$\pm$0.25 & (2.1--3.3)E36& -- & $\sim$10 & -- \\
 & & & & (2001) [129] & & [129] &  1--10 [2] & & [129] & B3 ? [129]\\
 & & & & Glob. spin-up $\rightarrow$ & (-2.84$\pm$0.02)E-9 [129] & & & & & \\
 & & & & MJD $\rightarrow$ & 51550--51950 & &  & & & \\
\hline
21 & GS 1843+00 & 18:45:36.84 & 00:51:47.44 & 29.477$\pm$0.001 & & 50--60 & 3E37 & -- & $\geq$10 &  --\\
 & & & & (1997)  [130] & & [130] & 0.3--100 [130] & & [130a] & B0--2 IV--Ve [130a]\\
 & & & & Loc. spin-up $\rightarrow$ & (-3.79$\pm$0.10)E-8 [130] & & & & & B 24.1, V 20.89$\pm$0.05\\
 & & & & MJD $\rightarrow$ & 50500--50540 & &  & & & R 18.80$\pm$0.05, I 16.79$\pm$0.05 [130a]\\
 & & & & & & &  & & & J 13.790$\pm$0.035, H 12.820$\pm$0.035\\
 & & & & & & &  & & & K 12.179$\pm$0.037 [13a]\\
 & & & & & & &  & & & E(B-V) $\sim$ 2.8 [130a]\\
\hline
22 & RX J0812.4-3114 & 08:12:28.36 & -31:14:52.14 & 31.908$\pm$0.009  &  &  $\sim$81.3  & $\sim$2.3E36  & -- & 6.76$\pm$1.2  &  V572 Pup\\
 & & & & (2018) [131] & & [132] & 3--30 [132] & & [131] & B0.5 V--IIIe [50]\\
 & & & & & & & & & & B 13.03$\pm$0.02, V 12.74$\pm$0.03\\
 & & & & & & & & & & R 12.44$\pm$0.03 [51a]  \\
 & & & & & & & & & & J 11.335$\pm$0.022, H 11.126$\pm$0.022 \\
 & & & & & & & & & & K 10.909$\pm$0.023 [13a]\\
\hline
23 & EXO 2030+375 & 20:32:15.27 & 37:38:14.84 & 41.4106$\pm$1E-4 &  &  46.0205$\pm$2E-4 & (0.092--6.4)E37  & $\sim$3.1  & 3$\pm$0.2  &  V2246 Cyg \\
 & & & & (2013) [133]  & & [134] & 1--10 [2] & [135] & [62a] & B0 Ve [136] \\
 & & & & Loc. spin-up $\rightarrow$ & (-3.11$\pm$0.01)E-8 [134] & & & & & B 22.16, V 19.41\\
 & & & & MJD $\rightarrow$ &  53948--53953 & &  & & & R 17.32, I 15.18 [62a]\\
& & & & &  & &  & & & J 12.050$\pm$0.020, H 10.835$\pm$0.018 \\
& & & & &  & &  & & & K 10.074$\pm$0.016 [13a]\\
& & & & &  & &  & & & E(B-V) 3$\pm$0.2 [62a]\\
\hline
24 & IGR J18219-1347  & 18:21:54.82 & -13:47:26.70 & 52.468$\pm$3E-4 & - & 72.44$\pm$0.3 & (2--5)E36  & -- & 10--15  &  -- \\
 & & & & (2020) [136a] & & [136b]  & 3--79 [136a] & & [136a] &   B0--2 [136a] \\
  & & & & & & & & & &  J 21.3$\pm$0.4, H 18.62$\pm$0.07 \\
 & & & & & & & & & & K 16.93$\pm$0.03 [136a]\\
\hline
 25 & AX J1700.2-4220  & 17:00:25.24 & -42:19:00.28 & 54.22$\pm$0.03  &  & 44.12$\pm$0.04  & $\sim$1E36  & -- & 1.7--2.6  &  HD 153295 \\
 & & & & (2010) [137] & & [138] & 2--50 [137] & & [139] & B0.5 IVe [139] \\
  & & & & & & & & & & B 9.05, R 9.52 [139a]\\
  & & & & & & & & & & J 7.445$\pm$0.027, H 7.093$\pm$0.036 \\
  & & & & & & & & & & K 6.729$\pm$0.016 [13a]\\
\hline
26 & GS 2138+56  & 21:39:30.69 & 56:59:10.42 & 65.3529$\pm$17E-4  & & 23--143 & (0.27--1.4)E36  & 3.4$\pm$0.2  & 7.5$\pm$0.6  &  V490 Cep \\
 & (Cep X-4) & & & (2018) [140] & & [141] & 2--10 [142] & [142] & [143] & B1--2 Ve [143] \\
 & & & & Loc. spin-up $\rightarrow$ & (-1.06$\pm$0.08)E-8 [145]& &  & & & J 11.829$\pm$0.024, H 11.414$\pm$0.028\\
 & & & & MJD $\rightarrow$ & 56813--56831& & & & & K 10.926$\pm$0.020 [13a] \\
 & & & & Glob. spin-down $\rightarrow$ & (1.92$\pm$0.04)E-10 [141] & & & & & \\
 & & & & MJD $\rightarrow$ & 49163--50641 & & & & & \\
\hline
27 & XTE J1906+090 & 19:04:47.48 & 09:02:41.80 & 89.66$\pm$0.03 & & 26--30  & (0.07--7)E36  & -- & $\geq$ 10 & -  \\
 & & & & (2003) [145a]  & & [145b] & 2--10 [145a] & & [145c] & Be [145a]\\
  & & & & & & & & & & J 15.18$\pm$0.07, H 14.17$\pm$0.11\\
    & & & & & & & & & & K 13.50$\pm$0.06.\\
\hline
28 & GRO J1008-57  & 10:09:46.96 & -58:17:35.60 & 93.283$\pm$0.001 & & 249.48$\pm$0.04  & (0.56--9.8)E36  & $\sim$7.8  & 9.7$\pm$0.8  & --\\
 & & & &  (2017) [146] & & [147] & 1--10 [2] & [146] & [150] & B0 III--Ve [150] \\
 & & & & Loc. spin-up $\rightarrow$ & (-4.45$\pm$0.04)E-8  [151] & &  & & & B 16.74, V 15.08\\
 & & & & MJD $\rightarrow$ & 56993--57032 & & & & & R  13.77, I 12.68 [150]\\
 & & & & Glob. spin-down $\rightarrow$ & (4.75$\pm$1.04)E-10 [152] & & & & & J 10.943$\pm$0.024, H 10.272$\pm$0.023\\
 & & & & MJD $\rightarrow$& 48988--55196 & & & & & K 9.718$\pm$0.021 [13a]\\
 & & & & Glob. spin-up $\rightarrow$ & $\sim$-2.31E-9 [152] & & & & & E(B-V)  1.96$\pm$0.03 [150]\\
 & & & & MJD $\rightarrow$& 55562--56292 & & & & & \\
\hline
29 & GS 1843-024 & 	18:48:17.70 & 56:59:10.42 & 94.7171$\pm$3E-4  & &  242.18$\pm$0.01  & $\sim$1E34 & -- & 10--15 & -- \\
 & & & & (2017) [153] & & [154] & 0.3--10 [153]& & [153],[154] & O--B (SG) [153]\\
 & & & & Glob. spin-up $\rightarrow$ & $\sim$-2.44E-9 [154] & & & & & H 17.82$\pm$0.04, K 15.52$\pm$0.03 [153]\\
 & & & & MJD $\rightarrow$ & 48362--50600 & & & & & \\
 & & & & Glob. spin-down $\rightarrow$ & $\sim$2.16E-9 [78] & & & & & \\
 & & & & MJD $\rightarrow$& 56154--58103 & & & & & \\
\hline
30 & 3A 0726-260 & 	07:28:53.58 & -26:06:28.88 &  103.145$\pm$0.001 &  & 34.548$\pm$0.01 & $\sim$2.8E35&  -- & 7$\pm$0.5  &  V441 Pup \\
 & & & & (2016) [155] & & [155a] & 2--20  [156] & & [150] & O8.5 Ve [150] \\
 & & & & & & & & & & B  12.12, V  11.61\\
 & & & & & & & & & & R 11.12, I  10.88 [150]\\
 & & & & & & & & & & J 10.366$\pm$0.029, H 10.093$\pm$0.024\\
 & & & & & & & & & & K 9.832$\pm$0.023 [13a]\\
 & & & & & & & & & & E(B-V) 0.83$\pm$0.03 [150]\\
\hline
31 & 1A 0535+262 & 05:38:54.57 & 26:18:56.84 & 103.24$\pm$0.02 &  & 110$\pm$0.5 & (0.0099--1.5)E37 & $\sim$4.3  & 2.1$\pm$0.5 & V725 Tau \\
 & & & & (2020) [157] & & [158] & 1--10 [2] & [159] & [62a] & O9.7 IIIe [161] \\
 & & & & Loc. spin-up $\rightarrow$  &  (-6.73$\pm$0.23)E-8 [162]& &  & & & B 9.74, V 10.73\\
 & & & & MJD $\rightarrow$ & 55161--55197 & & & & & R 10.28, I 9.76 [62a]\\
 & & & & Glob. spin-up $\rightarrow$ & $\sim$-1.6E-9 [163] & & & & & J 8.368$\pm$0.021, H 8.272$\pm$0.026\\
 & & & & MJD $\rightarrow$ & 42531--47624 & & & & & K 8.157$\pm$0.027 [13a]\\
 & & & & Glob. spin-down $\rightarrow$ & $\sim$1.5E-9 [164] & & & & & E(B-V) 0.77$\pm$0.04 [62a]\\
 & & & & MJD $\rightarrow$ & 49353--53735 & & & & & \\
\hline
32 & AX J1749.1-2733  & 17:49:06.79 & -27:32:32.40 & 131.95$\pm$0.24 &  & 185.5$\pm$1.1 & (0.41--1.3)E37 & - & 16$\pm$3.5 &   -- \\
 & & & & (2007) [164a]& & [164b]  & 1--10 [2] & &  [164c] &   B1--2 [164c] \\
  & & & & &  & & & & & J $>$18.7, H 17.43$\pm$0.14 \\
  & & & & &  & & & & & K 15.18$\pm$0.03 [164c] \\
 \hline
 33 & Swift J1816.7-1613  & 18:16:42.66 & -16:13:23.40 & 143.6863$\pm$0.0002 & - & 118.5$\pm$0.8 & (0.01--5.5)E36 & -- & 7--13  &  --\\
   & & & & (2017) [164d] & & [164e]  & 3--79 [164d] & &  [164d] &  B0--2 e [164d]  \\
  & & & & Loc. spin-up $\rightarrow$ & $\sim$-5.93E-7 [164f] & & & & &  H 17.56$\pm$0.1, K 14.85$\pm$0.02 [164d]\\
  & & & & MJD $\rightarrow$ & 54554--54563 & & & & &  \\
 \hline
34 & AX J1820.5–1434  & 18:20:29.50 & -14:34:24.00 & 152.26$\pm$0.04 & & 54$\pm$0.4  & (1.7--3.8)E36 & -- & 8.2$\pm$3.5  & --\\
 & & & & (1998) [165] & & [166] & 1--10 [2] & & [44a] & B0 III--V [166]\\
 & & & & spin-down $\rightarrow$ & (3.0$\pm$0.14)E-9 [166] & & & & & J 15.41, H 13.25 \\
 & & & & MJD $\rightarrow$ & ? & & & & & K 11.75 [166]\\
\hline
35 & MXB 0656-072 & 06:58:17.29 & -07:12:35.20 & 160.4$\pm$0.4  &  & $\sim$101.2  & $\sim$6.6E36  & 3.67$\pm$0.06  & 5.7$\pm$0.5  & \\
 & & & & (2003) [167] & & [168] & 2--10 [167] & [167] & [143] & O9.5 Ve [170] \\
 & & & & Loc. spin-up $\rightarrow$ & (-1.12$\pm$0.04)E-7 [167] & &  & & & B 13.25$\pm$0.02, V 12.25$\pm$0.02\\
 & & & & MJD $\rightarrow$ & 52930--52971 & & & & & R 11.63$\pm$0.02, I 10.97$\pm$0.02 [170]\\
  & & & &  &  & & & & & J 9.664$\pm$0.026, H 9.332$\pm$0.025\\
    & & & &  &  & & & & & K 9.013$\pm$0.024 [13a]\\
\hline
36 & IGR J11435-6109 & 11:44:00.29 & -61:07:36.48 & 161.76$\pm$0.01   &  & 52.46$\pm$0.06  &  (0.99--2.1)E36  & -- & $\sim$8.6 & --\\
 & & & & (2004) [170a] & & [172] & 1--10 [2] & & [172a] & B0 V -- B2 III [172a] \\
  & & & & & & & & & & J 13.003$\pm$0.022, H 12.338$\pm$0.021\\
  & & & & & & & & & & K 11.852$\pm$0.019 [13a]\\
\hline
37 & IGR J11215-5952 & 11:21:46.82 & -59:51:47.97 & 187$\pm$0.12 &  & $\sim$164.6  & (0.86--5.1)E36 & -- & 7$\pm$1  & HD 306414 \\
 & & & & (2017) [173] & & [174] & 1--10 [2] & & [176] & B0.5 Ia [176] \\
  & & & & & & & & & & B 10.69$\pm$0.01, V 10.23$\pm$0.01\\
  & & & & & & & & & & R 10.00$\pm$0.01 [51a]\\
  & & & & & & & & & & J 8.548$\pm$0.030, H 8.340$\pm$0.036\\
  & & & & & & & & & & K 8.185$\pm$0.023 [13a]\\
  & & & & & & & & & & E(B-V) 0.7 [176]\\
\hline
38 & 1H 1238-599 & 	12:42:01.70 & -60:12:06.00 &  191.196$\pm$0.084  & &  -- & --  &  -- & -- & -- \\
 & & & & (1976) [177] & & & & & & \\
\hline
39 & GRO J2058+42 & 20:58:47.54 & 41:46:37.18 &  194.2201$\pm$0.0016   &  &  55?  110?  & 1E34--4.2E37   &  1--2  & 9$\pm$2.5 & -- \\
 & & & & (2019) [179]& & [180] [181] & 3--80 [179][182] & [182] &  [62a] & O9.5--B0 IV--Ve [62a] \\
 & & & & Loc. spin-up $\rightarrow$ & (-9.52$\pm$0.08)E-7 [180] & &  & & & B 16.04, V 14.89\\
 & & & & MJD $\rightarrow$ & 49974--50020 & &  & & & R 14.16, I 13.35 [62a]\\
  & & & & &  & &  & & & J 11.740$\pm$0.022, H 11.282$\pm$0.018 \\
   & & & & &  & &  & & & K 10.930$\pm$0.017 [13a] \\
    & & & & &  & &  & & & E(B-V) 1.37$\pm$0.03 [62a]\\
\hline
40 & RX J0440.9+4431 & 	04:40:59.33 & 44:31:49.26 & 206.01$\pm$0.47 &  & 150$\pm$0.2  & (0.04--7.1)E36 & $\sim$3.2 & 2.2$\pm$0.5 &  LS V +44 17\\
   & & & & (2011) [183] &  & [183] & [184] & 3--100 &  [62a] & B0.2 Ve  [62a] \\
   & & & & Glob. spin-down $\rightarrow$ & (6.4$\pm$1.3)E-9 [186] & & [184]  & & & B 11.42, V 10.73\\
 & & & & MJD $\rightarrow$ & 51179--56292 & & & & & R 10.28, I 9.76 [62a]\\
 & & & & Loc. spin-up $\rightarrow$ & (-1.11$\pm$0.33)E-7 [186a] & & & & & J 9.500$\pm$0.023, H 9.317$\pm$0.030\\
 & & & & MJD $\rightarrow$ & 55706--55714 & & & & & K 9.182$\pm$0.018 [13a]\\
  & & & & & & & & & & E(B-V) 0.91$\pm$0.03 [62a]\\
\hline
41 & Swift J1845.7–0037 & 18:45:54.62 & -00:39:34.20 & 207.379$\pm$0.002 &   & -- & $\sim$2.4E35 & -- & 10?  &  -- \\
 & & & & (2019) [187] & & & 0.1--100 [187] & & [187] & Be [187a]\\
\hline
42 & XTE J1858+034  & 18:58:36.00 & 03:26:09.00 & 218.382$\pm$0.002  &  & $\sim$380  & (0.17--1.5)E37  & $\sim$5.2 & $\sim$6  &  -- \\
 & & & & (2019) [187b] & & [188] & 1--10 [2] & [187b] & [188] & Be [203] \\
 & & & & Loc. spin-up $\rightarrow$ & $\sim$-4.40E-7
[188] & &  & & & B 19.61$\pm$0.02, V 18.00$\pm$0.02\\
 & & & & MJD $\rightarrow$ & 52761--52773 & & & & & R 16.95$\pm$0.01, I  15.32$\pm$0.02 [203]\\
\hline
43 & AX J1749.2-2725  & 17:49:12.28 & -27:25:37.40 & 220.38$\pm$0.20  &  & -- & $\sim$7E34  & -- & 11--16  & -- \\
& & & & (1997) [187c] & & & 2--10 [164c] & & [164c] & B1--3 [164c] \\
 & & & & Glob. spin-up $\rightarrow$ & $\sim$-9E-9
[164c] & &  & & & J 18.58$\pm$0.21, H 16.57$\pm$0.07 \\
 & & & & MJD $\rightarrow$ & 50814--54831 & & & & & K 14.95$\pm$0.05 [164c]\\
\hline
44 & IGR J16465-4507 & 16:46:35.26 & -45:07:04.61 & 228$\pm$6  & & 30.243$\pm$0.035  & (2.1--4.9)E36  & -- & 2.7$\pm$0.5 &  -- \\
 & & & & (2004) [189] & & [190] & 1--10 [2] & & [190a] & B0.5--1 Ib [190b] \\
  & & & & & & & & & & B  16.85$\pm$0.01, V 14.62$\pm$0.01\\
  & & & & & & & & & & R 13.55$\pm$0.01, I 12.45$\pm$0.01\\
  & & & & & & & & & & J  10.53$\pm$0.05, H  10.01$\pm$0.06\\
  & & & & & & & & & & K  9.83$\pm$0.07 [190b]\\
  & & & & & & & & & & E(B-V) 1.82$\pm$0.02 [190b]\\
\hline
45 & 4U 1258-61 & 13:01:17.10 & -61:36:06.64 & 275$\pm$0.02  & &  132.189$\pm$0.02  & 1E34--2.2E37  &  $\sim$4.7  & 1.9$\pm$0.05  & V850 Cen \\
 & (GX 304-1) & & & (2017) [191] & & [192] & 3--79 [193] & [194] & [143] &  B0.7 Ve [143] \\
 & & & & Loc. spin-down $\rightarrow$ & (2.39$\pm$0.51)E-8 [195] & & & & & V 14.4 [143]\\
 & & & & MJD $\rightarrow$ & 55563--55677 & & & & & J 9.798$\pm$0.022, H 9.297$\pm$0.022 \\
 & & & & Loc. spin-up $\rightarrow$ & (-2.45$\pm$0.25)E-7 [195] & & & & & K 9.040$\pm$0.021 [13a]\\
 & & & & MJD $\rightarrow$ & 55811--55821 & & & & & \\
\hline
46 & 2S 1145-619 & 11:48:00.02 & -62:12:24.90 & 292.274$\pm$0.001 &  &  186.68$\pm$0.05 &  (0.84--1.9)E35  & -- &  3.1$\pm$0.5 & 
HD 102567 \\
 & & & & (1985) [196] & &  [197] & 1--10 [2]& & [197a] & B0.2 III [197b]\\
  & & & & & & & & & & B 9.06, V 9.00 [28a]\\
    & & & & & & & & & &  J 8.682$\pm$0.028, H 8.562$\pm$0.009 \\
    & & & & & & & & & & K 8.395$\pm$0.011 [197a] \\
    & & & & & & & & & & E(B-V) 0.29$\pm$0.02 [197a] \\
\hline
47 & 1E 1145.1-6141 & 11:47:28.56 & -61:57:13.43 &   296.653$\pm$0.021  & &  14.365$\pm$0.002 & (0.1--4.7)E37  & -- & 8.2$\pm$1.5  & 
 -- \\
 & & & & (2019) [198] & & [199] & 1--10 [2]  & & [200] & B2 Iae [200]  \\
 & & & & Glob. spin-up $\rightarrow$ & $\sim$-1.06E-9 [199] & &  & & & B 14.55, V 13.10\\
 & & & & MJD $\rightarrow$ & 42614--51570 & & & & & R 14.11 [200]\\
 & & & & & & & & & & J 9.607$\pm$0.026, H 9.110$\pm$0.022 \\
 & & & & & & & & & & K  8.810$\pm$0.021 [13a]\\
 & & & & & & & & & & E(B-V) 1.61$\pm$0.02 [200]\\
\hline
48 & SGR 0755-2933 & 07:55:32.00 & -29:33:08.00 &  308.26$\pm$0.02  & &  $\sim$260 & $\sim$1E34 & -- & 3.5$\pm$0.2   & --\\
 & & & & (2020) [200a] & & [200a] & 3--80 [200a] & & [200a] & Be [200b]\\
\hline
49 & IGR J21343+4738 & 	21:34:20.37 & 47:38:00.21 & 322.71$\pm$0.04  & & $\sim$34.26 & (0.5--1.5)E35  & -- & 10$\pm$2.5  &  --\\
 & & & & (2020) [200c] & & [200c] & 2--12 [201] & & [62a] & B1 IV shell [62a] \\
 & & & & & & &  & & & B 14.68, V 14.16 \\
  & & & & & & &  & & & R 13.80, I 13.42 [62a] \\
 & & & & & & &  & & & J 12.939$\pm$0.02, H 12.726$\pm$0.023 \\
 & & & & & & &  & & & K 12.529$\pm$0.027 [13a]\\
  & & & & & & &  & & & E(B-V) 0.75$\pm$0.03 [62a]\\
\hline
50 & IGR J17200-3116 & 17:20:05.91  & -31:16:59.60 & 327.878$\pm$0.024  &  & --  & (0.9--1.2)E35 & -- & 5--10  &   -- \\
 & & & & (2013) [201a] & &  & 1--10 [201a] & & [201a] & Be [201a] \\
   & & & & & & &  & & & J 13.581$\pm$0.056, H 12.334$\pm$0.057\\
    & & & & & & &  & & & K 11.983$\pm$0.043 [62a]\\
\hline
51 & IGR J00370+6122 & 00:37:09.64 & 61:21:36.49 & 346$\pm$6  &  & 15.6649$\pm$0.0014  & (0.4--3)E36 & -- & 3.4$\pm$0.3  &  BD +60 73 \\
 & & & & (2006) [202] & & [202a] & 3--60 [202] & & [202b] & B0.5 II--III [203] \\
   & & & & & & &  & & & B  10.24$\pm$0.02, V  9.70$\pm$0.02\\
   & & & & & & &  & & & R 9.36$\pm$0.02, I 8.91$\pm$0.03 [203]\\
   & & & & & & &  & & & J 8.389$\pm$0.024, H 8.265$\pm$0.046 [203]\\
   & & & & & & &  & & & K 8.166$\pm$0.020 [13a]\\
   & & & & & & &  & & & E(B-V) 0.75 [203]\\
\hline
52 & SAX J2103.5+4545 & 21:03:35.71 & 45:45:05.56 & 351.13$\pm$0.02  &  & $\sim$12.68  & (0.75--8.4)E36  & -- & 6$\pm$1.5 & -- \\
 & & & & (2013)  [204] & & [205] & 1--10 [2] & & [62a] & B0 Ve [207]\\
 & & & & Loc. spin-up $\rightarrow$ & $\sim$-2.9E-7 [208] & & & & & B 15.34, V 14.20\\
 & & & & MJD $\rightarrow$ & 55481--55496 & & & & & R 13.49, I 12.75 [62a]\\
 & & & & Glob. spin-down $\rightarrow$ & $\sim$1.8E-9 [208] & &  & & & J 11.842$\pm$0.021, H 11.535$\pm$0.021\\
 & & & & MJD $\rightarrow$ & 55927--57022 & & & & & K 11.362$\pm$0.023 [13a]\\
    & & & & & & &  & & & E(B-V) 1.36$\pm$0.05 [62a]\\
\hline
53 & IGR J06074+2205 & 06:07:26.61 & 22:05:47.76 & 373.226$\pm$0.013 &  & - & $\sim$1.5E34  & -- & 4.1$\pm$1 & GSC 01325-01064\\
& & & & (2017)  [208a] & & & 0.2--12 [208a] & & [62a] & B0.5 Ve \\
 & & & & & & & & & & B 12.85, V 12.21\\
  & & & & & & & & & & R 11.80, I 11.32 [62a]\\
 & & & & & & & & & & J 10.49$\pm$0.021, H 10.19$\pm$0.022 \\
  & & & & & & & & & & K 9.96$\pm$0.019 [13a]\\
 & & & & & & & & & & E(B-V) 0.86$\pm$0.03  [62a]\\
\hline
54 & 1A 1118-615 & 11:20:57.17 & -61:55:00.17 & 407.77$\pm$0.08  & & 24$\pm$0.4 & (0.23--2.5)E37  & $\sim$4.8 & 5$\pm$2  &  WRAY 15-793 \\
 & & & & (2009) [209] & & [210] & 3--30 [209] &  [212] & [213] & O9.5 Ve [213]\\
 & & & & Loc. spin-up $\rightarrow$ & (-4.6$\pm$0.02)E-7 [212] & &  & & & B 13.06, V 12.12 [211]\\
 & & & & MJD $\rightarrow$ & 54841--54865 & & & & & J 9.563$\pm$0.024, H 9.071$\pm$0.023\\
 & & & & Glob.spin-down $\rightarrow$ & $\sim$1.49E-8 [78] & &  & & & K 8.587$\pm$0.019 [13a]\\
 & & & & MJD $\rightarrow$ & 54832--55561 & &  & & & \\
\hline
55 & 4U 1907+09  & 19:09:37.14 & 09:49:55.28 &  442.92$\pm$0.03  & &  8.380$\pm$0.002 &  (0.36--4.8)E36  & 2  & 4.4$\pm$1.2 & -- \\
 & & & & (2018) [215] & & [216]  & 1--10 [2] & [34] &  [62a] & O8--9 Ia [217] \\
 & & & & Glob. spin-down $\rightarrow$ & (6.87$\pm$0.04)E-9 & &  & & & B 19.41, V 16.35\\
 & & & & MJD & 45576--51080 & & & & & R 14.40, I 12.53 [62a]\\
 & & & &   &   & & & & & J 8.63, K 8.80 [94] \\
  & & & &  &  & & & & & E(B-V) 3.31$\pm$0.1 [62a] \\
\hline
56 & IGR J01583+6713 & 01:58:18.49 & 67:13:23.46 &  469.2 & -  & 216--561  & (0.01--4.4)E35  & 4$\pm$0.4  & 3.4$\pm$0.8 & --\\
 & & & & (2005) [219] & & [219] & 20--100 [220] & [220] &  [62a] & B2 IVe [219] \\
 & & & & & & & & & & B 15.71, V 14.41 \\
 & & & & & & & & & & R 13.51, I 12.66 [62a] \\
 & & & & & & & & & & J 11.481$\pm$0.026, H 11.03$\pm$0.03 \\
  & & & & & & & & & & K 10.601$\pm$0.021 [13a]\\
   & & & & & & & & & & E(B-V) 1.44$\pm$0.04 [62a]\\
\hline
57 & MAXI J1409-619 & 14:08:02.56 & -61:59:00.30 & 506.93$\pm$0.05  &  & 14.7$\pm$0.4  & (0.07--2)E37  & $\sim$3.8  & $\sim$14.5  & -- \\
 & & & & (2010) [221] & & [221a] & 2--10 [222] & [223] & [223] & B0 III--V [223]\\
 & & & & Loc. spin-up $\rightarrow$ & (-4.07$\pm$0.18)E-6 [224]& &  & & & J 15.874$\pm$0.086, H 13.620$\pm$0.022 \\
 & & & & MJD $\rightarrow$ & 55530--55546 & & & & & K 12.560$\pm$0.021 [223]\\
\hline
58 &  4U 1909+07 & 19:10:48.21 & 07:35:51.71 & 603.6$\pm$0.1 &  & 4.4$\pm$0.001  & (0.35--3.5)E36  & $\sim$3.8  & 7$\pm$3  & --  \\
 & & & & (2017) [224a]  & & [226] & 1--10 [2] & [225] & [227] & O7.5--O9.5 I [227]\\
 & & & & Glob. spin-down $\rightarrow$ & $\sim$7.15E-9 [228]& &  & & & J 13.228$\pm$0.021, H 11.457$\pm$0.027\\
 & & & & MJD $\rightarrow$ & 54280--55600  & & & & & K 10.480$\pm$0.022 [227]\\
\hline
59 & IGR J13020-6359  & 13:01:58.72 & -63:58:08.83 &  642.90$\pm$0.01  &  & -- & (0.8--2.6)E35  & -- & 4--7 & --\\
 & & & & (2014) [228a] & & & 3--79 [228a] & & [230] & B0.5 Ve [231]  \\
 &  & & & Glob. spin-up $\rightarrow$ & $\sim$-2E-7 [229] & &  & & & J 12.962$\pm$1.339, H 12.047$\pm$0.031 \\
 & & & & MJD $\rightarrow$ & 51910--53370 & & & & & K 11.346$\pm$0.088 [13a]\\
\hline
60 &  IGR J18462-0223 & 18:46:12.79 & -02:22:26.04 & 997$\pm$1  &  & $\sim$2.13  & (0.011--2.2)E37 & -- & $\sim$11 & -- \\
 & & & & (2011) [233] & & [234] & 18--60 [234] & & [234] & O--B (SG) [234]\\
  & & & & & & & & & & J 14.48$\pm$0.05, H 13.75$\pm$0.07 [13a]\\
\hline
61 & IGR J16418-4532 & 16:41:50.80 & -45:32:25.37 & 1212$\pm$6  & - &  3.7389$\pm$4E-4 & (0.3--2.1)E37  & -- & $\sim$13 & --\\
 & & & & (2011) [235] & &  [236] & 1--10 [2] & & [235] & O9--9.5 Ia--III [235] \\
 & & & & & & & & & & J 13.867$\pm$0.049, H 12.305$\pm$0.040 \\
 & & & & & & & & & & K 11.483$\pm$0.032 [13a]\\
\hline
62 & SAX J2239.3+6116 & 22:39:20.84 & 61:16:26.61 & 1247.2$\pm$0.7 & -- & 262.6$\pm$0.7 & 2.3E36  & -- & $\sim$4.4 &  --\\
 & & & & (2001) [237] & & [237] & 2--28 [237] & &  [238] & B0 V--B2 III [238]\\
 & & & & & & & & & & B 16.5, V 15.1\\
 & & & & & & & & & & R 14.1 [238]\\
 & & & & & & & & & & J 11.450$\pm$0.026, H 10.955$\pm$0.031\\
 & & & & & & & & & & K 10.557$\pm$0.021 [13a]\\
 & & & & & & & & & & E(B-V) 1.8 [238]\\
\hline
63 & IGR J16320-4751 & 	16:32:01.87 & -47:52:28.30  & 1309$\pm$40  & & 8.99$\pm$0.01  & (0.18--2.5)E36  & -- & $\sim$3.5  & --\\
 & & & & (2003) [239] & & [240] & 1--10 [2] & & [102] & O8 I [102] \\
 & & & & Glob. spin-up $\rightarrow$ & $\sim$-8.57E-7 [239] & &  & & & \\
 & & & & MJD $\rightarrow$ & 50449--53735 & & & & & \\
 \hline
64 & AX J1910.7+0917 & 19:10:43.55 & 09:16:29.83  & 36200$\pm$110  & & --  & (0.017--1)E36  & -- & 16.0$\pm$0.5  & --\\
 & & & & (2011) [241] & & & 1--10 [241] & & [242] & B (sg) [242] \\
 & & & & &  & &  & & & J $>$17.1, H 14.43$\pm$0.05 \\
 & & & & &  & & & & & K 13.135$\pm$0.003 [242]\\
\enddata
\end{deluxetable*}
\end{longrotatetable}

References for Table \ref{tab:chartable2}: [71] - \citet{Cusumano...2000ApJ...528L..25C}, [72] - \citet{Kaaret...2000ApJ...542L..41K}, [73] - \citet{Kaaret...1999ApJ...523..197K}, [74] - \citet{Mereghetti...2009A&A...504..181M}, [75] - \citet{Belczynski...2009ApJ...707..870B}, [76] - \citet{Hemphill...2019ATel12556....1H}, [77] - \citet{Galloway...2005ApJ...635.1217G}, [78] - \citet{Malacaria...2020ApJ...896...90M}, [79] - \citet{Strader...2019ATel12554....1S}, [80] - \citet{Ding...2021MNRAS.503.6045D}, [81] - \citet{Boldin...2013AstL...39..375B}, [82] - \citet{Nagase...1991ApJ...375L..49N}, [83] - \citet{Negueruela...2001A&A...369..108N}, [84] - \citet{Li...2012MNRAS.423.2854L}, [86] - \citet{Zhang...2005ApJ...630L..65Z}, [87] - \citet{Makishima...1990ApJ...365L..59M}, [88] - \citet{Negueruela...1999MNRAS.307..695N}, [90] - \citet{Devaraj...2022MNRAS.514L..46D}, [91] - \citet{Scott...1997ApJ...488..831S}, [91a] - \citet{Lutovinov...2019MNRAS.485..770L}, [92] - \citet{Shaw...2009MNRAS.393..419S}, [93] - \citet{Bamba...2001PASJ...53.1179B}, [94] - \citet{Nespoli...2008A&A...486..911N}, [95] - \citet{Shtykovsky...2019MNRAS.482L..14S}, [95a] - \citet{Corbet...2022ATel15614....1C}, [96] - \citet{Halpern...2007ApJ...669..579H}, [96a] - \citet{UKIDSS...2012yCat.2316....0U}, [97] - \citet{Christodoulou...2022ApJ...929..137C}, [97a] - \citet{Malacaria...2022ApJ...927..194M}, [97b] - \citet{Lutovinov...2016MNRAS.462.3823L}, [97c] - \citet{Pahari...2012MNRAS.423.3352P}, [98] - \citet{Tsygankov...2016MNRAS.457..258T}, [98a] - \citet{Salganik...2022MNRAS.509.5955S}, [98b] - \citet{Corbet...2009ApJ...695...30C}, [98c] - \citet{Jenke...2017ATel10812....1J}, [98d] - \citet{Doroshenko...2018A&A...613A..19D}, [98e] - \citet{Doroshenko...2020MNRAS.491.1857D}, [98f] - \citet{Wilson-Hodge...2018ApJ...863....9W}, [98g] - \citet{GAIA_DR3}, [98h] - \citet{Reig...2020A&A...640A..35R}, [99] - \citet{Romano...2015A&A...576L...4R}, [100] - \citet{Clark...2009MNRAS.399L.113C}, [101] - \citet{Bhalerao...2015MNRAS.447.2274B}, [102] - \citet{Rahoui...2008A&A...484..801R}, [102a] - \citet{Bozzo...2016A&A...596A..16B}, [102b] - \citet{Esposito...2020ATel13625....1E}, [102c] - \citet{Li...2012MNRAS.426L..16L}, [103] - \citet{Nowak...2012ApJ...757..143N}, [104] - \citet{Jenke...2012ATel.4235....1J}, [105] - \citet{Wilson...1997ApJ...479..388W}, [106] - \citet{Israel...2000MNRAS.314...87I},[107] - \citet{Raman...2021MNRAS.508.5578R}, [108] - \citet{Corbet...2009ATel.2008....1C}, [109] - \citet{Tsygankov...2019A&A...621A.134T}, [109a] - \citet{Rodes_Roca...2018MNRAS.476.2110R}, [109b] - \citet{Fermi_GBM_IGR_J19294_1816}, [110] - \citet{Reig...2008A&A...485..797R}, [111] - \citet{Baykal...2010ApJ...711.1306B}, [112] - \citet{DeCesar...2013ApJ...762...61D}, [113] - \citet{Reig...2011A&A...533A..23R}, [114] - \citet{Deo...2023arXiv230110678D}, [115] - \citet{Wilson...2003ApJ...584..996W}, [116] - \citet{Muller...2012A&A...546A.125M}, [117] - \citet{Marcu...2015ApJ...815...44M}, [118] - \citet{Raichur...2010MNRAS.406.2663R}, [119] - \citet{Finger...1996A&AS..120C.209F}, [120] - \citet{Inam...2004MNRAS.349..173I}, [121] - \citet{Negueruela...1998A&A...338..505N}, [121a] - \citet{Grindlay...1984ApJ...276..621G}, [122] - \citet{Ballhausen...2016A&A...591A..65B}, [123] - \citet{Galloway...2004ApJ...613.1164G}, [124] - \citet{Furst...2014ApJ...784L..40F}, [125] - \citet{Negueruela...2003A&A...397..739N}, [126] - \citet{Fermi_GBM_KS_1947_300}, [127] - \citet{Sguera...2007A&A...467..249S}, [128] - \citet{Rahoui...2008A&A...492..163R}, [129] - \citet{Zand...2001ApJ...553L.165I}, [130] - \citet{Piraino...2000A&A...357..501P}, [130a] - \citet{Israel...2001A&A...371.1018I}, [131] - \citet{Zhao...2019MNRAS.488.4427Z}, [132] - \citet{Corbet...2000ApJ...530L..33C}, [133] - \citet{Naik...2013ApJ...764..158N}, [134] - \citet{Wilson...2008ApJ...678.1263W}, [135] - \citet{Reig...1999MNRAS.302..700R}, [136] - \citet{Reig...2014MNRAS.445.4235R}, [136a] - \citet{OConnor...2022ApJ...927..139O}, [136b] - \citet{Parola...2013ApJ...775L..24L}, [137] - \citet{Markwardt...2010ATel.2564....1M}, [138] - \citet{Corbet...2010ATel.2559....1C}, [139] - \citet{Negueruela...2007A&A...461..631N}, [139a] - \citet{Anderson...2014ApJS..212...13A}, [140] - \citet{Mukerjee...2021ApJ...920..139M}, [141] - \citet{Wilson...1999ApJ...511..367W}, [142] - \citet{McBride...2007A&A...470.1065M}, [143] - \citet{Reig...2022A&A...667A..18R}, [145] - \citet{Fermi_GBM_Cep_X-4}, [145a] - \citet{Gogus...2005ApJ...632.1069G}, [145b] - \citet{Wilson...2002ApJ...565.1150W}, [145c] - \citet{Marsden...1998ApJ...502L.129M}, [146] - \citet{Ge...2020ApJ...899L..19G}, [147] - \citet{Kuhnel...2013A&A...555A..95K}, [150] - \citet{Riquelme...2012A&A...539A.114R}, [151] - \citet{Fermi_GBM_GRO_J1008-57}, [152] - \citet{Wang...2014RAA....14..565W}, [153] - \citet{Nabizadeh...2022A&A...657A..58N}, [154] - \citet{Finger...1999ApJ...517..449F}, [155] - \citet{Roy...2020RAA....20..155R}, [155a] - \citet{Corbet...2016ATel.9823....1C}, [156] - \citet{Corbet...1997ApJ...489L..83C}, [157] - \citet{Coley...2023AAS...24142807C}, [158] - \citet{Coe...2006MNRAS.368..447C}, [159] - \citet{Kendziorra...1994A&A...291L..31K}, [161] - \citet{Giangrande...1980A&AS...40..289G}, [162] - \citet{Fermi_GBM_1A_0535_262}, [163] - \citet{Coe...1990MNRAS.243..475C}, [164] - \citet{Hill...2007MNRAS.381.1275H}, [164a] - \citet{Karasev...2008MNRAS.386L..10K}, [164b] - \citet{Zurita...2008A&A...489..657Z}, [164c] - \citet{Karasev...2010MNRAS.409L..69K}, [164d] - \citet{Nabizadeh...2019A&A...622A.198N}, [164e] - \citet{Parola...2014MNRAS.445L.119L}, [164f] - \citet{Krimm...2013ApJS..209...14K}, [165] - \citet{Kinugasa...1998ApJ...495..435K}, [166] - \citet{Segreto...2013A&A...558A..99S}, [167] - \citet{McBride...2006A&A...451..267M}, [168] - \citet{Yan...2012ApJ...753...73Y}, [170] - \citet{Nespoli...2012A&A...547A.103N}, [170a] - \citet{Zand...2004ATel..362....1I}, [171] - \citet{Torrejon...2007ESASP.622..503T}, [172] - \citet{Corbet...2005ATel..377....1C}, [172a] - \citet{Masetti...2009A&A...495..121M}, [173] - \citet{Sidoli...2020A&A...638A..71S}, [174] - \citet{Romano...2009ApJ...696.2068R},  [176] - \citet{Lorenzo...2014A&A...562A..18L}, [177] - \citet{Huckle...1977MNRAS.180P..21H}, [179] - \citet{Mukerjee...2020ApJ...897...73M}, [180] - \citet{Wilson...1998ApJ...499..820W}, [181] - \citet{Bildsten...1997ApJS..113..367B}, [182] - \citet{Molkov...2019ApJ...883L..11M}, [183] - \citet{Ferrigno...2013A&A...553A.103F}, [184] - \citet{Tsygankov...2012MNRAS.421.2407T}, [185] - \citet{Reig...2005A&A...440.1079R}, [186] - \citet{LaPalombara...2012A&A...539A..82L}, [186a] - \citet{Fermi_GBM_RX_J0440}, [187] - \citet{Doroshenko...2020A&A...634A..89D}, [187a] - \citet{McCollum...2019ATel13211....1M}, [187b] - \citet{Tsygankov...2021ApJ...909..154T}, [187c] - \citet{Torii...1998ApJ...508..854T}, [188] - \citet{Doroshenko...2008ARep...52..138D}, [189] - \citet{Lutovinov...2005A&A...444..821L}, [190] - \citet{LaParola...2010MNRAS.405L..66L}, [190a] -\citet{Arnason...2021MNRAS.502.5455A}, [190b] - \citet{Chaty...2016A&A...591A..87C}, [191] - \citet{Rouco...2018A&A...620L..13R}, [192] - \citet{Sugizaki...2015PASJ...67...73S}, [193] - \citet{Tsygankov...2019MNRAS.483L.144T}, [194] - \citet{Yamamoto...2011PASJ...63S.751Y}, [195] - \citet{Postnov...2015MNRAS.446.1013P}, [196] - \citet{Cook...1987MNRAS.225..369C}, [197] - \citet{Wilson...1999PhDT.........6W}, [197a] - \citet{Stevens...1997MNRAS.288..988S}, [197b] - \citet{Alfonso...2017A&A...607A..52A}, [198] - \citet{Ghising...2022MNRAS.517.4132G}, [199] - \citet{Ray...2002ApJ...581.1293R}, [200] - \citet{Densham...1982MNRAS.201..171D}, [200a] - \citet{Doroshenko...2021A&A...647A.165D}, [200b] - \citet{Chrimes...2022MNRAS.512.6093C}, [200c] - \citet{Gorban...2022AstL...48..798G}, [201] - \citet{Reig...2014MNRAS.442..472R}, [201a] - \citet{Esposito...2014MNRAS.441.1126E}, [202] - \citet{Zand...2007A&A...469.1063I}, [202a] - \citet{Uchida...2021PASJ...73.1389U}, [202b] - \citet{Hainich...2020A&A...634A..49H}, [203] - \citet{Reig...2005A&A...440..637R}, [204] - \citet{Reig...2014MNRAS.445.1314R}, [205] - \citet{Baykal...2002ApJ...569..903B}, [207] - \citet{Reig...2004A&A...421..673R}, [208] - \citet{Camero...2014A&A...568A.115C}, [208a] - \citet{Reig...2018A&A...613A..52R}, [209] - \citet{Nespoli...2011A&A...526A...7N}, [210] - \citet{Staubert...2011A&A...527A...7S}, [211] - \citet{Reed...2003AJ....125.2531R}, [212] - \citet{Doroshenko...2010A&A...515L...1D}, [213] - \citet{Janot-Pacheco...1981A&A....99..274J}, [215] - \citet{Tobrej...2023MNRAS.518.4861T}, [216] - \citet{Marshall...1980MNRAS.193P...7M}, [217] - \citet{Cox...2005A&A...436..661C}, [218] - \citet{Baykal...2001MNRAS.327.1269B}, [219] - \citet{Kaur...2008MNRAS.386.2253K}, [220] - \citet{Wang...2010A&A...516A..15W}, [221] - \citet{Camero-Arranz...2010ATel.3069....1C}, [221a] - \citet{Donmez...2020MNRAS.496.1768D}, [222] - \citet{Kaur...2010ATel.3082....1K}, [223] - \citet{Orlandini...2012ApJ...748...86O}, [224] - \citet{Fermi_GBM_MAXI}, [224a] - \citet{Jaisawal...2020MNRAS.498.4830J}, [225] - \citet{Jaisawal...2013ApJ...779...54J}, [226] - \citet{Wen...2000ApJ...532.1119W}, [227] - \citet{Morel...2005MNRAS.356..665M}, [228] - \citet{Sahiner...2012MNRAS.421.2079S}, [228a] - \citet{Krivonos...2015ApJ...809..140K}, [229] - \citet{Chernyakova...2005MNRAS.364..455C}, [230] - \citet{Chernyakova...2006IAUS..230...33C}, [231] - \citet{Coleiro...2013A&A...560A.108C}, [233] - \citet{Bodaghee...2012ApJ...753....3B}, [234] - \citet{Sguera...2013A&A...556A..27S}, [235] - \citet{Sidoli...2012MNRAS.420..554S}, [236] - \citet{Corbet...2006ATel..779....1C}, [237] - \citet{Zand...2001A&A...380L..26I}, [238] - \citet{Zand...2000A&A...361...85I}, [239] - \citet{Lutovinov...2005A&A...433L..41L}, [240] - \citet{Garcia...2018A&A...618A..61G}, [241] - \citet{Sidoli...2017MNRAS.469.3056S}, [242] - \citet{Rodes...2013A&A...555A.115R}

\bibliography{literature}{}

\begin{thebibliography}{}
\expandafter\ifx\csname natexlab\endcsname\relax\def\natexlab#1{#1}\fi
\providecommand{\url}[1]{\href{#1}{#1}}
\providecommand{\dodoi}[1]{doi:~\href{http://doi.org/#1}{\nolinkurl{#1}}}
\providecommand{\doeprint}[1]{\href{http://ascl.net/#1}{\nolinkurl{http://ascl.net/#1}}}
\providecommand{\doarXiv}[1]{\href{https://arxiv.org/abs/#1}{\nolinkurl{https://arxiv.org/abs/#1}}}

\bibitem[{{Alfonso-Garz{\'o}n} {et~al.}(2012){Alfonso-Garz{\'o}n}, {Domingo},
  {Mas-Hesse}, \& {Gim{\'e}nez}}]{Alfonso...2012A&A...548A..79A}
{Alfonso-Garz{\'o}n}, J., {Domingo}, A., {Mas-Hesse}, J.~M., \& {Gim{\'e}nez},
  A. 2012, \aap, 548, A79, \dodoi{10.1051/0004-6361/201220095}

\bibitem[{{Alfonso-Garz{\'o}n} {et~al.}(2017){Alfonso-Garz{\'o}n}, {Fabregat},
  {Reig}, {Kajava}, {S{\'a}nchez-Fern{\'a}ndez}, {Townsend}, {Mas-Hesse},
  {Crawford}, {Kretschmar}, \& {Coe}}]{Alfonso...2017A&A...607A..52A}
{Alfonso-Garz{\'o}n}, J., {Fabregat}, J., {Reig}, P., {et~al.} 2017, \aap, 607,
  A52, \dodoi{10.1051/0004-6361/201630211}

\bibitem[{{Anderson} {et~al.}(2014){Anderson}, {Gaensler}, {Kaplan}, {Slane},
  {Muno}, {Posselt}, {Hong}, {Murray}, {Steeghs}, {Brogan}, {Drake}, {Farrell},
  {Benjamin}, {Chakrabarty}, {Drew}, {Finley}, {Grindlay}, {Lazio}, {Lee},
  {Mauerhan}, \& {van Kerkwijk}}]{Anderson...2014ApJS..212...13A}
{Anderson}, G.~E., {Gaensler}, B.~M., {Kaplan}, D.~L., {et~al.} 2014, \apjs,
  212, 13, \dodoi{10.1088/0067-0049/212/1/13}

\bibitem[{{Angelini} {et~al.}(1998){Angelini}, {Church}, {Parmar},
  {Balucinska-Church}, \& {Mineo}}]{Angelini...1998AandAP..339..L41}
{Angelini}, L., {Church}, M.~J., {Parmar}, A.~N., {Balucinska-Church}, M., \&
  {Mineo}, T. 1998, \aap, 339, L41

\bibitem[{{Arnason} {et~al.}(2021){Arnason}, {Papei}, {Barmby}, {Bahramian}, \&
  {Gorski}}]{Arnason...2021MNRAS.502.5455A}
{Arnason}, R.~M., {Papei}, H., {Barmby}, P., {Bahramian}, A., \& {Gorski},
  M.~D. 2021, \mnras, 502, 5455, \dodoi{10.1093/mnras/stab345}

\bibitem[{{Arons} \& {Lea}(1976)}]{Arons...1976ApJ...210..792A}
{Arons}, J., \& {Lea}, S.~M. 1976, \apj, 210, 792, \dodoi{10.1086/154888}

\bibitem[{{Astropy Collaboration} {et~al.}(2013){Astropy Collaboration},
  {Robitaille}, {Tollerud}, {Greenfield}, {Droettboom}, {Bray}, {Aldcroft},
  {Davis}, {Ginsburg}, {Price-Whelan}, {Kerzendorf}, {Conley}, {Crighton},
  {Barbary}, {Muna}, {Ferguson}, {Grollier}, {Parikh}, {Nair}, {Unther},
  {Deil}, {Woillez}, {Conseil}, {Kramer}, {Turner}, {Singer}, {Fox}, {Weaver},
  {Zabalza}, {Edwards}, {Azalee Bostroem}, {Burke}, {Casey}, {Crawford},
  {Dencheva}, {Ely}, {Jenness}, {Labrie}, {Lim}, {Pierfederici}, {Pontzen},
  {Ptak}, {Refsdal}, {Servillat}, \& {Streicher}}]{astropy:2013}
{Astropy Collaboration}, {Robitaille}, T.~P., {Tollerud}, E.~J., {et~al.} 2013,
  \aap, 558, A33, \dodoi{10.1051/0004-6361/201322068}

\bibitem[{{Astropy Collaboration} {et~al.}(2018){Astropy Collaboration},
  {Price-Whelan}, {Sip{\H{o}}cz}, {G{\"u}nther}, {Lim}, {Crawford}, {Conseil},
  {Shupe}, {Craig}, {Dencheva}, {Ginsburg}, {Vand erPlas}, {Bradley},
  {P{\'e}rez-Su{\'a}rez}, {de Val-Borro}, {Aldcroft}, {Cruz}, {Robitaille},
  {Tollerud}, {Ardelean}, {Babej}, {Bach}, {Bachetti}, {Bakanov}, {Bamford},
  {Barentsen}, {Barmby}, {Baumbach}, {Berry}, {Biscani}, {Boquien}, {Bostroem},
  {Bouma}, {Brammer}, {Bray}, {Breytenbach}, {Buddelmeijer}, {Burke},
  {Calderone}, {Cano Rodr{\'\i}guez}, {Cara}, {Cardoso}, {Cheedella}, {Copin},
  {Corrales}, {Crichton}, {D'Avella}, {Deil}, {Depagne}, {Dietrich}, {Donath},
  {Droettboom}, {Earl}, {Erben}, {Fabbro}, {Ferreira}, {Finethy}, {Fox},
  {Garrison}, {Gibbons}, {Goldstein}, {Gommers}, {Greco}, {Greenfield},
  {Groener}, {Grollier}, {Hagen}, {Hirst}, {Homeier}, {Horton}, {Hosseinzadeh},
  {Hu}, {Hunkeler}, {Ivezi{\'c}}, {Jain}, {Jenness}, {Kanarek}, {Kendrew},
  {Kern}, {Kerzendorf}, {Khvalko}, {King}, {Kirkby}, {Kulkarni}, {Kumar},
  {Lee}, {Lenz}, {Littlefair}, {Ma}, {Macleod}, {Mastropietro}, {McCully},
  {Montagnac}, {Morris}, {Mueller}, {Mumford}, {Muna}, {Murphy}, {Nelson},
  {Nguyen}, {Ninan}, {N{\"o}the}, {Ogaz}, {Oh}, {Parejko}, {Parley}, {Pascual},
  {Patil}, {Patil}, {Plunkett}, {Prochaska}, {Rastogi}, {Reddy Janga},
  {Sabater}, {Sakurikar}, {Seifert}, {Sherbert}, {Sherwood-Taylor}, {Shih},
  {Sick}, {Silbiger}, {Singanamalla}, {Singer}, {Sladen}, {Sooley},
  {Sornarajah}, {Streicher}, {Teuben}, {Thomas}, {Tremblay}, {Turner},
  {Terr{\'o}n}, {van Kerkwijk}, {de la Vega}, {Watkins}, {Weaver}, {Whitmore},
  {Woillez}, {Zabalza}, \& {Astropy Contributors}}]{astropy:2018}
{Astropy Collaboration}, {Price-Whelan}, A.~M., {Sip{\H{o}}cz}, B.~M., {et~al.}
  2018, \aj, 156, 123, \dodoi{10.3847/1538-3881/aabc4f}

\bibitem[{{Astropy Collaboration} {et~al.}(2022){Astropy Collaboration},
  {Price-Whelan}, {Lim}, {Earl}, {Starkman}, {Bradley}, {Shupe}, {Patil},
  {Corrales}, {Brasseur}, {N{"o}the}, {Donath}, {Tollerud}, {Morris},
  {Ginsburg}, {Vaher}, {Weaver}, {Tocknell}, {Jamieson}, {van Kerkwijk},
  {Robitaille}, {Merry}, {Bachetti}, {G{"u}nther}, {Aldcroft},
  {Alvarado-Montes}, {Archibald}, {B{'o}di}, {Bapat}, {Barentsen}, {Baz{'a}n},
  {Biswas}, {Boquien}, {Burke}, {Cara}, {Cara}, {Conroy}, {Conseil}, {Craig},
  {Cross}, {Cruz}, {D'Eugenio}, {Dencheva}, {Devillepoix}, {Dietrich},
  {Eigenbrot}, {Erben}, {Ferreira}, {Foreman-Mackey}, {Fox}, {Freij}, {Garg},
  {Geda}, {Glattly}, {Gondhalekar}, {Gordon}, {Grant}, {Greenfield}, {Groener},
  {Guest}, {Gurovich}, {Handberg}, {Hart}, {Hatfield-Dodds}, {Homeier},
  {Hosseinzadeh}, {Jenness}, {Jones}, {Joseph}, {Kalmbach}, {Karamehmetoglu},
  {Ka{l}uszy{'n}ski}, {Kelley}, {Kern}, {Kerzendorf}, {Koch}, {Kulumani},
  {Lee}, {Ly}, {Ma}, {MacBride}, {Maljaars}, {Muna}, {Murphy}, {Norman},
  {O'Steen}, {Oman}, {Pacifici}, {Pascual}, {Pascual-Granado}, {Patil},
  {Perren}, {Pickering}, {Rastogi}, {Roulston}, {Ryan}, {Rykoff}, {Sabater},
  {Sakurikar}, {Salgado}, {Sanghi}, {Saunders}, {Savchenko}, {Schwardt},
  {Seifert-Eckert}, {Shih}, {Jain}, {Shukla}, {Sick}, {Simpson},
  {Singanamalla}, {Singer}, {Singhal}, {Sinha}, {Sip{H{o}}cz}, {Spitler},
  {Stansby}, {Streicher}, {{{S}}umak}, {Swinbank}, {Taranu}, {Tewary},
  {Tremblay}, {Val-Borro}, {Van Kooten}, {Vasovi{'c}}, {Verma}, {de Miranda
  Cardoso}, {Williams}, {Wilson}, {Winkel}, {Wood-Vasey}, {Xue}, {Yoachim},
  {Zhang}, {Zonca}, \& {Astropy Project Contributors}}]{astropy:2022}
{Astropy Collaboration}, {Price-Whelan}, A.~M., {Lim}, P.~L., {et~al.} 2022,
  apj, 935, 167, \dodoi{10.3847/1538-4357/ac7c74}

\bibitem[{{Ballhausen} {et~al.}(2016){Ballhausen}, {K{\"u}hnel}, {Pottschmidt},
  {F{\"u}rst}, {Hemphill}, {Falkner}, {Gottlieb}, {Grinberg}, {Kretschmar},
  {Kreykenbohm}, {Rothschild}, \& {Wilms}}]{Ballhausen...2016A&A...591A..65B}
{Ballhausen}, R., {K{\"u}hnel}, M., {Pottschmidt}, K., {et~al.} 2016, \aap,
  591, A65, \dodoi{10.1051/0004-6361/201527193}

\bibitem[{{Bamba} {et~al.}(2001){Bamba}, {Yokogawa}, {Ueno}, {Koyama}, \&
  {Yamauchi}}]{Bamba...2001PASJ...53.1179B}
{Bamba}, A., {Yokogawa}, J., {Ueno}, M., {Koyama}, K., \& {Yamauchi}, S. 2001,
  \pasj, 53, 1179, \dodoi{10.1093/pasj/53.6.1179}

\bibitem[{{Barnstedt} {et~al.}(2008){Barnstedt}, {Staubert}, {Santangelo},
  {Ferrigno}, {Horns}, {Klochkov}, {Kretschmar}, {Kreykenbohm}, {Segreto}, \&
  {Wilms}}]{Barnstedt...2008AandAP..486..P293}
{Barnstedt}, J., {Staubert}, R., {Santangelo}, A., {et~al.} 2008, \aap, 486,
  293–302, \dodoi{10.1051/0004-6361:20078707}

\bibitem[{{Barthelmy} {et~al.}(2016){Barthelmy}, {D'Elia}, {Gehrels}, {Izzo},
  {Kennea}, {Krimm}, {Palmer}, {Siegel}, \&
  {Ukwatta}}]{Barthelmy...2016GCN.19204....1B}
{Barthelmy}, S.~D., {D'Elia}, V., {Gehrels}, N., {et~al.} 2016, GRB Coordinates
  Network, 19204, 1

\bibitem[{{Baykal} {et~al.}(2010){Baykal}, {G{\"o}{\v{g}}{\"u}{\c{s}}},
  {{\c{C}}a{\v{g}}da{\c{s}} {\.I}nam}, \&
  {Belloni}}]{Baykal...2010ApJ...711.1306B}
{Baykal}, A., {G{\"o}{\v{g}}{\"u}{\c{s}}}, E., {{\c{C}}a{\v{g}}da{\c{s}}
  {\.I}nam}, S., \& {Belloni}, T. 2010, \apj, 711, 1306,
  \dodoi{10.1088/0004-637X/711/2/1306}

\bibitem[{{Baykal} {et~al.}(2001){Baykal}, {Inam}, {Ali Alpar}, {in't Zand}, \&
  {Strohmayer}}]{Baykal...2001MNRAS.327.1269B}
{Baykal}, A., {Inam}, {\c{C}}., {Ali Alpar}, M., {in't Zand}, J., \&
  {Strohmayer}, T. 2001, \mnras, 327, 1269,
  \dodoi{10.1046/j.1365-8711.2001.04804.x}

\bibitem[{{Baykal} {et~al.}(2002){Baykal}, {Stark}, \&
  {Swank}}]{Baykal...2002ApJ...569..903B}
{Baykal}, A., {Stark}, M.~J., \& {Swank}, J.~H. 2002, \apj, 569, 903,
  \dodoi{10.1086/339429}

\bibitem[{{Belczynski} \&
  {Ziolkowski}(2009)}]{Belczynski...2009ApJ...707..870B}
{Belczynski}, K., \& {Ziolkowski}, J. 2009, \apj, 707, 870,
  \dodoi{10.1088/0004-637X/707/2/870}

\bibitem[{{Bhalerao} {et~al.}(2015){Bhalerao}, {Romano}, {Tomsick},
  {Natalucci}, {Smith}, {Bellm}, {Boggs}, {Chakrabarty}, {Christensen},
  {Craig}, {Fuerst}, {Hailey}, {Harrison}, {Krivonos}, {Lu}, {Madsen}, {Stern},
  {Younes}, \& {Zhang}}]{Bhalerao...2015MNRAS.447.2274B}
{Bhalerao}, V., {Romano}, P., {Tomsick}, J., {et~al.} 2015, \mnras, 447, 2274,
  \dodoi{10.1093/mnras/stu2495}

\bibitem[{{Bildsten} {et~al.}(1997){Bildsten}, {Chakrabarty}, {Chiu}, {Finger},
  {Koh}, {Nelson}, {Prince}, {Rubin}, {Scott}, {Stollberg}, {Vaughan},
  {Wilson}, \& {Wilson}}]{Bildsten...1997ApJS..113..367B}
{Bildsten}, L., {Chakrabarty}, D., {Chiu}, J., {et~al.} 1997, \apjs, 113, 367,
  \dodoi{10.1086/313060}

\bibitem[{{Bird} {et~al.}(2016){Bird}, {Bazzano}, {Malizia}, {Fiocchi},
  {Sguera}, {Bassani}, {Hill}, {Ubertini}, \&
  {Winkler}}]{Bird...2016ApJS...223...15B}
{Bird}, A.~J., {Bazzano}, A., {Malizia}, A., {et~al.} 2016, \apjs, 223, 15,
  \dodoi{10.3847/0067-0049/223/1/15}

\bibitem[{{Bodaghee} {et~al.}(2012{\natexlab{a}}){Bodaghee}, {Tomsick}, \&
  {Rodriguez}}]{Bodaghee...2012ApJ...753....3B}
{Bodaghee}, A., {Tomsick}, J.~A., \& {Rodriguez}, J. 2012{\natexlab{a}}, \apj,
  753, 3, \dodoi{10.1088/0004-637X/753/1/3}

\bibitem[{{Bodaghee} {et~al.}(2012{\natexlab{b}}){Bodaghee}, {Tomsick},
  {Rodriguez}, \& {James}}]{Bodaghee...2012ApJ...744..108B}
{Bodaghee}, A., {Tomsick}, J.~A., {Rodriguez}, J., \& {James}, J.~B.
  2012{\natexlab{b}}, \apj, 744, 108, \dodoi{10.1088/0004-637X/744/2/108}

\bibitem[{{Bodaghee} {et~al.}(2006){Bodaghee}, {Walter}, {Zurita Heras},
  {Bird}, {Courvoisier}, {Malizia}, {Terrier}, \&
  {Ubertini}}]{Bodaghee...2006A&A...447.1027B}
{Bodaghee}, A., {Walter}, R., {Zurita Heras}, J.~A., {et~al.} 2006, \aap, 447,
  1027, \dodoi{10.1051/0004-6361:20053809}

\bibitem[{{Bodaghee} {et~al.}(2007){Bodaghee}, {Courvoisier}, {Rodriguez},
  {Beckmann}, {Produit}, {Hannikainen}, {Kuulkers}, {Willis}, \&
  {Wendt}}]{Bodaghee...2007A&A...467..585B}
{Bodaghee}, A., {Courvoisier}, T.~J.~L., {Rodriguez}, J., {et~al.} 2007, \aap,
  467, 585, \dodoi{10.1051/0004-6361:20077091}

\bibitem[{{Bodaghee} {et~al.}(2016){Bodaghee}, {Tomsick}, {Fornasini},
  {Krivonos}, {Stern}, {Mori}, {Rahoui}, {Boggs}, {Christensen}, {Craig},
  {Hailey}, {Harrison}, \& {Zhang}}]{Bodaghee...2016ApJ...823..146B}
{Bodaghee}, A., {Tomsick}, J.~A., {Fornasini}, F.~M., {et~al.} 2016, \apj, 823,
  146, \dodoi{10.3847/0004-637X/823/2/146}

\bibitem[{{Boldin} {et~al.}(2013){Boldin}, {Tsygankov}, \&
  {Lutovinov}}]{Boldin...2013AstL...39..375B}
{Boldin}, P.~A., {Tsygankov}, S.~S., \& {Lutovinov}, A.~A. 2013, Astronomy
  Letters, 39, 375, \dodoi{10.1134/S1063773713060029}

\bibitem[{{Bonning} \& {Falanga}(2005)}]{Bonning...2005A&A...436L..31B}
{Bonning}, E.~W., \& {Falanga}, M. 2005, \aap, 436, L31,
  \dodoi{10.1051/0004-6361:200500117}

\bibitem[{{Bozzo} {et~al.}(2016){Bozzo}, {Bhalerao}, {Pradhan}, {Tomsick},
  {Romano}, {Ferrigno}, {Chaty}, {Oskinova}, {Manousakis}, {Walter}, {Falanga},
  {Campana}, {Stella}, {Ramolla}, \& {Chini}}]{Bozzo...2016A&A...596A..16B}
{Bozzo}, E., {Bhalerao}, V., {Pradhan}, P., {et~al.} 2016, \aap, 596, A16,
  \dodoi{10.1051/0004-6361/201629311}

\bibitem[{{Bradt} \& {McClintock}(1983)}]{Bradt...1983ARAA...21...13B}
{Bradt}, H.~V.~D., \& {McClintock}, J.~E. 1983, \araa, 21, 13,
  \dodoi{10.1146/annurev.aa.21.090183.000305}

\bibitem[{{Camero} {et~al.}(2014){Camero}, {Zurita}, {Guti{\'e}rrez-Soto},
  {{\"O}zbey Arabac{\i}}, {Nespoli}, {Kiaeerad}, {Beklen}, {Garc{\'\i}a-Rojas},
  \& {Caballero-Garc{\'\i}a}}]{Camero...2014A&A...568A.115C}
{Camero}, A., {Zurita}, C., {Guti{\'e}rrez-Soto}, J., {et~al.} 2014, \aap, 568,
  A115, \dodoi{10.1051/0004-6361/201423452}

\bibitem[{{Camero-Arranz} {et~al.}(2010){Camero-Arranz}, {Finger}, \&
  {Jenke}}]{Camero-Arranz...2010ATel.3069....1C}
{Camero-Arranz}, A., {Finger}, M.~H., \& {Jenke}, P. 2010, The Astronomer's
  Telegram, 3069, 1

\bibitem[{{CGRO-Collaboration}(2000)}]{Batse}
{CGRO-Collaboration}. 2000, The Burst And Transient Source Experiment (BATSE)
  database.
\newblock
  \url{https://heasarc.gsfc.nasa.gov/FTP/compton/data/batse/pulsar/histories/}

\bibitem[{{Chakrabarty} {et~al.}(2002){Chakrabarty}, {Wang}, {Lee}, \&
  {Roche}}]{Chakrabarty...2002ApJ...573..P789}
{Chakrabarty}, D., {Wang}, Z., {Lee}, J., \& {Roche}, P. 2002, \apj, 573, 789,
  \dodoi{10.1086/340746}

\bibitem[{{Chaty}(2008)}]{Chaty...2008ChJAS...8..197C}
{Chaty}, S. 2008, Chinese Journal of Astronomy and Astrophysics Supplement, 8,
  197, \dodoi{10.48550/arXiv.0710.0292}

\bibitem[{{Chaty}(2022)}]{Chaty...2022abn..book.....C}
---. 2022, {Accreting Binaries; Nature, formation, and evolution} (IOP
  Publishing Ltd), \dodoi{10.1088/2514-3433/ac595f}

\bibitem[{{Chaty} {et~al.}(2016){Chaty}, {LeReun}, {Negueruela}, {Coleiro},
  {Castro}, {Sim{\'o}n-D{\'\i}az}, {Zurita Heras}, {Goldoni}, \&
  {Goldwurm}}]{Chaty...2016A&A...591A..87C}
{Chaty}, S., {LeReun}, A., {Negueruela}, I., {et~al.} 2016, \aap, 591, A87,
  \dodoi{10.1051/0004-6361/201628110}

\bibitem[{{Chernyakova} {et~al.}(2005){Chernyakova}, {Lutovinov},
  {Rodr{\'\i}guez}, \& {Revnivtsev}}]{Chernyakova...2005MNRAS.364..455C}
{Chernyakova}, M., {Lutovinov}, A., {Rodr{\'\i}guez}, J., \& {Revnivtsev}, M.
  2005, \mnras, 364, 455, \dodoi{10.1111/j.1365-2966.2005.09548.x}

\bibitem[{{Chernyakova} {et~al.}(2006){Chernyakova}, {Lutovinov}, {Rodriguez},
  \& {Revnivtsev}}]{Chernyakova...2006IAUS..230...33C}
{Chernyakova}, M., {Lutovinov}, A., {Rodriguez}, J., \& {Revnivtsev}, M. 2006,
  in Populations of High Energy Sources in Galaxies, ed. E.~J.~A. {Meurs} \&
  G.~{Fabbiano}, Vol. 230, 33--34, \dodoi{10.1017/S1743921306007745}

\bibitem[{{Chrimes} {et~al.}(2022){Chrimes}, {Levan}, {Fruchter}, {Groot},
  {Kouveliotou}, {Lyman}, {Tanvir}, \&
  {Wiersema}}]{Chrimes...2022MNRAS.512.6093C}
{Chrimes}, A.~A., {Levan}, A.~J., {Fruchter}, A.~S., {et~al.} 2022, \mnras,
  512, 6093, \dodoi{10.1093/mnras/stac870}

\bibitem[{{Christodoulou} {et~al.}(2022){Christodoulou}, {Bhattacharya},
  {Laycock}, \& {Kazanas}}]{Christodoulou...2022ApJ...929..137C}
{Christodoulou}, D.~M., {Bhattacharya}, S., {Laycock}, S. G.~T., \& {Kazanas},
  D. 2022, \apj, 929, 137, \dodoi{10.3847/1538-4357/ac5f0a}

\bibitem[{{Clark} {et~al.}(2009){Clark}, {Hill}, {Bird}, {McBride}, {Scaringi},
  \& {Dean}}]{Clark...2009MNRAS.399L.113C}
{Clark}, D.~J., {Hill}, A.~B., {Bird}, A.~J., {et~al.} 2009, \mnras, 399, L113,
  \dodoi{10.1111/j.1745-3933.2009.00737.x}

\bibitem[{{Coburn} {et~al.}(2002){Coburn}, {Heindl}, {Rothschild}, {Gruber},
  {Kreykenbohm}, {Wilms}, {Kretschmar}, \&
  {Staubert}}]{Coburn...2002ApJ...580..394C}
{Coburn}, W., {Heindl}, W.~A., {Rothschild}, R.~E., {et~al.} 2002, \apj, 580,
  394, \dodoi{10.1086/343033}

\bibitem[{{Coe} {et~al.}(2006){Coe}, {Reig}, {McBride}, {Galache}, \&
  {Fabregat}}]{Coe...2006MNRAS.368..447C}
{Coe}, M.~J., {Reig}, P., {McBride}, V.~A., {Galache}, J.~L., \& {Fabregat}, J.
  2006, \mnras, 368, 447, \dodoi{10.1111/j.1365-2966.2006.10127.x}

\bibitem[{{Coe} {et~al.}(1990){Coe}, {Carstairs}, {Court}, {Davies}, {Dean},
  {Dipper}, {Lewis}, {Perotti}, {Quadrini}, {Bazzano}, {Ubertini}, \&
  {Stephen}}]{Coe...1990MNRAS.243..475C}
{Coe}, M.~J., {Carstairs}, I.~R., {Court}, A.~J., {et~al.} 1990, \mnras, 243,
  475

\bibitem[{{Coleiro} \& {Chaty}(2013)}]{Coleiro...2013ApJ...764..185C}
{Coleiro}, A., \& {Chaty}, S. 2013, \apj, 764, 185,
  \dodoi{10.1088/0004-637X/764/2/185}

\bibitem[{{Coleiro} {et~al.}(2013){Coleiro}, {Chaty}, {Zurita Heras}, {Rahoui},
  \& {Tomsick}}]{Coleiro...2013A&A...560A.108C}
{Coleiro}, A., {Chaty}, S., {Zurita Heras}, J.~A., {Rahoui}, F., \& {Tomsick},
  J.~A. 2013, \aap, 560, A108, \dodoi{10.1051/0004-6361/201322382}

\bibitem[{{Coley} {et~al.}(2023){Coley}, {Jaisawal}, {Arzoumanian},
  {Ballhausen}, {Chakrabarty}, {Fuerst}, {Gendreau}, {Islam}, {Kretschmar},
  {Malacaria}, {Naik}, {Pottschmidt}, {Pradhan}, {Rothschild}, {Thalhammer},
  {Vasilopoulos}, {Wolff}, {Wood}, \& {Wilms}}]{Coley...2023AAS...24142807C}
{Coley}, J., {Jaisawal}, G.~K., {Arzoumanian}, Z., {et~al.} 2023, in American
  Astronomical Society Meeting Abstracts, Vol.~55, American Astronomical
  Society Meeting Abstracts, 428.07

\bibitem[{{Coley} {et~al.}(2019){Coley}, {Corbet}, {F{\"u}rst}, {Huxtable},
  {Krimm}, {Pearlman}, \& {Pottschmidt}}]{Coley...2019ApJ...879...34C}
{Coley}, J.~B., {Corbet}, R. H.~D., {F{\"u}rst}, F., {et~al.} 2019, \apj, 879,
  34, \dodoi{10.3847/1538-4357/ab223c}

\bibitem[{{Cook} \& {Warwick}(1987)}]{Cook...1987MNRAS.225..369C}
{Cook}, M.~C., \& {Warwick}, R.~S. 1987, \mnras, 225, 369,
  \dodoi{10.1093/mnras/225.2.369}

\bibitem[{{Corbet} {et~al.}(2006){Corbet}, {Barbier}, {Barthelmy}, {Cummings},
  {Fenimore}, {Gehrels}, {Hullinger}, {Krimm}, {Markwardt}, {Palmer},
  {Parsons}, {Sakamoto}, {Sato}, {Tueller}, \&
  {Remillard}}]{Corbet...2006ATel..779....1C}
{Corbet}, R., {Barbier}, L., {Barthelmy}, S., {et~al.} 2006, The Astronomer's
  Telegram, 779, 1

\bibitem[{{Corbet} {et~al.}(2010{\natexlab{a}}){Corbet}, {Barthelmy},
  {Baumgartner}, {Krimm}, {Markwardt}, {Skinner}, \&
  {Tueller}}]{Corbet...2010ATel.2598....1C}
{Corbet}, R.~H.~D., {Barthelmy}, S.~D., {Baumgartner}, W.~H., {et~al.}
  2010{\natexlab{a}}, The Astronomer's Telegram, 2598, 1

\bibitem[{{Corbet} {et~al.}(2016){Corbet}, {Coley}, \&
  {Krimm}}]{Corbet...2016ATel.9823....1C}
{Corbet}, R. H.~D., {Coley}, J.~B., \& {Krimm}, H.~A. 2016, The Astronomer's
  Telegram, 9823, 1

\bibitem[{{Corbet} {et~al.}(2017){Corbet}, {Coley}, \&
  {Krimm}}]{Corbet...2017ApJ...846..161C}
---. 2017, \apj, 846, 161, \dodoi{10.3847/1538-4357/aa8638}

\bibitem[{{Corbet} {et~al.}(2009){Corbet}, {in't Zand}, {Levine}, \&
  {Marshall}}]{Corbet...2009ApJ...695...30C}
{Corbet}, R.~H.~D., {in't Zand}, J.~J.~M., {Levine}, A.~M., \& {Marshall},
  F.~E. 2009, \apj, 695, 30, \dodoi{10.1088/0004-637X/695/1/30}

\bibitem[{{Corbet} \& {Krimm}(2009)}]{Corbet...2009ATel.2008....1C}
{Corbet}, R.~H.~D., \& {Krimm}, H.~A. 2009, The Astronomer's Telegram, 2008, 1

\bibitem[{{Corbet} \& {Krimm}(2013)}]{Corbet...2013ApJ...778...45C}
{Corbet}, R. H.~D., \& {Krimm}, H.~A. 2013, \apj, 778, 45,
  \dodoi{10.1088/0004-637X/778/1/45}

\bibitem[{{Corbet} {et~al.}(2010{\natexlab{b}}){Corbet}, {Krimm}, \&
  {Skinner}}]{Corbet...2010ATel.2559....1C}
{Corbet}, R. H.~D., {Krimm}, H.~A., \& {Skinner}, G.~K. 2010{\natexlab{b}}, The
  Astronomer's Telegram, 2559, 1

\bibitem[{{Corbet} {et~al.}(2007){Corbet}, {Markwardt}, \&
  {Tueller}}]{Corbet...2007ApJ...655..458C}
{Corbet}, R.~H.~D., {Markwardt}, C.~B., \& {Tueller}, J. 2007, \apj, 655, 458,
  \dodoi{10.1086/509319}

\bibitem[{{Corbet} {et~al.}(1999){Corbet}, {Marshall}, {Peele}, \&
  {Takeshima}}]{Corbet...1999ApJ...517..956C}
{Corbet}, R.~H.~D., {Marshall}, F.~E., {Peele}, A.~G., \& {Takeshima}, T. 1999,
  \apj, 517, 956, \dodoi{10.1086/307235}

\bibitem[{{Corbet} \& {Mukai}(2002)}]{Corbet...2002ApJ...577..923C}
{Corbet}, R. H.~D., \& {Mukai}, K. 2002, \apj, 577, 923, \dodoi{10.1086/342244}

\bibitem[{{Corbet} \& {Peele}(1997)}]{Corbet...1997ApJ...489L..83C}
{Corbet}, R. H.~D., \& {Peele}, A.~G. 1997, \apjl, 489, L83,
  \dodoi{10.1086/310972}

\bibitem[{{Corbet} \& {Peele}(2000)}]{Corbet...2000ApJ...530L..33C}
---. 2000, \apjl, 530, L33, \dodoi{10.1086/312485}

\bibitem[{{Corbet} \& {Remillard}(2005)}]{Corbet...2005ATel..377....1C}
{Corbet}, R.~H.~D., \& {Remillard}, R. 2005, The Astronomer's Telegram, 377, 1

\bibitem[{{Corbet} {et~al.}(2022){Corbet}, {Coley}, {Gendreau}, {Guillot},
  {Islam}, {Jaisawal}, {Malacaria}, {Ng}, {Pottschmidt}, {Pradhan}, {Ray},
  {Sanna}, {Wilms}, \& {Wolff}}]{Corbet...2022ATel15614....1C}
{Corbet}, R. H.~D., {Coley}, J.~B., {Gendreau}, K.~C., {et~al.} 2022, The
  Astronomer's Telegram, 15614, 1

\bibitem[{{Cox} {et~al.}(2005){Cox}, {Kaper}, \&
  {Mokiem}}]{Cox...2005A&A...436..661C}
{Cox}, N.~L.~J., {Kaper}, L., \& {Mokiem}, M.~R. 2005, \aap, 436, 661,
  \dodoi{10.1051/0004-6361:20040511}

\bibitem[{{{\c{S}}ahiner} {et~al.}(2012){{\c{S}}ahiner}, {Inam}, \&
  {Baykal}}]{Sahiner...2012MNRAS.421.2079S}
{{\c{S}}ahiner}, {\c{S}}., {Inam}, S.~{\c{C}}., \& {Baykal}, A. 2012, \mnras,
  421, 2079, \dodoi{10.1111/j.1365-2966.2012.20455.x}

\bibitem[{{Cusumano} {et~al.}(2000){Cusumano}, {Maccarone}, {Nicastro},
  {Sacco}, \& {Kaaret}}]{Cusumano...2000ApJ...528L..25C}
{Cusumano}, G., {Maccarone}, M.~C., {Nicastro}, L., {Sacco}, B., \& {Kaaret},
  P. 2000, \apjl, 528, L25, \dodoi{10.1086/312413}

\bibitem[{{Cusumano} {et~al.}(2013){Cusumano}, {Segreto}, {La Parola},
  {Masetti}, {D'Ai}, \& {Tagliaferri}}]{Cusumano...2013MNRAS.436L..74C}
{Cusumano}, G., {Segreto}, A., {La Parola}, V., {et~al.} 2013, \mnras, 436,
  L74, \dodoi{10.1093/mnrasl/slt116}

\bibitem[{{Cutri} {et~al.}(2003){Cutri}, {Skrutskie}, {van Dyk}, {Beichman},
  {Carpenter}, {Chester}, {Cambresy}, {Evans}, {Fowler}, {Gizis}, {Howard},
  {Huchra}, {Jarrett}, {Kopan}, {Kirkpatrick}, {Light}, {Marsh}, {McCallon},
  {Schneider}, {Stiening}, {Sykes}, {Weinberg}, {Wheaton}, {Wheelock}, \&
  {Zacarias}}]{Cutri...2003yCat.2246....0C}
{Cutri}, R.~M., {Skrutskie}, M.~F., {van Dyk}, S., {et~al.} 2003, VizieR Online
  Data Catalog, II/246

\bibitem[{{D'A{\`\i}} {et~al.}(2011){D'A{\`\i}}, {Cusumano}, {La Parola},
  {Segreto}, {di Salvo}, {Iaria}, \& {Robba}}]{DAi...2011A&A...532A..73D}
{D'A{\`\i}}, A., {Cusumano}, G., {La Parola}, V., {et~al.} 2011, \aap, 532,
  A73, \dodoi{10.1051/0004-6361/201117035}

\bibitem[{{D'Amico} {et~al.}(2006){D'Amico}, {Jablonski}, {Rodrigues},
  {Cieslinski}, \& {Hickel}}]{Amico...2006AIPC..840...97D}
{D'Amico}, F., {Jablonski}, F., {Rodrigues}, C.~V., {Cieslinski}, D., \&
  {Hickel}, G. 2006, in American Institute of Physics Conference Series, Vol.
  840, The Transient Milky Way: A Perspective for MIRAX, ed. F.~{D'Amico},
  J.~{Braga}, \& R.~E. {Rothschild}, 97--101, \dodoi{10.1063/1.2216611}

\bibitem[{{Davidson} \& {Ostriker}(1973)}]{Davidson...1973ApJ...179..585D}
{Davidson}, K., \& {Ostriker}, J.~P. 1973, \apj, 179, 585,
  \dodoi{10.1086/151897}

\bibitem[{{de Burgos} {et~al.}(2023){de Burgos}, {Sim{\'o}n-D{\'\i}az},
  {Urbaneja}, \& {Negueruela}}]{Burgos...2023arXiv230500305D}
{de Burgos}, A., {Sim{\'o}n-D{\'\i}az}, S., {Urbaneja}, M.~A., \& {Negueruela},
  I. 2023, arXiv e-prints, arXiv:2305.00305, \dodoi{10.48550/arXiv.2305.00305}

\bibitem[{{DeCesar} {et~al.}(2013){DeCesar}, {Boyd}, {Pottschmidt}, {Wilms},
  {Suchy}, \& {Miller}}]{DeCesar...2013ApJ...762...61D}
{DeCesar}, M.~E., {Boyd}, P.~T., {Pottschmidt}, K., {et~al.} 2013, \apj, 762,
  61, \dodoi{10.1088/0004-637X/762/1/61}

\bibitem[{{Delgado-Mart{\'\i}} {et~al.}(2001){Delgado-Mart{\'\i}}, {Levine},
  {Pfahl}, \& {Rappaport}}]{Delgado-Mart...2001ApJ...546..455D}
{Delgado-Mart{\'\i}}, H., {Levine}, A.~M., {Pfahl}, E., \& {Rappaport}, S.~A.
  2001, \apj, 546, 455, \dodoi{10.1086/318236}

\bibitem[{{Densham} \& {Charles}(1982)}]{Densham...1982MNRAS.201..171D}
{Densham}, R.~H., \& {Charles}, P.~A. 1982, \mnras, 201, 171,
  \dodoi{10.1093/mnras/201.1.171}

\bibitem[{{Deo Chandra} {et~al.}(2023){Deo Chandra}, {Roy}, \&
  {Agrawal}}]{Deo...2023arXiv230110678D}
{Deo Chandra}, A., {Roy}, J., \& {Agrawal}, P.~C. 2023, arXiv e-prints,
  arXiv:2301.10678, \dodoi{10.48550/arXiv.2301.10678}

\bibitem[{{Devaraj} \& {Paul}(2022)}]{Devaraj...2022MNRAS.514L..46D}
{Devaraj}, A., \& {Paul}, B. 2022, \mnras, 514, L46,
  \dodoi{10.1093/mnrasl/slac052}

\bibitem[{{Di Salvo} {et~al.}(1998){Di Salvo}, {Burderi}, {Robba}, \&
  {Guainazzi}}]{DiSalvo...1998ApJ...509..897D}
{Di Salvo}, T., {Burderi}, L., {Robba}, N.~R., \& {Guainazzi}, M. 1998, \apj,
  509, 897, \dodoi{10.1086/306525}

\bibitem[{{Ding} {et~al.}(2021){Ding}, {Wang}, {Zhang}, {Bu}, {Cai}, {Cao},
  {Zhi}, {Chen}, {Chen}, {Chen}, {Chen}, {Chen}, {Cui}, {Du}, {Gao}, {Gao},
  {Ge}, {Gu}, {Guan}, {Guo}, {Han}, {Huang}, {Huo}, {Jia}, {Jiang}, {Jin},
  {Kong}, {Li}, {Li}, {Li}, {Li}, {Li}, {Li}, {Li}, {Li}, {Li}, {Liang},
  {Liao}, {Liu}, {Liu}, {Liu}, {Liu}, {Liu}, {Lu}, {Lu}, {Lou}, {Luo}, {Ma},
  {Ma}, {Meng}, {Nang}, {Nie}, {Qu}, {Ren}, {Sai}, {Song}, {Song}, {Sun},
  {Tan}, {Tao}, {Tuo}, {Wang}, {Wang}, {Wang}, {Wang}, {Wang}, {Wen}, {Wu},
  {Wu}, {Wu}, {Xiao}, {Xiao}, {Xiong}, {Xu}, {Yang}, {Yang}, {Yang}, {Yi},
  {Yin}, {You}, {Zhang}, {Zhang}, {Zhang}, {Zhang}, {Zhang}, {Zhang}, {Zhang},
  {Zhang}, {Zhang}, {Zhang}, {Zhao}, {Zhao}, {Zheng}, {Zheng}, \&
  {Zhou}}]{Ding...2021MNRAS.503.6045D}
{Ding}, Y.~Z., {Wang}, W., {Zhang}, P., {et~al.} 2021, \mnras, 503, 6045,
  \dodoi{10.1093/mnras/stab835}

\bibitem[{{D{\"o}nmez} {et~al.}(2020){D{\"o}nmez}, {Serim}, {{\.I}nam},
  {{\c{S}}ahiner}, {Serim}, \& {Baykal}}]{Donmez...2020MNRAS.496.1768D}
{D{\"o}nmez}, {\c{C}}.~K., {Serim}, M.~M., {{\.I}nam}, S.~{\c{C}}., {et~al.}
  2020, \mnras, 496, 1768, \dodoi{10.1093/mnras/staa1562}

\bibitem[{{Doroshenko} {et~al.}(2021){Doroshenko}, {Santangelo}, {Tsygankov},
  \& {Ji}}]{Doroshenko...2021A&A...647A.165D}
{Doroshenko}, V., {Santangelo}, A., {Tsygankov}, S.~S., \& {Ji}, L. 2021, \aap,
  647, A165, \dodoi{10.1051/0004-6361/202039785}

\bibitem[{{Doroshenko} {et~al.}(2010){Doroshenko}, {Suchy}, {Santangelo},
  {Staubert}, {Kreykenbohm}, {Rothschild}, {Pottschmidt}, \&
  {Wilms}}]{Doroshenko...2010A&A...515L...1D}
{Doroshenko}, V., {Suchy}, S., {Santangelo}, A., {et~al.} 2010, \aap, 515, L1,
  \dodoi{10.1051/0004-6361/201014858}

\bibitem[{{Doroshenko} {et~al.}(2020{\natexlab{a}}){Doroshenko}, {Tsygankov},
  {Long}, {Santangelo}, {Molkov}, {Lutovinov}, {Kong}, \&
  {Zhang}}]{Doroshenko...2020A&A...634A..89D}
{Doroshenko}, V., {Tsygankov}, S., {Long}, J., {et~al.} 2020{\natexlab{a}},
  \aap, 634, A89, \dodoi{10.1051/0004-6361/201937036}

\bibitem[{{Doroshenko} {et~al.}(2018){Doroshenko}, {Tsygankov}, \&
  {Santangelo}}]{Doroshenko...2018A&A...613A..19D}
{Doroshenko}, V., {Tsygankov}, S., \& {Santangelo}, A. 2018, \aap, 613, A19,
  \dodoi{10.1051/0004-6361/201732208}

\bibitem[{{Doroshenko} {et~al.}(2020{\natexlab{b}}){Doroshenko}, {Zhang},
  {Santangelo}, {Ji}, {Tsygankov}, {Mushtukov}, {Qu}, {Zhang}, {Ge}, {Chen},
  {Bu}, {Cao}, {Chang}, {Chen}, {Chen}, {Chen}, {Chen}, {Chen}, {Cui}, {Cui},
  {Deng}, {Dong}, {Du}, {Fu}, {Gao}, {Gao}, {Gao}, {Gu}, {Guan}, {Guo}, {Han},
  {Hu}, {Huang}, {Huo}, {Jia}, {Jiang}, {Jiang}, {Jin}, {Jin}, {Kong}, {Li},
  {Li}, {Li}, {Li}, {Li}, {Li}, {Li}, {Li}, {Li}, {Li}, {Li}, {Li}, {Liang},
  {Liao}, {Liu}, {Liu}, {Liu}, {Liu}, {Liu}, {Liu}, {Liu}, {Lu}, {Lu}, {Lu},
  {Luo}, {Ma}, {Meng}, {Nang}, {Nie}, {Ou}, {Sai}, {Shang}, {Song}, {Song},
  {Sun}, {Tan}, {Tao}, {Tuo}, {Wang}, {Wang}, {Wang}, {Wang}, {Wen}, {Wu},
  {Wu}, {Xiao}, {Xiong}, {Xu}, {Xu}, {Yang}, {Yang}, {Yang}, {Yang}, {Zhang},
  {Zhang}, {Zhang}, {Zhang}, {Zhang}, {Zhang}, {Zhang}, {Zhang}, {Zhang},
  {Zhang}, {Zhang}, {Zhang}, {Zhang}, {Zhang}, {Zhang}, {Zhang}, {Zhang},
  {Zhao}, {Zhao}, {Zhao}, {Zheng}, {Zhu}, {Zhu}, {Zou}, \&
  {Zhang}}]{Doroshenko...2020MNRAS.491.1857D}
{Doroshenko}, V., {Zhang}, S.~N., {Santangelo}, A., {et~al.}
  2020{\natexlab{b}}, \mnras, 491, 1857, \dodoi{10.1093/mnras/stz2879}

\bibitem[{{Doroshenko} {et~al.}(2008){Doroshenko}, {Doroshenko}, {Postnov},
  {Cherepashchuk}, \& {Tsygankov}}]{Doroshenko...2008ARep...52..138D}
{Doroshenko}, V.~A., {Doroshenko}, R.~F., {Postnov}, K.~A., {Cherepashchuk},
  A.~M., \& {Tsygankov}, S.~S. 2008, Astronomy Reports, 52, 138,
  \dodoi{10.1134/S1063772908020054}

\bibitem[{{Ducati}(2002)}]{Ducati...2002yCat.2237....0D}
{Ducati}, J.~R. 2002, VizieR Online Data Catalog

\bibitem[{{Esposito} {et~al.}(2013){Esposito}, {Israel}, {Sidoli}, {Mason},
  {Rodriguez}, {Halpern}, {Moretti}, \&
  {Gotz}}]{Esposito...2013MNRAS..433..P2028}
{Esposito}, P., {Israel}, G., {Sidoli}, L., {et~al.} 2013, \mnras, 433,
  2028–2035, \dodoi{10.1093/mnras/stt870}

\bibitem[{{Esposito} {et~al.}(2020){Esposito}, {Israel}, {Rea}, {Borghese}, \&
  {Coti Zelati}}]{Esposito...2020ATel13625....1E}
{Esposito}, P., {Israel}, G.~L., {Rea}, N., {Borghese}, A., \& {Coti Zelati},
  F. 2020, The Astronomer's Telegram, 13625, 1

\bibitem[{{Esposito} {et~al.}(2014){Esposito}, {Israel}, {Sidoli}, {Tiengo},
  {Campana}, \& {Moretti}}]{Esposito...2014MNRAS.441.1126E}
{Esposito}, P., {Israel}, G.~L., {Sidoli}, L., {et~al.} 2014, \mnras, 441,
  1126, \dodoi{10.1093/mnras/stu659}

\bibitem[{{Falanga} {et~al.}(2015){Falanga}, {Bozzo}, {Lutovinov},
  {Bonnet-Bidaud}, {Fetisova}, \& {Puls}}]{Falanga...2015AandAP..577..P1}
{Falanga}, M., {Bozzo}, E., {Lutovinov}, A., {et~al.} 2015, \aap, 577, 1–16,
  \dodoi{10.1051/0004-6361/201425191}

\bibitem[{Fermi-Collaboration(2022)}]{Fermi_pulsars}
Fermi-Collaboration. 2022, GBM Accreting Pulsar Histories.
\newblock \url{https://gammaray.nsstc.nasa.gov/gbm/science/pulsars.html}

\bibitem[{{Fermi Collaboration (1A 0535$+$262)}(2021)}]{Fermi_GBM_1A_0535_262}
{Fermi Collaboration (1A 0535$+$262)}. 2021, {Fermi GBM Accreting Pulsar
  Histories (1A 0535$+$262)}.
\newblock
  \url{https://gammaray.nsstc.nasa.gov/gbm/science/pulsars/lightcurves/a0535.html}

\bibitem[{{Fermi Collaboration (4U 1538-52)}(2021)}]{Fermi_GBM_4U_1538_52}
{Fermi Collaboration (4U 1538-52)}. 2021, Fermi GBM Accreting Pulsar Histories
  (4U 1538-52).
\newblock
  \url{https://gammaray.nsstc.nasa.gov/gbm/science/pulsars/lightcurves/4u1538.html}

\bibitem[{{Fermi Collaboration (Cen X-3)}(2021)}]{Fermi_GBM_Cen_X_3}
{Fermi Collaboration (Cen X-3)}. 2021, Fermi GBM Accreting Pulsar Histories
  (Cen X-3).
\newblock
  \url{https://gammaray.nsstc.nasa.gov/gbm/science/pulsars/lightcurves/cenx3.html}

\bibitem[{{Fermi Collaboration (Cep X-4)}(2021)}]{Fermi_GBM_Cep_X-4}
{Fermi Collaboration (Cep X-4)}. 2021, Fermi GBM Accreting Pulsar Histories
  (Cep X-4).
\newblock
  \url{https://gammaray.nsstc.nasa.gov/gbm/science/pulsars/lightcurves/cepx4.html}

\bibitem[{{Fermi Collaboration (GRO J1008-57)}(2021)}]{Fermi_GBM_GRO_J1008-57}
{Fermi Collaboration (GRO J1008-57)}. 2021, Fermi GBM Accreting Pulsar
  Histories (GRO J1008-57).
\newblock
  \url{https://gammaray.nsstc.nasa.gov/gbm/science/pulsars/lightcurves/groj1008.html}

\bibitem[{{Fermi Collaboration (GX 301-2)}(2021)}]{Fermi_GBM_GX_301_2}
{Fermi Collaboration (GX 301-2)}. 2021, Fermi GBM Accreting Pulsar Histories
  (GX 301-2).
\newblock
  \url{https://gammaray.nsstc.nasa.gov/gbm/science/pulsars/lightcurves/gx301m2.html}

\bibitem[{{Fermi Collaboration (IGR
  J19294+1816}(2021)}]{Fermi_GBM_IGR_J19294_1816}
{Fermi Collaboration (IGR J19294+1816}. 2021, Fermi GBM Accreting Pulsar
  Histories (IGR J19294+1816).
\newblock
  \url{https://gammaray.msfc.nasa.gov/gbm/science/pulsars/lightcurves/igrj19294.html}

\bibitem[{{Fermi Collaboration (KS 1947+300)}(2021)}]{Fermi_GBM_KS_1947_300}
{Fermi Collaboration (KS 1947+300)}. 2021, Fermi GBM Accreting Pulsar Histories
  (KS 1947+300).
\newblock
  \url{https://gammaray.nsstc.nasa.gov/gbm/science/pulsars/lightcurves/ks1947.html}

\bibitem[{{Fermi Collaboration (MAXI J1409-619)}(2021)}]{Fermi_GBM_MAXI}
{Fermi Collaboration (MAXI J1409-619)}. 2021, Fermi GBM Accreting Pulsar
  Histories (MAXI J1409-619).
\newblock
  \url{https://gammaray.nsstc.nasa.gov/gbm/science/pulsars/lightcurves/maxij1409.html}

\bibitem[{{Fermi Collaboration (OAO 1657-415)}(2021)}]{Fermi_GBM_OAO-1657}
{Fermi Collaboration (OAO 1657-415)}. 2021, Fermi GBM Accreting Pulsar
  Histories (OAO 1657-415).
\newblock
  \url{http://gammaray.msfc.nasa.gov/\\gbm/science/pulsars/lightcurves/oao1657.fits.gz}

\bibitem[{{Fermi Collaboration (RX J0440.9+4431)}(2021)}]{Fermi_GBM_RX_J0440}
{Fermi Collaboration (RX J0440.9+4431)}. 2021, Fermi GBM Accreting Pulsar
  Histories (RX J0440.9+4431).
\newblock
  \url{https://gammaray.msfc.nasa.gov/gbm/science/pulsars/lightcurves/rxj0440.html}

\bibitem[{{Fermi Collaboration (Vela X-1)}(2021)}]{Fermi_GBM_Vela_X_1}
{Fermi Collaboration (Vela X-1)}. 2021, Fermi GBM Accreting Pulsar Histories
  (Vela X-1).
\newblock
  \url{https://gammaray.nsstc.nasa.gov/gbm/science/pulsars/lightcurves/velax1.html}

\bibitem[{{Ferrigno} {et~al.}(2013){Ferrigno}, {Farinelli}, {Bozzo},
  {Pottschmidt}, {Klochkov}, \& {Kretschmar}}]{Ferrigno...2013A&A...553A.103F}
{Ferrigno}, C., {Farinelli}, R., {Bozzo}, E., {et~al.} 2013, \aap, 553, A103,
  \dodoi{10.1051/0004-6361/201321053}

\bibitem[{{Finger} {et~al.}(1999){Finger}, {Bildsten}, {Chakrabarty}, {Prince},
  {Scott}, {Wilson}, {Wilson}, \& {Zhang}}]{Finger...1999ApJ...517..449F}
{Finger}, M.~H., {Bildsten}, L., {Chakrabarty}, D., {et~al.} 1999, \apj, 517,
  449, \dodoi{10.1086/307191}

\bibitem[{{Finger} {et~al.}(2010){Finger}, {Ikhsanov}, {Wilson-Hodge}, \&
  {Patel}}]{Finger...2010ApJ...709.1249F}
{Finger}, M.~H., {Ikhsanov}, N.~R., {Wilson-Hodge}, C.~A., \& {Patel}, S.~K.
  2010, \apj, 709, 1249, \dodoi{10.1088/0004-637X/709/2/1249}

\bibitem[{{Finger} {et~al.}(1996){Finger}, {Wilson}, \&
  {Chakrabarty}}]{Finger...1996A&AS..120C.209F}
{Finger}, M.~H., {Wilson}, R.~B., \& {Chakrabarty}, D. 1996, \aaps, 120, 209

\bibitem[{{Fortin} {et~al.}(2023){Fortin}, {Garc{\'\i}a}, {Simaz Bunzel}, \&
  {Chaty}}]{Fortin...2023A&A...671A.149F}
{Fortin}, F., {Garc{\'\i}a}, F., {Simaz Bunzel}, A., \& {Chaty}, S. 2023, \aap,
  671, A149, \dodoi{10.1051/0004-6361/202245236}

\bibitem[{{F{\"u}rst} {et~al.}(2014{\natexlab{a}}){F{\"u}rst}, {Pottschmidt},
  {Wilms}, {Tomsick}, {Bachetti}, {Boggs}, {Christensen}, {Craig},
  {Grefenstette}, {Hailey}, {Harrison}, {Madsen}, {Miller}, {Stern}, {Walton},
  \& {Zhang}}]{Furst...2014ApJ...780..133F}
{F{\"u}rst}, F., {Pottschmidt}, K., {Wilms}, J., {et~al.} 2014{\natexlab{a}},
  \apj, 780, 133, \dodoi{10.1088/0004-637X/780/2/133}

\bibitem[{{F{\"u}rst} {et~al.}(2014{\natexlab{b}}){F{\"u}rst}, {Pottschmidt},
  {Wilms}, {Kennea}, {Bachetti}, {Bellm}, {Boggs}, {Chakrabarty},
  {Christensen}, {Craig}, {Hailey}, {Harrison}, {Stern}, {Tomsick}, {Walton},
  \& {Zhang}}]{Furst...2014ApJ...784L..40F}
---. 2014{\natexlab{b}}, \apjl, 784, L40, \dodoi{10.1088/2041-8205/784/2/L40}

\bibitem[{GAIA-Collaboration(2022)}]{GAIA_DR3}
GAIA-Collaboration. 2022, GAIA DR3 catalog.
\newblock \url{https://www.cosmos.esa.int/web/gaia/data-release-3}

\bibitem[{{Galloway} {et~al.}(2004){Galloway}, {Morgan}, \&
  {Levine}}]{Galloway...2004ApJ...613.1164G}
{Galloway}, D.~K., {Morgan}, E.~H., \& {Levine}, A.~M. 2004, \apj, 613, 1164,
  \dodoi{10.1086/423265}

\bibitem[{{Galloway} {et~al.}(2005){Galloway}, {Wang}, \&
  {Morgan}}]{Galloway...2005ApJ...635.1217G}
{Galloway}, D.~K., {Wang}, Z., \& {Morgan}, E.~H. 2005, \apj, 635, 1217,
  \dodoi{10.1086/497573}

\bibitem[{{Garc{\'\i}a} {et~al.}(2018){Garc{\'\i}a}, {Fogantini}, {Chaty}, \&
  {Combi}}]{Garcia...2018A&A...618A..61G}
{Garc{\'\i}a}, F., {Fogantini}, F.~A., {Chaty}, S., \& {Combi}, J.~A. 2018,
  \aap, 618, A61, \dodoi{10.1051/0004-6361/201833365}

\bibitem[{{Ge} {et~al.}(2020){Ge}, {Ji}, {Zhang}, {Santangelo}, {Liu},
  {Doroshenko}, {Staubert}, {Qu}, {Zhang}, {Lu}, {Song}, {Li}, {Tao}, {Xu},
  {Cao}, {Chen}, {Bu}, {Cai}, {Chang}, {Chen}, {Chen}, {Chen}, {Chen}, {Chen},
  {Cui}, {Cui}, {Deng}, {Dong}, {Du}, {Fu}, {Gao}, {Gao}, {Gao}, {Gu}, {Guan},
  {Guo}, {Han}, {Huang}, {Huo}, {Jia}, {Jiang}, {Jiang}, {Jin}, {Jin}, {Kong},
  {Li}, {Li}, {Li}, {Li}, {Li}, {Li}, {Li}, {Li}, {Li}, {Li}, {Liang}, {Liao},
  {Liu}, {Liu}, {Liu}, {Liu}, {Liu}, {Lu}, {Lu}, {Luo}, {Luo}, {Ma}, {Meng},
  {Nang}, {Nie}, {Ou}, {Sai}, {Shang}, {Song}, {Sun}, {Tan}, {Tuo}, {Wang},
  {Wang}, {Wang}, {Wang}, {Wang}, {Wang}, {Wang}, {Wen}, {Wu}, {Wu}, {Wu},
  {Xiao}, {Xiao}, {Xiong}, {Xu}, {Yang}, {Yang}, {Yang}, {Yang}, {Yi}, {Yin},
  {You}, {Zhang}, {Zhang}, {Zhang}, {Zhang}, {Zhang}, {Zhang}, {Zhang},
  {Zhang}, {Zhang}, {Zhang}, {Zhang}, {Zhang}, {Zhang}, {Zhang}, {Zhang},
  {Zhang}, {Zhao}, {Zhao}, {Zheng}, {Zheng}, {Zhou}, {Zhou}, {Zhuang}, {Zhu},
  \& {Zhu}}]{Ge...2020ApJ...899L..19G}
{Ge}, M.~Y., {Ji}, L., {Zhang}, S.~N., {et~al.} 2020, \apjl, 899, L19,
  \dodoi{10.3847/2041-8213/abac05}

\bibitem[{{Ghising} {et~al.}(2022){Ghising}, {Tobrej}, {Rai}, {Tamang}, \&
  {Paul}}]{Ghising...2022MNRAS.517.4132G}
{Ghising}, M., {Tobrej}, M., {Rai}, B., {Tamang}, R., \& {Paul}, B.~C. 2022,
  \mnras, 517, 4132, \dodoi{10.1093/mnras/stac2890}

\bibitem[{{Ghosh} \& {Lamb}(1979{\natexlab{a}})}]{Ghosh...1979ApJ...232..259G}
{Ghosh}, P., \& {Lamb}, F.~K. 1979{\natexlab{a}}, \apj, 232, 259,
  \dodoi{10.1086/157285}

\bibitem[{{Ghosh} \& {Lamb}(1979{\natexlab{b}})}]{Ghosh...1979ApJ...234..296G}
---. 1979{\natexlab{b}}, \apj, 234, 296, \dodoi{10.1086/157498}

\bibitem[{{Giangrande} {et~al.}(1980){Giangrande}, {Giovannelli}, {Bartolini},
  {Guarnieri}, \& {Piccioni}}]{Giangrande...1980A&AS...40..289G}
{Giangrande}, A., {Giovannelli}, F., {Bartolini}, C., {Guarnieri}, A., \&
  {Piccioni}, A. 1980, \aaps, 40, 289

\bibitem[{{Gnedin} \& {Sunyaev}(1974)}]{Gnedin...1974A&A....36..379G}
{Gnedin}, I.~N., \& {Sunyaev}, R.~A. 1974, \aap, 36, 379

\bibitem[{{Gorban} {et~al.}(2022){Gorban}, {Molkov}, {Lutovinov}, \&
  {Semena}}]{Gorban...2022AstL...48..798G}
{Gorban}, A.~S., {Molkov}, S.~V., {Lutovinov}, A.~A., \& {Semena}, A.~N. 2022,
  Astronomy Letters, 48, 798, \dodoi{10.1134/S106377372211007X}

\bibitem[{{G{\"o}{\v{g}}{\"u}{\c{s}}}
  {et~al.}(2005){G{\"o}{\v{g}}{\"u}{\c{s}}}, {Patel}, {Wilson}, {Woods},
  {Finger}, \& {Kouveliotou}}]{Gogus...2005ApJ...632.1069G}
{G{\"o}{\v{g}}{\"u}{\c{s}}}, E., {Patel}, S.~K., {Wilson}, C.~A., {et~al.}
  2005, \apj, 632, 1069, \dodoi{10.1086/444373}

\bibitem[{{Grindlay} {et~al.}(1984){Grindlay}, {Petro}, \&
  {McClintock}}]{Grindlay...1984ApJ...276..621G}
{Grindlay}, J.~E., {Petro}, L.~D., \& {McClintock}, J.~E. 1984, \apj, 276, 621,
  \dodoi{10.1086/161650}

\bibitem[{{Hainich} {et~al.}(2020){Hainich}, {Oskinova}, {Torrej{\'o}n},
  {Fuerst}, {Bodaghee}, {Shenar}, {Sander}, {Todt}, {Spetzer}, \&
  {Hamann}}]{Hainich...2020A&A...634A..49H}
{Hainich}, R., {Oskinova}, L.~M., {Torrej{\'o}n}, J.~M., {et~al.} 2020, \aap,
  634, A49, \dodoi{10.1051/0004-6361/201935498}

\bibitem[{{Halpern} \& {Gotthelf}(2007)}]{Halpern...2007ApJ...669..579H}
{Halpern}, J.~P., \& {Gotthelf}, E.~V. 2007, \apj, 669, 579,
  \dodoi{10.1086/521704}

\bibitem[{Harris {et~al.}(2020)Harris, Millman, van~der Walt, Gommers,
  Virtanen, Cournapeau, Wieser, Taylor, Berg, Smith, Kern, Picus, Hoyer, van
  Kerkwijk, Brett, Haldane, del R{\'{i}}o, Wiebe, Peterson,
  G{\'{e}}rard-Marchant, Sheppard, Reddy, Weckesser, Abbasi, Gohlke, \&
  Oliphant}]{Harris...Nature2020...585...7825H}
Harris, C.~R., Millman, K.~J., van~der Walt, S.~J., {et~al.} 2020, Nature, 585,
  357, \dodoi{10.1038/s41586-020-2649-2}

\bibitem[{{Hemphill} {et~al.}(2019{\natexlab{a}}){Hemphill}, {Coley}, {Fuerst},
  {Kretschmar}, {Kuehnel}, {Malacaria}, \&
  {Pottschmidt}}]{Hemphill...2019ATel12556....1H}
{Hemphill}, P., {Coley}, J., {Fuerst}, F., {et~al.} 2019{\natexlab{a}}, The
  Astronomer's Telegram, 12556, 1

\bibitem[{{Hemphill} {et~al.}(2013){Hemphill}, {Rothschild}, {Caballero},
  {Pottschmidt}, {K{\"u}hnel}, {F{\"u}rst}, \&
  {Wilms}}]{Hemphill...2013ApJ...777...61H}
{Hemphill}, P.~B., {Rothschild}, R.~E., {Caballero}, I., {et~al.} 2013, \apj,
  777, 61, \dodoi{10.1088/0004-637X/777/1/61}

\bibitem[{{Hemphill} {et~al.}(2019{\natexlab{b}}){Hemphill}, {Rothschild},
  {Cheatham}, {F{\"u}rst}, {Kretschmar}, {K{\"u}hnel}, {Pottschmidt},
  {Staubert}, {Wilms}, \& {Wolff}}]{Hemphill...2019ApJ...873...62H}
{Hemphill}, P.~B., {Rothschild}, R.~E., {Cheatham}, D.~M., {et~al.}
  2019{\natexlab{b}}, \apj, 873, 62, \dodoi{10.3847/1538-4357/ab03d3}

\bibitem[{{Hill} {et~al.}(2005){Hill}, {Walter}, {Knigge}, {Bazzano},
  {B{\'e}langer}, {Bird}, {Dean}, {Galache}, {Malizia}, {Renaud}, {Stephen}, \&
  {Ubertini}}]{Hill...2005A&A...439..255H}
{Hill}, A.~B., {Walter}, R., {Knigge}, C., {et~al.} 2005, \aap, 439, 255,
  \dodoi{10.1051/0004-6361:20052937}

\bibitem[{{Hill} {et~al.}(2007){Hill}, {Bird}, {Dean}, {McBride}, {Sguera},
  {Clark}, {Molina}, {Scaringi}, \& {Shaw}}]{Hill...2007MNRAS.381.1275H}
{Hill}, A.~B., {Bird}, A.~J., {Dean}, A.~J., {et~al.} 2007, \mnras, 381, 1275,
  \dodoi{10.1111/j.1365-2966.2007.12326.x}

\bibitem[{{H{\o}g} {et~al.}(2000){H{\o}g}, {Fabricius}, {Makarov}, {Urban},
  {Corbin}, {Wycoff}, {Bastian}, {Schwekendiek}, \&
  {Wicenec}}]{Hog...2000A&A...355L..27H}
{H{\o}g}, E., {Fabricius}, C., {Makarov}, V.~V., {et~al.} 2000, \aap, 355, L27

\bibitem[{{Houk}(1978)}]{Houk...1978mcts.book.....H}
{Houk}, N. 1978, {Michigan catalogue of two-dimensional spectral types for the
  HD stars} (Department of Astronomy, University of Michigan00)

\bibitem[{{Huckle} {et~al.}(1977){Huckle}, {Mason}, {White}, {Sanford},
  {Maraschi}, {Tarenghi}, \& {Tapia}}]{Huckle...1977MNRAS.180P..21H}
{Huckle}, H.~E., {Mason}, K.~O., {White}, N.~E., {et~al.} 1977, \mnras, 180,
  21P, \dodoi{10.1093/mnras/180.1.21P}

\bibitem[{{Hunter}(2007)}]{Hunter...CSE2007...9...90}
{Hunter}, J.~D. 2007, Computing in Science \& Engineering, 9, 90,
  \dodoi{10.1109/MCSE.2007.55}

\bibitem[{{Hutchings} {et~al.}(1979){Hutchings}, {Crampton}, {van Paradijs}, \&
  {White}}]{Hutchings...1979ApJ...229..P1079}
{Hutchings}, J., {Crampton}, D., {van Paradijs}, J., \& {White}, N. 1979, \apj,
  229, 1079, \dodoi{10.1086/157042}

\bibitem[{{Ikhsanov} \& {Finger}(2012)}]{Ikhsanov...2012ApJ...753....1I}
{Ikhsanov}, N.~R., \& {Finger}, M.~H. 2012, \apj, 753, 1,
  \dodoi{10.1088/0004-637X/753/1/1}

\bibitem[{{Ikhsanov} {et~al.}(2014){Ikhsanov}, {Likh}, \&
  {Beskrovnaya}}]{Ikhsanov...2014ARep...58..376I}
{Ikhsanov}, N.~R., {Likh}, Y.~S., \& {Beskrovnaya}, N.~G. 2014, Astronomy
  Reports, 58, 376, \dodoi{10.1134/S1063772914050035}

\bibitem[{{Ikhsanov} \& {Mereghetti}(2015)}]{Ikhsanov...2015MNRAS.454.3760I}
{Ikhsanov}, N.~R., \& {Mereghetti}, S. 2015, \mnras, 454, 3760,
  \dodoi{10.1093/mnras/stv2108}

\bibitem[{{{\.I}nam} {et~al.}(2004){{\.I}nam}, {Baykal}, {Matthew Scott},
  {Finger}, \& {Swank}}]{Inam...2004MNRAS.349..173I}
{{\.I}nam}, S.~{\c{C}}., {Baykal}, A., {Matthew Scott}, D., {Finger}, M., \&
  {Swank}, J. 2004, \mnras, 349, 173, \dodoi{10.1111/j.1365-2966.2004.07478.x}

\bibitem[{{in't Zand} \& {Heise}(2004)}]{Zand...2004ATel..362....1I}
{in't Zand}, J., \& {Heise}, J. 2004, The Astronomer's Telegram, 362, 1

\bibitem[{{in't Zand} {et~al.}(2001{\natexlab{a}}){in't Zand}, {Corbet}, \&
  {Marshall}}]{Zand...2001ApJ...553L.165I}
{in't Zand}, J.~J.~M., {Corbet}, R.~H.~D., \& {Marshall}, F.~E.
  2001{\natexlab{a}}, \apjl, 553, L165, \dodoi{10.1086/320688}

\bibitem[{{in't Zand} {et~al.}(2000){in't Zand}, {Halpern}, {Eracleous},
  {McCollough}, {Augusteijn}, {Remillard}, \&
  {Heise}}]{Zand...2000A&A...361...85I}
{in't Zand}, J.~J.~M., {Halpern}, J., {Eracleous}, M., {et~al.} 2000, \aap,
  361, 85

\bibitem[{{in't Zand} {et~al.}(2007){in't Zand}, {Kuiper}, {den Hartog},
  {Hermsen}, \& {Corbet}}]{Zand...2007A&A...469.1063I}
{in't Zand}, J.~J.~M., {Kuiper}, L., {den Hartog}, P.~R., {Hermsen}, W., \&
  {Corbet}, R.~H.~D. 2007, \aap, 469, 1063, \dodoi{10.1051/0004-6361:20077189}

\bibitem[{{in't Zand} {et~al.}(2001{\natexlab{b}}){in't Zand}, {Swank},
  {Corbet}, \& {Markwardt}}]{Zand...2001A&A...380L..26I}
{in't Zand}, J.~J.~M., {Swank}, J., {Corbet}, R.~H.~D., \& {Markwardt}, C.~B.
  2001{\natexlab{b}}, \aap, 380, L26, \dodoi{10.1051/0004-6361:20011512}

\bibitem[{Integral-Collaboration(2022)}]{Integral_pulsars}
Integral-Collaboration. 2022, INTEGRAL Data Archives.
\newblock \url{https://www.cosmos.esa.int/web/integral/integral-data-archives}

\bibitem[{{Islam} {et~al.}(2015){Islam}, {Maitra}, {Pradhan}, \&
  {Paul}}]{Nazma...2015MNRAS.446.4148I}
{Islam}, N., {Maitra}, C., {Pradhan}, P., \& {Paul}, B. 2015, \mnras, 446,
  4148, \dodoi{10.1093/mnras/stu2395}

\bibitem[{{Israel} {et~al.}(2000){Israel}, {Covino}, {Campana}, {Polcaro},
  {Roche}, {Stella}, {Di Paola}, {Lazzati}, {Mereghetti}, {Giallongo},
  {Fontana}, \& {Verrecchia}}]{Israel...2000MNRAS.314...87I}
{Israel}, G.~L., {Covino}, S., {Campana}, S., {et~al.} 2000, \mnras, 314, 87,
  \dodoi{10.1046/j.1365-8711.2000.03404.x}

\bibitem[{{Israel} {et~al.}(2001){Israel}, {Negueruela}, {Campana}, {Covino},
  {Di Paola}, {Maxwell}, {Norton}, {Speziali}, {Verrecchia}, \&
  {Stella}}]{Israel...2001A&A...371.1018I}
{Israel}, G.~L., {Negueruela}, I., {Campana}, S., {et~al.} 2001, \aap, 371,
  1018, \dodoi{10.1051/0004-6361:20010417}

\bibitem[{{Israel} {et~al.}(2017){Israel}, {Belfiore}, {Stella}, {Esposito},
  {Casella}, {De Luca}, {Marelli}, {Papitto}, {Perri}, {Puccetti}, {Castillo},
  {Salvetti}, {Tiengo}, {Zampieri}, {D'Agostino}, {Greiner}, {Haberl},
  {Novara}, {Salvaterra}, {Turolla}, {Watson}, {Wilms}, \&
  {Wolter}}]{Israel...2017Sci...355..817I}
{Israel}, G.~L., {Belfiore}, A., {Stella}, L., {et~al.} 2017, Science, 355,
  817, \dodoi{10.1126/science.aai8635}

\bibitem[{{Jaisawal} {et~al.}(2020){Jaisawal}, {Naik}, {Ho}, {Kumari}, {Epili},
  \& {Vasilopoulos}}]{Jaisawal...2020MNRAS.498.4830J}
{Jaisawal}, G.~K., {Naik}, S., {Ho}, W. C.~G., {et~al.} 2020, \mnras, 498,
  4830, \dodoi{10.1093/mnras/staa2604}

\bibitem[{{Jaisawal} {et~al.}(2013){Jaisawal}, {Naik}, \&
  {Paul}}]{Jaisawal...2013ApJ...779...54J}
{Jaisawal}, G.~K., {Naik}, S., \& {Paul}, B. 2013, \apj, 779, 54,
  \dodoi{10.1088/0004-637X/779/1/54}

\bibitem[{{Janot-Pacheco} {et~al.}(1981){Janot-Pacheco}, {Ilovaisky}, \&
  {Chevalier}}]{Janot-Pacheco...1981A&A....99..274J}
{Janot-Pacheco}, E., {Ilovaisky}, S.~A., \& {Chevalier}, C. 1981, \aap, 99, 274

\bibitem[{{Jenke} {et~al.}(2012{\natexlab{a}}){Jenke}, {Finger},
  {Wilson-Hodge}, \& {Camero-Arranz}}]{Jenke...2012ApJ...759..P124}
{Jenke}, P., {Finger}, M., {Wilson-Hodge}, C., \& {Camero-Arranz}, A.
  2012{\natexlab{a}}, \apj, 759, 124, \dodoi{10.1088/0004-637X/759/2/124}

\bibitem[{{Jenke} {et~al.}(2012{\natexlab{b}}){Jenke}, {Finger}, \&
  {Connaughton}}]{Jenke...2012ATel.4235....1J}
{Jenke}, P., {Finger}, M.~H., \& {Connaughton}, V. 2012{\natexlab{b}}, The
  Astronomer's Telegram, 4235, 1

\bibitem[{{Jenke} \& {Wilson-Hodge}(2017)}]{Jenke...2017ATel10812....1J}
{Jenke}, P., \& {Wilson-Hodge}, C.~A. 2017, The Astronomer's Telegram, 10812, 1

\bibitem[{{Kaaret} {et~al.}(2000){Kaaret}, {Cusumano}, \&
  {Sacco}}]{Kaaret...2000ApJ...542L..41K}
{Kaaret}, P., {Cusumano}, G., \& {Sacco}, B. 2000, \apjl, 542, L41,
  \dodoi{10.1086/312918}

\bibitem[{{Kaaret} {et~al.}(1999){Kaaret}, {Piraino}, {Halpern}, \&
  {Eracleous}}]{Kaaret...1999ApJ...523..197K}
{Kaaret}, P., {Piraino}, S., {Halpern}, J., \& {Eracleous}, M. 1999, \apj, 523,
  197, \dodoi{10.1086/307711}

\bibitem[{{Kaper} {et~al.}(1995){Kaper}, {Lamers}, {Ruymaekers}, {van den
  Heuvel}, \& {Zuiderwijk}}]{Kaper...1995A&A...300..446K}
{Kaper}, L., {Lamers}, H.~J.~G.~L.~M., {Ruymaekers}, E., {van den Heuvel},
  E.~P.~J., \& {Zuiderwijk}, E.~J. 1995, \aap, 300, 446

\bibitem[{{Karasev} {et~al.}(2010){Karasev}, {Lutovinov}, \&
  {Burenin}}]{Karasev...2010MNRAS.409L..69K}
{Karasev}, D.~I., {Lutovinov}, A.~A., \& {Burenin}, R.~A. 2010, \mnras, 409,
  L69, \dodoi{10.1111/j.1745-3933.2010.00949.x}

\bibitem[{{Karasev} {et~al.}(2008){Karasev}, {Tsygankov}, \&
  {Lutovinov}}]{Karasev...2008MNRAS.386L..10K}
{Karasev}, D.~I., {Tsygankov}, S.~S., \& {Lutovinov}, A.~A. 2008, \mnras, 386,
  L10, \dodoi{10.1111/j.1745-3933.2008.00449.x}

\bibitem[{{Kaur} {et~al.}(2008){Kaur}, {Paul}, {Kumar}, \&
  {Sagar}}]{Kaur...2008MNRAS.386.2253K}
{Kaur}, R., {Paul}, B., {Kumar}, B., \& {Sagar}, R. 2008, \mnras, 386, 2253,
  \dodoi{10.1111/j.1365-2966.2008.13233.x}

\bibitem[{{Kaur} {et~al.}(2009){Kaur}, {Wijnands}, {Patruno}, {Testa},
  {Israel}, {Degenaar}, {Paul}, \& {Kumar}}]{Kaur...2009MNRAS..394..P1597}
{Kaur}, R., {Wijnands}, R., {Patruno}, A., {et~al.} 2009, \mnras, 394,
  1597–1604, \dodoi{10.1111/j.1365-2966.2009.14438.x}

\bibitem[{{Kaur} {et~al.}(2010{\natexlab{a}}){Kaur}, {Wijnands}, {Paul},
  {Patruno}, \& {Degenaar}}]{Kaur...2010MNRAS.402.2388K}
{Kaur}, R., {Wijnands}, R., {Paul}, B., {Patruno}, A., \& {Degenaar}, N.
  2010{\natexlab{a}}, \mnras, 402, 2388,
  \dodoi{10.1111/j.1365-2966.2009.15919.x}

\bibitem[{{Kaur} {et~al.}(2010{\natexlab{b}}){Kaur}, {Casella}, {Linares},
  {Altamirano}, {Patruno}, {Armas-Padilla}, {Cavecchi}, {Degenaar}, {Russell},
  {Kalamkar}, {van der Klis}, {Watts}, {Wijnands}, \&
  {Rea}}]{Kaur...2010ATel.3082....1K}
{Kaur}, R., {Casella}, P., {Linares}, M., {et~al.} 2010{\natexlab{b}}, The
  Astronomer's Telegram, 3082, 1

\bibitem[{{Kendziorra} {et~al.}(1994){Kendziorra}, {Kretschmar}, {Pan}, {Kunz},
  {Maisack}, {Staubert}, {Pietsch}, {Truemper}, {Efremov}, \&
  {Sunyaev}}]{Kendziorra...1994A&A...291L..31K}
{Kendziorra}, E., {Kretschmar}, P., {Pan}, H.~C., {et~al.} 1994, \aap, 291, L31

\bibitem[{{Kennea} {et~al.}(2017){Kennea}, {Lien}, {Krimm}, {Cenko}, \&
  {Siegel}}]{Kennea...2017ATel10809....1K}
{Kennea}, J.~A., {Lien}, A.~Y., {Krimm}, H.~A., {Cenko}, S.~B., \& {Siegel},
  M.~H. 2017, The Astronomer's Telegram, 10809, 1

\bibitem[{{Kim} \& {Ikhsanov}(2017)}]{Kim...2017JPhCS.929a2005K}
{Kim}, V.~Y., \& {Ikhsanov}, N.~R. 2017, in Journal of Physics Conference
  Series, Vol. 929, Journal of Physics Conference Series, 012005,
  \dodoi{10.1088/1742-6596/929/1/012005}

\bibitem[{{Kinugasa} {et~al.}(1998){Kinugasa}, {Torii}, {Hashimoto}, {Tsunemi},
  {Hayashida}, {Kitamoto}, {Kamata}, {Dotani}, {Nagase}, {Sugizaki}, {Ueda},
  {Kawai}, {Makishima}, \& {Yamauchi}}]{Kinugasa...1998ApJ...495..435K}
{Kinugasa}, K., {Torii}, K., {Hashimoto}, Y., {et~al.} 1998, \apj, 495, 435,
  \dodoi{10.1086/305291}

\bibitem[{{Klus} {et~al.}(2014){Klus}, {Ho}, {Coe}, {Corbet}, \&
  {Townsend}}]{Klus...2014MNRAS.437.3863K}
{Klus}, H., {Ho}, W.~C.~G., {Coe}, M.~J., {Corbet}, R.~H.~D., \& {Townsend},
  L.~J. 2014, \mnras, 437, 3863, \dodoi{10.1093/mnras/stt2192}

\bibitem[{{Kouveliotou}(2002)}]{Kouveliotou...2002cxo..prop.1360K}
{Kouveliotou}, C. 2002, {Determining the Nature of the Peculiar X-ray Transient
  IGR J16358-4726}, Chandra Proposal ID 04408146

\bibitem[{{Kretschmar} {et~al.}(2017){Kretschmar},
  {Mart{\'\i}nez-N{\'u}{\~n}ez}, {Bozzo}, {Oskinova}, {Puls}, {Sidoli},
  {Sundqvist}, {Blay}, {Falanga}, {F{\"u}rst}, {G{\'\i}menez-Garc{\'\i}a},
  {Kreykenbohm}, {K{\"u}hnel}, {Sander}, {Torrej{\'o}n}, {Wilms},
  {Podsiadlowski}, \& {Manousakis}}]{Kretschmar...2017IAUS..329..355K}
{Kretschmar}, P., {Mart{\'\i}nez-N{\'u}{\~n}ez}, S., {Bozzo}, E., {et~al.}
  2017, in The Lives and Death-Throes of Massive Stars, ed. J.~J. {Eldridge},
  J.~C. {Bray}, L.~A.~S. {McClelland}, \& L.~{Xiao}, Vol. 329, 355--358,
  \dodoi{10.1017/S1743921317002411}

\bibitem[{{Kretschmar} {et~al.}(2019){Kretschmar}, {F{\"u}rst}, {Sidoli},
  {Bozzo}, {Alfonso-Garz{\'o}n}, {Bodaghee}, {Chaty}, {Chernyakova},
  {Ferrigno}, {Manousakis}, {Negueruela}, {Postnov}, {Paizis}, {Reig},
  {Rodes-Roca}, {Tsygankov}, {Bird}, {Bissinger n{\'e} K{\"u}hnel}, {Blay},
  {Caballero}, {Coe}, {Domingo}, {Doroshenko}, {Ducci}, {Falanga}, {Grebenev},
  {Grinberg}, {Hemphill}, {Kreykenbohm}, {Kreykenbohm n{\'e} Fritz}, {Li},
  {Lutovinov}, {Mart{\'\i}nez-N{\'u}{\~n}ez}, {Mas-Hesse}, {Masetti},
  {McBride}, {Neronov}, {Pottschmidt}, {Rodriguez}, {Romano}, {Rothschild},
  {Santangelo}, {Sguera}, {Staubert}, {Tomsick}, {Torrej{\'o}n}, {Torres},
  {Walter}, {Wilms}, {Wilson-Hodge}, \&
  {Zhang}}]{Kretschmar...2019NewAR...8601546K}
{Kretschmar}, P., {F{\"u}rst}, F., {Sidoli}, L., {et~al.} 2019, New Astronomy
  Reviews, 86, 101546, \dodoi{10.1016/j.newar.2020.101546}

\bibitem[{{Kretschmar} {et~al.}(2021){Kretschmar}, {El Mellah},
  {Mart{\'\i}nez-N{\'u}{\~n}ez}, {F{\"u}rst}, {Grinberg}, {Sander}, {van den
  Eijnden}, {Degenaar}, {Ma{\'\i}z Apell{\'a}niz}, {Jim{\'e}nez Esteban},
  {Ramos-Lerate}, \& {Utrilla}}]{Kretschmar...2021A&A...652A..95K}
{Kretschmar}, P., {El Mellah}, I., {Mart{\'\i}nez-N{\'u}{\~n}ez}, S., {et~al.}
  2021, \aap, 652, A95, \dodoi{10.1051/0004-6361/202040272}

\bibitem[{{Kreykenbohm} {et~al.}(2004){Kreykenbohm}, {Wilms}, {Coburn},
  {Kuster}, {Rothschild}, {Heindl}, {Kretschmar}, \&
  {Staubert}}]{Kreykenbohm...2004A&A...427..975K}
{Kreykenbohm}, I., {Wilms}, J., {Coburn}, W., {et~al.} 2004, \aap, 427, 975,
  \dodoi{10.1051/0004-6361:20035836}

\bibitem[{{Krimm} {et~al.}(2013){Krimm}, {Holland}, {Corbet}, {Pearlman},
  {Romano}, {Kennea}, {Bloom}, {Barthelmy}, {Baumgartner}, {Cummings},
  {Gehrels}, {Lien}, {Markwardt}, {Palmer}, {Sakamoto}, {Stamatikos}, \&
  {Ukwatta}}]{Krimm...2013ApJS..209...14K}
{Krimm}, H.~A., {Holland}, S.~T., {Corbet}, R.~H.~D., {et~al.} 2013, \apjs,
  209, 14, \dodoi{10.1088/0067-0049/209/1/14}

\bibitem[{{Krivonos} {et~al.}(2022){Krivonos}, {Sazonov}, {Kuznetsova},
  {Lutovinov}, {Mereminskiy}, \& {Tsygankov}}]{Krivonos...2022MNRAS.510.4796K}
{Krivonos}, R.~A., {Sazonov}, S.~Y., {Kuznetsova}, E.~A., {et~al.} 2022,
  \mnras, 510, 4796, \dodoi{10.1093/mnras/stab3751}

\bibitem[{{Krivonos} {et~al.}(2015){Krivonos}, {Tsygankov}, {Lutovinov},
  {Tomsick}, {Chakrabarty}, {Bachetti}, {Boggs}, {Chernyakova}, {Christensen},
  {Craig}, {F{\"u}rst}, {Hailey}, {Harrison}, {Lansbury}, {Rahoui}, {Stern}, \&
  {Zhang}}]{Krivonos...2015ApJ...809..140K}
{Krivonos}, R.~A., {Tsygankov}, S.~S., {Lutovinov}, A.~A., {et~al.} 2015, \apj,
  809, 140, \dodoi{10.1088/0004-637X/809/2/140}

\bibitem[{{Krti{\v{c}}ka}(2014)}]{Krticka...2014A&A...564A..70K}
{Krti{\v{c}}ka}, J. 2014, \aap, 564, A70, \dodoi{10.1051/0004-6361/201321980}

\bibitem[{{K{\"u}hnel} {et~al.}(2013){K{\"u}hnel}, {M{\"u}ller}, {Kreykenbohm},
  {F{\"u}rst}, {Pottschmidt}, {Rothschild}, {Caballero}, {Grinberg},
  {Sch{\"o}nherr}, {Shrader}, {Klochkov}, {Staubert}, {Ferrigno},
  {Torrej{\'o}n}, {Mart{\'\i}nez-N{\'u}{\~n}ez}, \&
  {Wilms}}]{Kuhnel...2013A&A...555A..95K}
{K{\"u}hnel}, M., {M{\"u}ller}, S., {Kreykenbohm}, I., {et~al.} 2013, \aap,
  555, A95, \dodoi{10.1051/0004-6361/201321203}

\bibitem[{{La Palombara} \&
  {Mereghetti}(2006)}]{LaPalombara...2006A&A...455..283L}
{La Palombara}, N., \& {Mereghetti}, S. 2006, \aap, 455, 283,
  \dodoi{10.1051/0004-6361:20065107}

\bibitem[{{La Palombara} {et~al.}(2009){La Palombara}, {Sidoli}, {Esposito},
  {Tiengo}, \& {Mereghetti}}]{LaPalombara...2009A&A...505..947L}
{La Palombara}, N., {Sidoli}, L., {Esposito}, P., {Tiengo}, A., \&
  {Mereghetti}, S. 2009, \aap, 505, 947, \dodoi{10.1051/0004-6361/200912538}

\bibitem[{{La Palombara} {et~al.}(2012){La Palombara}, {Sidoli}, {Esposito},
  {Tiengo}, \& {Mereghetti}}]{LaPalombara...2012A&A...539A..82L}
---. 2012, \aap, 539, A82, \dodoi{10.1051/0004-6361/201118221}

\bibitem[{{La Parola} {et~al.}(2010){La Parola}, {Cusumano}, {Romano},
  {Segreto}, {Vercellone}, \& {Chincarini}}]{LaParola...2010MNRAS.405L..66L}
{La Parola}, V., {Cusumano}, G., {Romano}, P., {et~al.} 2010, \mnras, 405, L66,
  \dodoi{10.1111/j.1745-3933.2010.00860.x}

\bibitem[{{La Parola} {et~al.}(2013){La Parola}, {Cusumano}, {Segreto},
  {D'A{\`\i}}, {Masetti}, \& {D'Elia}}]{Parola...2013ApJ...775L..24L}
{La Parola}, V., {Cusumano}, G., {Segreto}, A., {et~al.} 2013, \apjl, 775, L24,
  \dodoi{10.1088/2041-8205/775/1/L24}

\bibitem[{{La Parola} {et~al.}(2014){La Parola}, {Segreto}, {Cusumano},
  {Masetti}, {D'Ai}, \& {Melandri}}]{Parola...2014MNRAS.445L.119L}
{La Parola}, V., {Segreto}, A., {Cusumano}, G., {et~al.} 2014, \mnras, 445,
  L119, \dodoi{10.1093/mnrasl/slu147}

\bibitem[{{Landi} {et~al.}(2009){Landi}, {Masetti}, {Capitanio}, {Fiocchi}, \&
  {Bird}}]{Landi...2009ATel.2355....1L}
{Landi}, R., {Masetti}, N., {Capitanio}, F., {Fiocchi}, M., \& {Bird}, A.~J.
  2009, The Astronomer's Telegram, 2355, 1

\bibitem[{{Li} {et~al.}(2012{\natexlab{a}}){Li}, {Wang}, \&
  {Zhao}}]{Li...2012MNRAS.423.2854L}
{Li}, J., {Wang}, W., \& {Zhao}, Y. 2012{\natexlab{a}}, \mnras, 423, 2854,
  \dodoi{10.1111/j.1365-2966.2012.21096.x}

\bibitem[{{Li} {et~al.}(2012{\natexlab{b}}){Li}, {Zhang}, {Torres}, {Papitto},
  {Chen}, \& {Wang}}]{Li...2012MNRAS.426L..16L}
{Li}, J., {Zhang}, S., {Torres}, D.~F., {et~al.} 2012{\natexlab{b}}, \mnras,
  426, L16, \dodoi{10.1111/j.1745-3933.2012.01313.x}

\bibitem[{{Li} {et~al.}(2017){Li}, {Zhang}, \&
  {Feng}}]{Li...2017MNRAS.472..289L}
{Li}, K.~J., {Zhang}, J., \& {Feng}, W. 2017, \mnras, 472, 289,
  \dodoi{10.1093/mnras/stx1904}

\bibitem[{{Lin} {et~al.}(2002){Lin}, {Church}, {Nagase}, \&
  {Ba{\l}uci{\'n}ska-Church}}]{Lin...2002MNRAS.337.1245L}
{Lin}, X.~B., {Church}, M.~J., {Nagase}, F., \& {Ba{\l}uci{\'n}ska-Church}, M.
  2002, \mnras, 337, 1245, \dodoi{10.1046/j.1365-8711.2002.05952.x}

\bibitem[{{Lipunov}(1982)}]{Lipunov...1982SvA....26..537L}
{Lipunov}, V.~M. 1982, \sovast, 26, 537

\bibitem[{{Lipunov}(1992)}]{Lipunov...1992ans..book.....L}
---. 1992, {Astrophysics of Neutron Stars} ({Springer-Verlag Berlin Heidelberg
  New York})

\bibitem[{{Liu} {et~al.}(2000){Liu}, {van Paradijs}, \& {van den
  Heuvel}}]{Liu...2000AAS...147...25L}
{Liu}, Q.~Z., {van Paradijs}, J., \& {van den Heuvel}, E.~P.~J. 2000, \aaps,
  147, 25, \dodoi{10.1051/aas:2000288}

\bibitem[{{Liu} {et~al.}(2006){Liu}, {van Paradijs}, \& {van den
  Heuvel}}]{Liu...2006A&A...455.1165L}
---. 2006, \aap, 455, 1165, \dodoi{10.1051/0004-6361:20064987}

\bibitem[{{Lorenzo} {et~al.}(2014){Lorenzo}, {Negueruela}, {Castro}, {Norton},
  {Vilardell}, \& {Herrero}}]{Lorenzo...2014A&A...562A..18L}
{Lorenzo}, J., {Negueruela}, I., {Castro}, N., {et~al.} 2014, \aap, 562, A18,
  \dodoi{10.1051/0004-6361/201321913}

\bibitem[{{Lutovinov} {et~al.}(2005{\natexlab{a}}){Lutovinov}, {Revnivtsev},
  {Gilfanov}, {Shtykovskiy}, {Molkov}, \&
  {Sunyaev}}]{Lutovinov...2005A&A...444..821L}
{Lutovinov}, A., {Revnivtsev}, M., {Gilfanov}, M., {et~al.} 2005{\natexlab{a}},
  \aap, 444, 821, \dodoi{10.1051/0004-6361:20042392}

\bibitem[{{Lutovinov} {et~al.}(2005{\natexlab{b}}){Lutovinov}, {Rodriguez},
  {Revnivtsev}, \& {Shtykovskiy}}]{Lutovinov...2005A&A...433L..41L}
{Lutovinov}, A., {Rodriguez}, J., {Revnivtsev}, M., \& {Shtykovskiy}, P.
  2005{\natexlab{b}}, \aap, 433, L41, \dodoi{10.1051/0004-6361:200500092}

\bibitem[{{Lutovinov} {et~al.}(2016){Lutovinov}, {Buckley}, {Townsend},
  {Tsygankov}, \& {Kennea}}]{Lutovinov...2016MNRAS.462.3823L}
{Lutovinov}, A.~A., {Buckley}, D. A.~H., {Townsend}, L.~J., {Tsygankov}, S.~S.,
  \& {Kennea}, J. 2016, \mnras, 462, 3823, \dodoi{10.1093/mnras/stw1889}

\bibitem[{{Lutovinov} {et~al.}(2019){Lutovinov}, {Tsygankov}, {Karasev},
  {Molkov}, \& {Doroshenko}}]{Lutovinov...2019MNRAS.485..770L}
{Lutovinov}, A.~A., {Tsygankov}, S.~S., {Karasev}, D.~I., {Molkov}, S.~V., \&
  {Doroshenko}, V. 2019, \mnras, 485, 770, \dodoi{10.1093/mnras/stz437}

\bibitem[{{Lutovinov} {et~al.}(2017){Lutovinov}, {Tsygankov}, {Postnov},
  {Krivonos}, {Molkov}, \& {Tomsick}}]{Lutovinov...2017MNRAS.466..593L}
{Lutovinov}, A.~A., {Tsygankov}, S.~S., {Postnov}, K.~A., {et~al.} 2017,
  \mnras, 466, 593, \dodoi{10.1093/mnras/stw3058}

\bibitem[{{Lyne} \& {Graham-Smith}(2012)}]{Lyne...2012puas.book.....L}
{Lyne}, A., \& {Graham-Smith}, F. 2012, {Pulsar Astronomy} ({Cambridge
  University Press})

\bibitem[{{Lyubimkov} {et~al.}(1997){Lyubimkov}, {Rostopchin}, {Roche}, \&
  {Tarasov}}]{Lyubimkov...1997MNRAS.286..549L}
{Lyubimkov}, L.~S., {Rostopchin}, S.~I., {Roche}, P., \& {Tarasov}, A.~E. 1997,
  \mnras, 286, 549, \dodoi{10.1093/mnras/286.3.549}

\bibitem[{{Maccarone} {et~al.}(2014){Maccarone}, {Girard}, \&
  {Casetti-Dinescu}}]{Maccarone...2014MNRAS.440.1626M}
{Maccarone}, T.~J., {Girard}, T.~M., \& {Casetti-Dinescu}, D.~I. 2014, \mnras,
  440, 1626, \dodoi{10.1093/mnras/stu320}

\bibitem[{{Maitra} {et~al.}(2017){Maitra}, {Raichur}, {Pradhan}, \&
  {Paul}}]{Maitra...2017MNRAS.470..713M}
{Maitra}, C., {Raichur}, H., {Pradhan}, P., \& {Paul}, B. 2017, \mnras, 470,
  713, \dodoi{10.1093/mnras/stx1281}

\bibitem[{{Ma{\'\i}z Apell{\'a}niz} {et~al.}(2013){Ma{\'\i}z Apell{\'a}niz},
  {Sota}, {Morrell}, {Barb{\'a}}, {Walborn}, {Alfaro}, {Gamen}, {Arias}, \&
  {Gallego Calvente}}]{Apellaniz...2013msao.confE.198M}
{Ma{\'\i}z Apell{\'a}niz}, J., {Sota}, A., {Morrell}, N.~I., {et~al.} 2013, in
  Massive Stars: From alpha to Omega, 198, \dodoi{10.48550/arXiv.1306.6417}

\bibitem[{{Makishima} {et~al.}(1990){Makishima}, {Mihara}, {Ishida}, {Ohashi},
  {Sakao}, {Tashiro}, {Tsuru}, {Kii}, {Makino}, {Murakami}, {Nagase}, {Tanaka},
  {Kunieda}, {Tawara}, {Kitamoto}, {Miyamoto}, {Yoshida}, \&
  {Turner}}]{Makishima...1990ApJ...365L..59M}
{Makishima}, K., {Mihara}, T., {Ishida}, M., {et~al.} 1990, \apjl, 365, L59,
  \dodoi{10.1086/185888}

\bibitem[{{Malacaria} {et~al.}(2020){Malacaria}, {Jenke}, {Roberts},
  {Wilson-Hodge}, {Cleveland}, {Mailyan}, \& {GBM Accreting Pulsars Program
  Team}}]{Malacaria...2020ApJ...896...90M}
{Malacaria}, C., {Jenke}, P., {Roberts}, O.~J., {et~al.} 2020, \apj, 896, 90,
  \dodoi{10.3847/1538-4357/ab855c}

\bibitem[{{Malacaria} {et~al.}(2022){Malacaria}, {Bhargava}, {Coley}, {Ducci},
  {Pradhan}, {Ballhausen}, {Fuerst}, {Islam}, {Jaisawal}, {Jenke},
  {Kretschmar}, {Kreykenbohm}, {Pottschmidt}, {Sokolova-Lapa}, {Staubert},
  {Wilms}, {Wilson-Hodge}, \& {Wolff}}]{Malacaria...2022ApJ...927..194M}
{Malacaria}, C., {Bhargava}, Y., {Coley}, J.~B., {et~al.} 2022, \apj, 927, 194,
  \dodoi{10.3847/1538-4357/ac524f}

\bibitem[{{Marcu-Cheatham} {et~al.}(2015){Marcu-Cheatham}, {Pottschmidt},
  {K{\"u}hnel}, {M{\"u}ller}, {Falkner}, {Caballero}, {Finger}, {Jenke},
  {Wilson-Hodge}, {F{\"u}rst}, {Grinberg}, {Hemphill}, {Kreykenbohm},
  {Klochkov}, {Rothschild}, {Terada}, {Enoto}, {Iwakiri}, {Wolff}, {Becker},
  {Wood}, \& {Wilms}}]{Marcu...2015ApJ...815...44M}
{Marcu-Cheatham}, D.~M., {Pottschmidt}, K., {K{\"u}hnel}, M., {et~al.} 2015,
  \apj, 815, 44, \dodoi{10.1088/0004-637X/815/1/44}

\bibitem[{{Markwardt} {et~al.}(2010){Markwardt}, {Baumgartner}, {Skinner}, \&
  {Corbet}}]{Markwardt...2010ATel.2564....1M}
{Markwardt}, C.~B., {Baumgartner}, W.~H., {Skinner}, G.~K., \& {Corbet}, R.
  H.~D. 2010, The Astronomer's Telegram, 2564, 1

\bibitem[{{Marsden} {et~al.}(1998){Marsden}, {Gruber}, {Heindl}, {Pelling}, \&
  {Rothschild}}]{Marsden...1998ApJ...502L.129M}
{Marsden}, D., {Gruber}, D.~E., {Heindl}, W.~A., {Pelling}, M.~R., \&
  {Rothschild}, R.~E. 1998, \apjl, 502, L129, \dodoi{10.1086/311510}

\bibitem[{{Marshall} \& {Ricketts}(1980)}]{Marshall...1980MNRAS.193P...7M}
{Marshall}, N., \& {Ricketts}, M.~J. 1980, \mnras, 193, 7P,
  \dodoi{10.1093/mnras/193.1.7P}

\bibitem[{{Masetti} {et~al.}(2008){Masetti}, {Mason}, {Morelli}, {Cellone},
  {McBride}, {Palazzi}, {Bassani}, {Bazzano}, {Bird}, {Charles}, {Dean},
  {Galaz}, {Gehrels}, {Landi}, {Malizia}, {Minniti}, {Panessa}, {Romero},
  {Stephen}, {Ubertini}, \& {Walter}}]{Masetti...2008A&A...482..113M}
{Masetti}, N., {Mason}, E., {Morelli}, L., {et~al.} 2008, \aap, 482, 113,
  \dodoi{10.1051/0004-6361:20079332}

\bibitem[{{Masetti} {et~al.}(2009){Masetti}, {Parisi}, {Palazzi},
  {Jim{\'e}nez-Bail{\'o}n}, {Morelli}, {Chavushyan}, {Mason}, {McBride},
  {Bassani}, {Bazzano}, {Bird}, {Dean}, {Galaz}, {Gehrels}, {Landi}, {Malizia},
  {Minniti}, {Schiavone}, {Stephen}, \&
  {Ubertini}}]{Masetti...2009A&A...495..121M}
{Masetti}, N., {Parisi}, P., {Palazzi}, E., {et~al.} 2009, \aap, 495, 121,
  \dodoi{10.1051/0004-6361:200811322}

\bibitem[{{Mason} {et~al.}(2009){Mason}, {Clark}, {Norton}, {Negueruela}, \&
  {Roche}}]{Mason...2009A&A...505..281M}
{Mason}, A.~B., {Clark}, J.~S., {Norton}, A.~J., {Negueruela}, I., \& {Roche},
  P. 2009, \aap, 505, 281, \dodoi{10.1051/0004-6361/200912480}

\bibitem[{{Mason} {et~al.}(2010){Mason}, {Norton}, {Clark}, {Negueruela}, \&
  {Roche}}]{Mason...2010A&A...509A..79M}
{Mason}, A.~B., {Norton}, A.~J., {Clark}, J.~S., {Negueruela}, I., \& {Roche},
  P. 2010, \aap, 509, A79, \dodoi{10.1051/0004-6361/200913394}

\bibitem[{{McBride} {et~al.}(2006){McBride}, {Wilms}, {Coe}, {Kreykenbohm},
  {Rothschild}, {Coburn}, {Galache}, {Kretschmar}, {Edge}, \&
  {Staubert}}]{McBride...2006A&A...451..267M}
{McBride}, V.~A., {Wilms}, J., {Coe}, M.~J., {et~al.} 2006, \aap, 451, 267,
  \dodoi{10.1051/0004-6361:20054239}

\bibitem[{{McBride} {et~al.}(2007){McBride}, {Wilms}, {Kreykenbohm}, {Coe},
  {Rothschild}, {Kretschmar}, {Pottschmidt}, {Fisher}, \&
  {Hamson}}]{McBride...2007A&A...470.1065M}
{McBride}, V.~A., {Wilms}, J., {Kreykenbohm}, I., {et~al.} 2007, \aap, 470,
  1065, \dodoi{10.1051/0004-6361:20077238}

\bibitem[{{McCollum} \& {Laine}(2019)}]{McCollum...2019ATel13211....1M}
{McCollum}, B., \& {Laine}, S. 2019, The Astronomer's Telegram, 13211, 1

\bibitem[{{M}c{K}inney(2010)}]{Mckinney...PPSC2010...56M}
{M}c{K}inney. 2010, in {P}roceedings of the 9th {P}ython in {S}cience
  {C}onference, ed. {S}t\'efan van~der {W}alt \& {J}arrod {M}illman, 56 -- 61,
  \dodoi{10.25080/Majora-92bf1922-00a}

\bibitem[{{Mereghetti} \& {La
  Palombara}(2009)}]{Mereghetti...2009A&A...504..181M}
{Mereghetti}, S., \& {La Palombara}, N. 2009, \aap, 504, 181,
  \dodoi{10.1051/0004-6361/200911944}

\bibitem[{{Molkov} {et~al.}(2019){Molkov}, {Lutovinov}, {Tsygankov},
  {Mereminskiy}, \& {Mushtukov}}]{Molkov...2019ApJ...883L..11M}
{Molkov}, S., {Lutovinov}, A., {Tsygankov}, S., {Mereminskiy}, I., \&
  {Mushtukov}, A. 2019, \apjl, 883, L11, \dodoi{10.3847/2041-8213/ab3e4d}

\bibitem[{{Morel} \& {Grosdidier}(2005)}]{Morel...2005MNRAS.356..665M}
{Morel}, T., \& {Grosdidier}, Y. 2005, \mnras, 356, 665,
  \dodoi{10.1111/j.1365-2966.2004.08488.x}

\bibitem[{{Motch} {et~al.}(1997){Motch}, {Haberl}, {Dennerl}, {Pakull}, \&
  {Janot-Pacheco}}]{Motch...1997A&A...323..853M}
{Motch}, C., {Haberl}, F., {Dennerl}, K., {Pakull}, M., \& {Janot-Pacheco}, E.
  1997, \aap, 323, 853

\bibitem[{{Mukerjee} \& {Antia}(2021)}]{Mukerjee...2021ApJ...920..139M}
{Mukerjee}, K., \& {Antia}, H.~M. 2021, \apj, 920, 139,
  \dodoi{10.3847/1538-4357/ac11f1}

\bibitem[{{Mukerjee} {et~al.}(2020){Mukerjee}, {Antia}, \&
  {Katoch}}]{Mukerjee...2020ApJ...897...73M}
{Mukerjee}, K., {Antia}, H.~M., \& {Katoch}, T. 2020, \apj, 897, 73,
  \dodoi{10.3847/1538-4357/ab97b6}

\bibitem[{{M{\"u}ller} {et~al.}(2012){M{\"u}ller}, {K{\"u}hnel}, {Caballero},
  {Pottschmidt}, {F{\"u}rst}, {Kreykenbohm}, {Sagredo}, {Obst}, {Wilms},
  {Ferrigno}, {Rothschild}, \& {Staubert}}]{Muller...2012A&A...546A.125M}
{M{\"u}ller}, S., {K{\"u}hnel}, M., {Caballero}, I., {et~al.} 2012, \aap, 546,
  A125, \dodoi{10.1051/0004-6361/201219580}

\bibitem[{{Mushtukov} {et~al.}(2015){Mushtukov}, {Suleimanov}, {Tsygankov}, \&
  {Poutanen}}]{Mushtukov...2015MNRAS.454.2539M}
{Mushtukov}, A.~A., {Suleimanov}, V.~F., {Tsygankov}, S.~S., \& {Poutanen}, J.
  2015, \mnras, 454, 2539, \dodoi{10.1093/mnras/stv2087}

\bibitem[{{Nabizadeh} {et~al.}(2019{\natexlab{a}}){Nabizadeh},
  {M{\"o}nkk{\"o}nen}, {Tsygankov}, {Doroshenko}, {Molkov}, \&
  {Poutanen}}]{Nabizadeh...2019A&A...629A.101N}
{Nabizadeh}, A., {M{\"o}nkk{\"o}nen}, J., {Tsygankov}, S.~S., {et~al.}
  2019{\natexlab{a}}, \aap, 629, A101, \dodoi{10.1051/0004-6361/201936045}

\bibitem[{{Nabizadeh} {et~al.}(2019{\natexlab{b}}){Nabizadeh}, {Tsygankov},
  {Karasev}, {M{\"o}nkk{\"o}nen}, {Lutovinov}, {Nagirner}, \&
  {Poutanen}}]{Nabizadeh...2019A&A...622A.198N}
{Nabizadeh}, A., {Tsygankov}, S.~S., {Karasev}, D.~I., {et~al.}
  2019{\natexlab{b}}, \aap, 622, A198, \dodoi{10.1051/0004-6361/201834635}

\bibitem[{{Nabizadeh} {et~al.}(2022){Nabizadeh}, {Tsygankov}, {Molkov},
  {Karasev}, {Ji}, {Lutovinov}, \&
  {Poutanen}}]{Nabizadeh...2022A&A...657A..58N}
{Nabizadeh}, A., {Tsygankov}, S.~S., {Molkov}, S.~V., {et~al.} 2022, \aap, 657,
  A58, \dodoi{10.1051/0004-6361/202141608}

\bibitem[{{Nagase}(1989)}]{Nagase...1989PasJ...396..P147}
{Nagase}, F. 1989, \pasj, 41, 1, \dodoi{1989PASJ...41....1N}

\bibitem[{{Nagase} {et~al.}(1991){Nagase}, {Dotani}, {Tanaka}, {Makishima},
  {Mihara}, {Sakao}, {Tsunemi}, {Kitamoto}, {Tamura}, {Yoshida}, \&
  {Nakamura}}]{Nagase...1991ApJ...375L..49N}
{Nagase}, F., {Dotani}, T., {Tanaka}, Y., {et~al.} 1991, \apjl, 375, L49,
  \dodoi{10.1086/186085}

\bibitem[{{Naik} {et~al.}(2013){Naik}, {Maitra}, {Jaisawal}, \&
  {Paul}}]{Naik...2013ApJ...764..158N}
{Naik}, S., {Maitra}, C., {Jaisawal}, G.~K., \& {Paul}, B. 2013, \apj, 764,
  158, \dodoi{10.1088/0004-637X/764/2/158}

\bibitem[{{Negueruela}(1998)}]{Negueruela...1998A&A...338..505N}
{Negueruela}, I. 1998, \aap, 338, 505

\bibitem[{{Negueruela} {et~al.}(2008){Negueruela}, {Casares}, {Verrecchia},
  {Blay}, {Israel}, \& {Covino}}]{Negueruela...2008ATel.1876....1N}
{Negueruela}, I., {Casares}, J., {Verrecchia}, F., {et~al.} 2008, The
  Astronomer's Telegram, 1876, 1

\bibitem[{{Negueruela} {et~al.}(2003){Negueruela}, {Israel}, {Marco}, {Norton},
  \& {Speziali}}]{Negueruela...2003A&A...397..739N}
{Negueruela}, I., {Israel}, G.~L., {Marco}, A., {Norton}, A.~J., \& {Speziali},
  R. 2003, \aap, 397, 739, \dodoi{10.1051/0004-6361:20021529}

\bibitem[{{Negueruela} \& {Okazaki}(2001)}]{Negueruela...2001A&A...369..108N}
{Negueruela}, I., \& {Okazaki}, A.~T. 2001, \aap, 369, 108,
  \dodoi{10.1051/0004-6361:20010146}

\bibitem[{{Negueruela} {et~al.}(1999){Negueruela}, {Roche}, {Fabregat}, \&
  {Coe}}]{Negueruela...1999MNRAS.307..695N}
{Negueruela}, I., {Roche}, P., {Fabregat}, J., \& {Coe}, M.~J. 1999, \mnras,
  307, 695, \dodoi{10.1046/j.1365-8711.1999.02682.x}

\bibitem[{{Negueruela} \& {Schurch}(2007)}]{Negueruela...2007A&A...461..631N}
{Negueruela}, I., \& {Schurch}, M.~P.~E. 2007, \aap, 461, 631,
  \dodoi{10.1051/0004-6361:20066054}

\bibitem[{{Nespoli} {et~al.}(2008){Nespoli}, {Fabregat}, \&
  {Mennickent}}]{Nespoli...2008A&A...486..911N}
{Nespoli}, E., {Fabregat}, J., \& {Mennickent}, R.~E. 2008, \aap, 486, 911,
  \dodoi{10.1051/0004-6361:200809645}

\bibitem[{{Nespoli} {et~al.}(2010){Nespoli}, {Fabregat}, \&
  {Mennickent}}]{Nespoli...2010A&A...516A..94N}
---. 2010, \aap, 516, A94, \dodoi{10.1051/0004-6361/200913410}

\bibitem[{{Nespoli} \& {Reig}(2011)}]{Nespoli...2011A&A...526A...7N}
{Nespoli}, E., \& {Reig}, P. 2011, \aap, 526, A7,
  \dodoi{10.1051/0004-6361/201015303}

\bibitem[{{Nespoli} {et~al.}(2012){Nespoli}, {Reig}, \&
  {Zezas}}]{Nespoli...2012A&A...547A.103N}
{Nespoli}, E., {Reig}, P., \& {Zezas}, A. 2012, \aap, 547, A103,
  \dodoi{10.1051/0004-6361/201219586}

\bibitem[{{Neumann} {et~al.}(2023){Neumann}, {Avakyan}, {Doroshenko}, \&
  {Santangelo}}]{Neumann...2023arXiv230316137N}
{Neumann}, M., {Avakyan}, A., {Doroshenko}, V., \& {Santangelo}, A. 2023, arXiv
  e-prints, arXiv:2303.16137, \dodoi{10.48550/arXiv.2303.16137}

\bibitem[{{Nikolov} {et~al.}(2017){Nikolov}, {Zamanov}, {Stoyanov}, \&
  {Mart{\'\i}}}]{Nikolov...2017BlgAJ..27...10N}
{Nikolov}, Y.~M., {Zamanov}, R.~K., {Stoyanov}, K.~A., \& {Mart{\'\i}}, J.
  2017, Bulgarian Astronomical Journal, 27, 10

\bibitem[{{Nowak} {et~al.}(2012){Nowak}, {Paizis}, {Rodriguez}, {Chaty}, {Del
  Santo}, {Grinberg}, {Wilms}, {Ubertini}, \&
  {Chini}}]{Nowak...2012ApJ...757..143N}
{Nowak}, M.~A., {Paizis}, A., {Rodriguez}, J., {et~al.} 2012, \apj, 757, 143,
  \dodoi{10.1088/0004-637X/757/2/143}

\bibitem[{{O'Connor} {et~al.}(2022){O'Connor}, {G{\"o}{\u{g}}{\"u}{\c{s}}},
  {Huppenkothen}, {Kouveliotou}, {Gorgone}, {Townsend}, {Calamida}, {Fruchter},
  {Buckley}, {Baring}, {Kennea}, {Younes}, {Arzoumanian}, {Bellm}, {Cenko},
  {Gendreau}, {Granot}, {Hailey}, {Harrison}, {Hartmann}, {Kaper}, {Kutyrev},
  {Slane}, {Stern}, {Troja}, {van der Horst}, {Wijers}, \&
  {Woudt}}]{OConnor...2022ApJ...927..139O}
{O'Connor}, B., {G{\"o}{\u{g}}{\"u}{\c{s}}}, E., {Huppenkothen}, D., {et~al.}
  2022, \apj, 927, 139, \dodoi{10.3847/1538-4357/ac5032}

\bibitem[{{Oh} \& {Kroupa}(2016)}]{Oh...2016A&A...590A.107O}
{Oh}, S., \& {Kroupa}, P. 2016, \aap, 590, A107,
  \dodoi{10.1051/0004-6361/201628233}

\bibitem[{{Oja}(1991)}]{Oja...1991A&AS...89..415O}
{Oja}, T. 1991, \aaps, 89, 415

\bibitem[{{Orlandini} {et~al.}(2012){Orlandini}, {Frontera}, {Masetti},
  {Sguera}, \& {Sidoli}}]{Orlandini...2012ApJ...748...86O}
{Orlandini}, M., {Frontera}, F., {Masetti}, N., {Sguera}, V., \& {Sidoli}, L.
  2012, \apj, 748, 86, \dodoi{10.1088/0004-637X/748/2/86}

\bibitem[{{Pacini}(1970)}]{Pacini...1970Natur.226..622P}
{Pacini}, F. 1970, \nat, 226, 622, \dodoi{10.1038/226622a0}

\bibitem[{{Pahari} \& {Pal}(2012)}]{Pahari...2012MNRAS.423.3352P}
{Pahari}, M., \& {Pal}, S. 2012, \mnras, 423, 3352,
  \dodoi{10.1111/j.1365-2966.2012.21128.x}

\bibitem[{{Parkes} {et~al.}(1978){Parkes}, {Murdin}, \&
  {Mason}}]{Parkes...1978MNRAS.184P..73P}
{Parkes}, G.~E., {Murdin}, P.~G., \& {Mason}, K.~O. 1978, \mnras, 184, 73P,
  \dodoi{10.1093/mnras/184.1.73P}

\bibitem[{{Pearlman} {et~al.}(2019){Pearlman}, {Coley}, {Corbet}, \&
  {Pottschmidt}}]{Pearlman...2019ApJ...873...86P}
{Pearlman}, A.~B., {Coley}, J.~B., {Corbet}, R. H.~D., \& {Pottschmidt}, K.
  2019, \apj, 873, 86, \dodoi{10.3847/1538-4357/aaf001}

\bibitem[{{Piraino} {et~al.}(2000){Piraino}, {Santangelo}, {Segreto},
  {Giarrusso}, {Cusumano}, {Del Sordo}, {Robba}, {Dal Fiume}, {Orlandini},
  {Oosterbroek}, \& {Parmar}}]{Piraino...2000A&A...357..501P}
{Piraino}, S., {Santangelo}, A., {Segreto}, A., {et~al.} 2000, \aap, 357, 501

\bibitem[{{Postnov} {et~al.}(2015){Postnov}, {Mironov}, {Lutovinov}, {Shakura},
  {Kochetkova}, \& {Tsygankov}}]{Postnov...2015MNRAS.446.1013P}
{Postnov}, K.~A., {Mironov}, A.~I., {Lutovinov}, A.~A., {et~al.} 2015, \mnras,
  446, 1013, \dodoi{10.1093/mnras/stu2155}

\bibitem[{{Pradhan} {et~al.}(2013){Pradhan}, {Maitra}, {Paul}, \&
  {Paul}}]{Pradhan...2013MNRAS.436..945P}
{Pradhan}, P., {Maitra}, C., {Paul}, B., \& {Paul}, B.~C. 2013, \mnras, 436,
  945, \dodoi{10.1093/mnras/stt1504}

\bibitem[{{Pringle} \& {Rees}(1972)}]{Pringle...1972A&A....21....1P}
{Pringle}, J.~E., \& {Rees}, M.~J. 1972, \aap, 21, 1

\bibitem[{{Puls} {et~al.}(2008){Puls}, {Vink}, \&
  {Najarro}}]{Puls...2008A&ARv..16..209P}
{Puls}, J., {Vink}, J.~S., \& {Najarro}, F. 2008, \aapr, 16, 209,
  \dodoi{10.1007/s00159-008-0015-8}

\bibitem[{{Quaintrell} {et~al.}(2003){Quaintrell}, {Norton}, {Ash}, {Roche},
  {Willems}, {Bedding}, {Baldry}, \&
  {Fender}}]{Quaintrell...2003A&A...401..313Q}
{Quaintrell}, H., {Norton}, A.~J., {Ash}, T.~D.~C., {et~al.} 2003, \aap, 401,
  313, \dodoi{10.1051/0004-6361:20030120}

\bibitem[{{Raddi} {et~al.}(2021){Raddi}, {Irrgang}, {Heber}, {Schneider}, \&
  {Kreuzer}}]{Raddi...2021A&A...645A.108R}
{Raddi}, R., {Irrgang}, A., {Heber}, U., {Schneider}, D., \& {Kreuzer}, S.
  2021, \aap, 645, A108, \dodoi{10.1051/0004-6361/202037872}

\bibitem[{{Rahoui} \& {Chaty}(2008)}]{Rahoui...2008A&A...492..163R}
{Rahoui}, F., \& {Chaty}, S. 2008, \aap, 492, 163,
  \dodoi{10.1051/0004-6361:200810695}

\bibitem[{{Rahoui} {et~al.}(2008){Rahoui}, {Chaty}, {Lagage}, \&
  {Pantin}}]{Rahoui...2008A&A...484..801R}
{Rahoui}, F., {Chaty}, S., {Lagage}, P.~O., \& {Pantin}, E. 2008, \aap, 484,
  801, \dodoi{10.1051/0004-6361:20078774}

\bibitem[{{Raichur} \& {Paul}(2010)}]{Raichur...2010MNRAS.406.2663R}
{Raichur}, H., \& {Paul}, B. 2010, \mnras, 406, 2663,
  \dodoi{10.1111/j.1365-2966.2010.16862.x}

\bibitem[{{Raman} {et~al.}(2021){Raman}, {Varun}, {Paul}, \&
  {Bhattacharya}}]{Raman...2021MNRAS.508.5578R}
{Raman}, G., {Varun}, {Paul}, B., \& {Bhattacharya}, D. 2021, \mnras, 508,
  5578, \dodoi{10.1093/mnras/stab2835}

\bibitem[{{Ray} \& {Chakrabarty}(2002)}]{Ray...2002ApJ...581.1293R}
{Ray}, P.~S., \& {Chakrabarty}, D. 2002, \apj, 581, 1293,
  \dodoi{10.1086/344300}

\bibitem[{{Reed}(2003)}]{Reed...2003AJ....125.2531R}
{Reed}, B.~C. 2003, \aj, 125, 2531, \dodoi{10.1086/374771}

\bibitem[{{Reig} {et~al.}(2008){Reig}, {Belloni}, {Israel}, {Campana},
  {Gehrels}, \& {Homan}}]{Reig...2008A&A...485..797R}
{Reig}, P., {Belloni}, T., {Israel}, G.~L., {et~al.} 2008, \aap, 485, 797,
  \dodoi{10.1051/0004-6361:200809457}

\bibitem[{{Reig} {et~al.}(2014{\natexlab{a}}){Reig}, {Blinov}, {Papadakis},
  {Kylafis}, \& {Tassis}}]{Reig...2014MNRAS.445.4235R}
{Reig}, P., {Blinov}, D., {Papadakis}, I., {Kylafis}, N., \& {Tassis}, K.
  2014{\natexlab{a}}, \mnras, 445, 4235, \dodoi{10.1093/mnras/stu2322}

\bibitem[{{Reig} \& {Coe}(1999)}]{Reig...1999MNRAS.302..700R}
{Reig}, P., \& {Coe}, M.~J. 1999, \mnras, 302, 700,
  \dodoi{10.1046/j.1365-8711.1999.02179.x}

\bibitem[{{Reig} {et~al.}(2014{\natexlab{b}}){Reig}, {Doroshenko}, \&
  {Zezas}}]{Reig...2014MNRAS.445.1314R}
{Reig}, P., {Doroshenko}, V., \& {Zezas}, A. 2014{\natexlab{b}}, \mnras, 445,
  1314, \dodoi{10.1093/mnras/stu1840}

\bibitem[{{Reig} \& {Fabregat}(2015)}]{Reig...2015A&A...574A..33R}
{Reig}, P., \& {Fabregat}, J. 2015, \aap, 574, A33,
  \dodoi{10.1051/0004-6361/201425008}

\bibitem[{{Reig} \& {Fabregat}(2022)}]{Reig...2022A&A...667A..18R}
---. 2022, \aap, 667, A18, \dodoi{10.1051/0004-6361/202243664}

\bibitem[{{Reig} {et~al.}(2020){Reig}, {Fabregat}, \&
  {Alfonso-Garz{\'o}n}}]{Reig...2020A&A...640A..35R}
{Reig}, P., {Fabregat}, J., \& {Alfonso-Garz{\'o}n}, J. 2020, \aap, 640, A35,
  \dodoi{10.1051/0004-6361/202038333}

\bibitem[{{Reig} {et~al.}(1997){Reig}, {Fabregat}, {Coe}, {Roche},
  {Chakrabarty}, {Negueruela}, \& {Steele}}]{Reig...1997A&A...322..183R}
{Reig}, P., {Fabregat}, J., {Coe}, M.~J., {et~al.} 1997, \aap, 322, 183

\bibitem[{{Reig} {et~al.}(2004){Reig}, {Negueruela}, {Fabregat}, {Chato},
  {Blay}, \& {Mavromatakis}}]{Reig...2004A&A...421..673R}
{Reig}, P., {Negueruela}, I., {Fabregat}, J., {et~al.} 2004, \aap, 421, 673,
  \dodoi{10.1051/0004-6361:20035786}

\bibitem[{{Reig} {et~al.}(2005{\natexlab{a}}){Reig}, {Negueruela}, {Fabregat},
  {Chato}, \& {Coe}}]{Reig...2005A&A...440.1079R}
{Reig}, P., {Negueruela}, I., {Fabregat}, J., {Chato}, R., \& {Coe}, M.~J.
  2005{\natexlab{a}}, \aap, 440, 1079, \dodoi{10.1051/0004-6361:20053124}

\bibitem[{{Reig} {et~al.}(2005{\natexlab{b}}){Reig}, {Negueruela},
  {Papamastorakis}, {Manousakis}, \&
  {Kougentakis}}]{Reig...2005A&A...440..637R}
{Reig}, P., {Negueruela}, I., {Papamastorakis}, G., {Manousakis}, A., \&
  {Kougentakis}, T. 2005{\natexlab{b}}, \aap, 440, 637,
  \dodoi{10.1051/0004-6361:20052684}

\bibitem[{{Reig} {et~al.}(2011){Reig}, {Nespoli}, {Fabregat}, \&
  {Mennickent}}]{Reig...2011A&A...533A..23R}
{Reig}, P., {Nespoli}, E., {Fabregat}, J., \& {Mennickent}, R.~E. 2011, \aap,
  533, A23, \dodoi{10.1051/0004-6361/201117301}

\bibitem[{{Reig} \& {Roche}(1999)}]{Reig...1999MNRAS.306..100R}
{Reig}, P., \& {Roche}, P. 1999, \mnras, 306, 100,
  \dodoi{10.1046/j.1365-8711.1999.02473.x}

\bibitem[{{Reig} {et~al.}(1996){Reig}, {Roche}, {Fabregat}, \&
  {Coe}}]{Reig...1996rftu.proc..181R}
{Reig}, P., {Roche}, P., {Fabregat}, J., \& {Coe}, M.~J. 1996, in
  Roentgenstrahlung from the Universe, ed. H.~U. {Zimmermann},
  J.~{Tr{\"u}mper}, \& H.~{Yorke}, 181--182

\bibitem[{{Reig} \& {Zezas}(2014)}]{Reig...2014MNRAS.442..472R}
{Reig}, P., \& {Zezas}, A. 2014, \mnras, 442, 472, \dodoi{10.1093/mnras/stu898}

\bibitem[{{Reig} \& {Zezas}(2018)}]{Reig...2018A&A...613A..52R}
---. 2018, \aap, 613, A52, \dodoi{10.1051/0004-6361/201732533}

\bibitem[{{Riquelme} {et~al.}(2012){Riquelme}, {Torrej{\'o}n}, \&
  {Negueruela}}]{Riquelme...2012A&A...539A.114R}
{Riquelme}, M.~S., {Torrej{\'o}n}, J.~M., \& {Negueruela}, I. 2012, \aap, 539,
  A114, \dodoi{10.1051/0004-6361/201117738}

\bibitem[{{Rodes-Roca} {et~al.}(2018){Rodes-Roca}, {Bernabeu}, {Magazz{\`u}},
  {Torrej{\'o}n}, \& {Solano}}]{Rodes_Roca...2018MNRAS.476.2110R}
{Rodes-Roca}, J.~J., {Bernabeu}, G., {Magazz{\`u}}, A., {Torrej{\'o}n}, J.~M.,
  \& {Solano}, E. 2018, \mnras, 476, 2110, \dodoi{10.1093/mnras/sty333}

\bibitem[{{Rodes-Roca} {et~al.}(2013){Rodes-Roca}, {Torrej{\'o}n},
  {Mart{\'\i}nez-N{\'u}{\~n}ez}, {Bernab{\'e}u}, \&
  {Magazz{\'u}}}]{Rodes...2013A&A...555A.115R}
{Rodes-Roca}, J.~J., {Torrej{\'o}n}, J.~M., {Mart{\'\i}nez-N{\'u}{\~n}ez}, S.,
  {Bernab{\'e}u}, G., \& {Magazz{\'u}}, A. 2013, \aap, 555, A115,
  \dodoi{10.1051/0004-6361/201321923}

\bibitem[{{Romano} {et~al.}(2009){Romano}, {Sidoli}, {Cusumano}, {Vercellone},
  {Mangano}, \& {Krimm}}]{Romano...2009ApJ...696.2068R}
{Romano}, P., {Sidoli}, L., {Cusumano}, G., {et~al.} 2009, \apj, 696, 2068,
  \dodoi{10.1088/0004-637X/696/2/2068}

\bibitem[{{Romano} {et~al.}(2015){Romano}, {Bozzo}, {Mangano}, {Esposito},
  {Israel}, {Tiengo}, {Campana}, {Ducci}, {Ferrigno}, \&
  {Kennea}}]{Romano...2015A&A...576L...4R}
{Romano}, P., {Bozzo}, E., {Mangano}, V., {et~al.} 2015, \aap, 576, L4,
  \dodoi{10.1051/0004-6361/201525749}

\bibitem[{{Rouco Escorial} {et~al.}(2018){Rouco Escorial}, {van den Eijnden},
  \& {Wijnands}}]{Rouco...2018A&A...620L..13R}
{Rouco Escorial}, A., {van den Eijnden}, J., \& {Wijnands}, R. 2018, \aap, 620,
  L13, \dodoi{10.1051/0004-6361/201834572}

\bibitem[{{Roy} {et~al.}(2020){Roy}, {Agrawal}, {Singari}, \&
  {Misra}}]{Roy...2020RAA....20..155R}
{Roy}, J., {Agrawal}, P.~C., {Singari}, B., \& {Misra}, R. 2020, Research in
  Astronomy and Astrophysics, 20, 155, \dodoi{10.1088/1674-4527/20/9/155}

\bibitem[{{Sadakane} {et~al.}(1985){Sadakane}, {Hirata}, {Jugaku}, {Kondo},
  {Matsuoka}, {Tanaka}, \&
  {Hammerschlag-Hensberge}}]{Sadakane...1985ApJ...288..284S}
{Sadakane}, K., {Hirata}, R., {Jugaku}, J., {et~al.} 1985, \apj, 288, 284,
  \dodoi{10.1086/162791}

\bibitem[{{Salganik} {et~al.}(2022){Salganik}, {Tsygankov}, {Djupvik},
  {Karasev}, {Lutovinov}, {Buckley}, {Gromadzki}, \&
  {Poutanen}}]{Salganik...2022MNRAS.509.5955S}
{Salganik}, A., {Tsygankov}, S.~S., {Djupvik}, A.~A., {et~al.} 2022, \mnras,
  509, 5955, \dodoi{10.1093/mnras/stab3362}

\bibitem[{{Santangelo} {et~al.}(1998){Santangelo}, {del Sordo}, {Segreto}, {dal
  Fiume}, {Orlandini}, \& {Piraino}}]{Santangelo...1998AandAP..340..L55}
{Santangelo}, A., {del Sordo}, S., {Segreto}, A., {et~al.} 1998, \aap, 340,
  L55–L59, \dodoi{10.1051/aas:1996164}

\bibitem[{{Sarty} {et~al.}(2009){Sarty}, {Kiss}, {Huziak}, {Catalan}, {Luciuk},
  {Crawford}, {Lane}, {Pickard}, {Grzybowski}, {Closas}, {Johnston}, {Balam},
  \& {Wu}}]{Sarty...2009MNRAS.392.1242S}
{Sarty}, G.~E., {Kiss}, L.~L., {Huziak}, R., {et~al.} 2009, \mnras, 392, 1242,
  \dodoi{10.1111/j.1365-2966.2008.14138.x}

\bibitem[{{Sarty} {et~al.}(2011){Sarty}, {Pilecki}, {Reichart}, {Ivarsen},
  {Haislip}, {Nysewander}, {LaCluyze}, {Johnston}, {Shobbrook}, {Kiss}, \&
  {Wu}}]{Sarty...2011RAA....11..947S}
{Sarty}, G.~E., {Pilecki}, B., {Reichart}, D.~E., {et~al.} 2011, Research in
  Astronomy and Astrophysics, 11, 947, \dodoi{10.1088/1674-4527/11/8/007}

\bibitem[{{Sato} {et~al.}(1986){Sato}, {Nagase}, {Kawai}, {Kelley},
  {Rappaport}, \& {White}}]{Sato...1986ApJ...304..241S}
{Sato}, N., {Nagase}, F., {Kawai}, N., {et~al.} 1986, \apj, 304, 241,
  \dodoi{10.1086/164157}

\bibitem[{{Scott} {et~al.}(1997){Scott}, {Finger}, {Wilson}, {Koh}, {Prince},
  {Vaughan}, \& {Chakrabarty}}]{Scott...1997ApJ...488..831S}
{Scott}, D.~M., {Finger}, M.~H., {Wilson}, R.~B., {et~al.} 1997, \apj, 488,
  831, \dodoi{10.1086/304740}

\bibitem[{{Segreto} {et~al.}(2013){Segreto}, {La Parola}, {Cusumano},
  {D'A{\`\i}}, {Masetti}, \& {Campana}}]{Segreto...2013A&A...558A..99S}
{Segreto}, A., {La Parola}, V., {Cusumano}, G., {et~al.} 2013, \aap, 558, A99,
  \dodoi{10.1051/0004-6361/201321892}

\bibitem[{{Sguera} {et~al.}(2013){Sguera}, {Drave}, {Sidoli}, {Masetti},
  {Landi}, {Bird}, \& {Bazzano}}]{Sguera...2013A&A...556A..27S}
{Sguera}, V., {Drave}, S.~P., {Sidoli}, L., {et~al.} 2013, \aap, 556, A27,
  \dodoi{10.1051/0004-6361/201220785}

\bibitem[{{Sguera} {et~al.}(2007){Sguera}, {Hill}, {Bird}, {Dean}, {Bazzano},
  {Ubertini}, {Masetti}, {Landi}, {Malizia}, {Clark}, \&
  {Molina}}]{Sguera...2007A&A...467..249S}
{Sguera}, V., {Hill}, A.~B., {Bird}, A.~J., {et~al.} 2007, \aap, 467, 249,
  \dodoi{10.1051/0004-6361:20066762}

\bibitem[{{Shakura}(1973)}]{Shakura...1973SvA....16..756S}
{Shakura}, N.~I. 1973, \sovast, 16, 756

\bibitem[{{Shakura}(1975)}]{Shakura...1975SvAL....1..223S}
---. 1975, Soviet Astronomy Letters, 1, 223

\bibitem[{{Shaw} {et~al.}(2009){Shaw}, {Hill}, {Kuulkers}, {Brandt},
  {Chenevez}, \& {Kretschmar}}]{Shaw...2009MNRAS.393..419S}
{Shaw}, S.~E., {Hill}, A.~B., {Kuulkers}, E., {et~al.} 2009, \mnras, 393, 419,
  \dodoi{10.1111/j.1365-2966.2008.14212.x}

\bibitem[{{Shirke} {et~al.}(2021){Shirke}, {Bala}, {Roy}, \&
  {Bhattacharya}}]{Shirke...2021JApA...42...58S}
{Shirke}, P., {Bala}, S., {Roy}, J., \& {Bhattacharya}, D. 2021, Journal of
  Astrophysics and Astronomy, 42, 58, \dodoi{10.1007/s12036-021-09710-w}

\bibitem[{{Shtykovsky} {et~al.}(2019){Shtykovsky}, {Lutovinov}, {Tsygankov}, \&
  {Molkov}}]{Shtykovsky...2019MNRAS.482L..14S}
{Shtykovsky}, A.~E., {Lutovinov}, A.~A., {Tsygankov}, S.~S., \& {Molkov}, S.~V.
  2019, \mnras, 482, L14, \dodoi{10.1093/mnrasl/sly182}

\bibitem[{{Sidoli} {et~al.}(2016){Sidoli}, {Esposito}, {Motta}, {Israel}, \&
  {Rodr{\'\i}guez Castillo}}]{Sidoli...2016MNRAS.460.3637S}
{Sidoli}, L., {Esposito}, P., {Motta}, S.~E., {Israel}, G.~L., \&
  {Rodr{\'\i}guez Castillo}, G.~A. 2016, \mnras, 460, 3637,
  \dodoi{10.1093/mnras/stw1246}

\bibitem[{{Sidoli} {et~al.}(2017){Sidoli}, {Israel}, {Esposito},
  {Rodr{\'\i}guez Castillo}, \& {Postnov}}]{Sidoli...2017MNRAS.469.3056S}
{Sidoli}, L., {Israel}, G.~L., {Esposito}, P., {Rodr{\'\i}guez Castillo},
  G.~A., \& {Postnov}, K. 2017, \mnras, 469, 3056,
  \dodoi{10.1093/mnras/stx1105}

\bibitem[{{Sidoli} {et~al.}(2012){Sidoli}, {Mereghetti}, {Sguera}, \&
  {Pizzolato}}]{Sidoli...2012MNRAS.420..554S}
{Sidoli}, L., {Mereghetti}, S., {Sguera}, V., \& {Pizzolato}, F. 2012, \mnras,
  420, 554, \dodoi{10.1111/j.1365-2966.2011.20063.x}

\bibitem[{{Sidoli} \& {Paizis}(2018)}]{Sidoli...2018MNRAS.481.2779S}
{Sidoli}, L., \& {Paizis}, A. 2018, \mnras, 481, 2779,
  \dodoi{10.1093/mnras/sty2428}

\bibitem[{{Sidoli} {et~al.}(2020){Sidoli}, {Postnov}, {Tiengo}, {Esposito},
  {Sguera}, {Paizis}, \& {Rodr{\'\i}guez
  Castillo}}]{Sidoli...2020A&A...638A..71S}
{Sidoli}, L., {Postnov}, K., {Tiengo}, A., {et~al.} 2020, \aap, 638, A71,
  \dodoi{10.1051/0004-6361/202038078}

\bibitem[{{Siuniaev} \& {Shakura}(1977)}]{Siuniaev...1977SvAL....3..114S}
{Siuniaev}, R.~A., \& {Shakura}, N.~I. 1977, Soviet Astronomy Letters, 3, 114

\bibitem[{{Staubert} {et~al.}(2011){Staubert}, {Pottschmidt}, {Doroshenko},
  {Wilms}, {Suchy}, {Rothschild}, \&
  {Santangelo}}]{Staubert...2011A&A...527A...7S}
{Staubert}, R., {Pottschmidt}, K., {Doroshenko}, V., {et~al.} 2011, \aap, 527,
  A7, \dodoi{10.1051/0004-6361/201015737}

\bibitem[{{Staubert} {et~al.}(2019){Staubert}, {Tr{\"u}mper}, {Kendziorra},
  {Klochkov}, {Postnov}, {Kretschmar}, {Pottschmidt}, {Haberl}, {Rothschild},
  {Santangelo}, {Wilms}, {Kreykenbohm}, \&
  {F{\"u}rst}}]{Staubert...2019A&A...622A..61S}
{Staubert}, R., {Tr{\"u}mper}, J., {Kendziorra}, E., {et~al.} 2019, \aap, 622,
  A61, \dodoi{10.1051/0004-6361/201834479}

\bibitem[{{Stevens} {et~al.}(1997){Stevens}, {Reig}, {Coe}, {Buckley},
  {Fabregat}, \& {Steele}}]{Stevens...1997MNRAS.288..988S}
{Stevens}, J.~B., {Reig}, P., {Coe}, M.~J., {et~al.} 1997, \mnras, 288, 988,
  \dodoi{10.1093/mnras/288.4.988}

\bibitem[{{Strader} {et~al.}(2019){Strader}, {Chomiuk}, {Swihart}, \&
  {Aydi}}]{Strader...2019ATel12554....1S}
{Strader}, J., {Chomiuk}, L., {Swihart}, S., \& {Aydi}, E. 2019, The
  Astronomer's Telegram, 12554, 1

\bibitem[{{Sugizaki} {et~al.}(2001){Sugizaki}, {Mitsuda}, {Kaneda},
  {Matsuzaki}, {Yamauchi}, \& {Koyama}}]{Sugizaki...2001ApJS..134...77S}
{Sugizaki}, M., {Mitsuda}, K., {Kaneda}, H., {et~al.} 2001, \apjs, 134, 77,
  \dodoi{10.1086/320358}

\bibitem[{{Sugizaki} {et~al.}(2015){Sugizaki}, {Yamamoto}, {Mihara},
  {Nakajima}, \& {Makishima}}]{Sugizaki...2015PASJ...67...73S}
{Sugizaki}, M., {Yamamoto}, T., {Mihara}, T., {Nakajima}, M., \& {Makishima},
  K. 2015, \pasj, 67, 73, \dodoi{10.1093/pasj/psv039}

\bibitem[{{Swank} \& {Markwardt}(2003)}]{Swank...2003ATel..128....1S}
{Swank}, J.~H., \& {Markwardt}, C.~B. 2003, The Astronomer's Telegram, 128, 1

\bibitem[{{Taylor}(2005)}]{Taylor...2005ASPC..347...29T}
{Taylor}, M.~B. 2005, in Astronomical Society of the Pacific Conference Series,
  Vol. 347, Astronomical Data Analysis Software and Systems XIV, ed.
  P.~{Shopbell}, M.~{Britton}, \& R.~{Ebert}, 29

\bibitem[{{Thompson} \& {Rothschild}(2008)}]{Thompson...2009ApJ...691..P1744}
{Thompson}, T., \& {Rothschild}, R. 2008, \apj, 691, 1744–1753,
  \dodoi{10.1088/0004-637X/691/2/1744}

\bibitem[{{Thompson} {et~al.}(2007){Thompson}, {Tomsick}, {in 't Zand},
  {Rothschild}, \& {Walter}}]{Thompson...2007ApJ...661..447T}
{Thompson}, T. W.~J., {Tomsick}, J.~A., {in 't Zand}, J.~J.~M., {Rothschild},
  R.~E., \& {Walter}, R. 2007, \apj, 661, 447, \dodoi{10.1086/513458}

\bibitem[{{Thompson} {et~al.}(2006){Thompson}, {Tomsick}, {Rothschild}, {in't
  Zand}, \& {Walter}}]{Thompson...2006ApJ...649..373T}
{Thompson}, T. W.~J., {Tomsick}, J.~A., {Rothschild}, R.~E., {in't Zand},
  J.~J.~M., \& {Walter}, R. 2006, \apj, 649, 373, \dodoi{10.1086/506251}

\bibitem[{{Tobrej} {et~al.}(2023){Tobrej}, {Rai}, {Ghising}, {Tamang}, \&
  {Paul}}]{Tobrej...2023MNRAS.518.4861T}
{Tobrej}, M., {Rai}, B., {Ghising}, M., {Tamang}, R., \& {Paul}, B.~C. 2023,
  \mnras, 518, 4861, \dodoi{10.1093/mnras/stac3203}

\bibitem[{{Torii} {et~al.}(1998){Torii}, {Kinugasa}, {Katayama}, {Kohmura},
  {Tsunemi}, {Sakano}, {Nishiuchi}, {Koyama}, {Yamauchi}, \&
  {Shigeo}}]{Torii...1998ApJ...508..854T}
{Torii}, K., {Kinugasa}, K., {Katayama}, K., {et~al.} 1998, \apj, 508, 854,
  \dodoi{10.1086/306455}

\bibitem[{{Torrejon} {et~al.}(2010){Torrejon}, {Negueruela}, {Smith}, \&
  {Harrison}}]{Torrejon...2010AandAP..510..P61}
{Torrejon}, J., {Negueruela}, I., {Smith}, D., \& {Harrison}, T. 2010, \aap,
  510, 1, \dodoi{10.1051/0004-6361/200912619}

\bibitem[{{Torrej{\'o}n} {et~al.}(2004){Torrej{\'o}n}, {Kreykenbohm}, {Orr},
  {Titarchuk}, \& {Negueruela}}]{Torrejon...2004A&A...423..301T}
{Torrej{\'o}n}, J.~M., {Kreykenbohm}, I., {Orr}, A., {Titarchuk}, L., \&
  {Negueruela}, I. 2004, \aap, 423, 301, \dodoi{10.1051/0004-6361:20035743}

\bibitem[{{Torrej{\'o}n} \&
  {Negueruela}(2007)}]{Torrejon...2007ESASP.622..503T}
{Torrej{\'o}n}, J.~M., \& {Negueruela}, I. 2007, in ESA Special Publication,
  Vol. 622, The Obscured Universe. Proceedings of the VI INTEGRAL Workshop, 503

\bibitem[{{Torrej{\'o}n} {et~al.}(2010){Torrej{\'o}n}, {Negueruela}, {Smith},
  \& {Harrison}}]{Torrejon...2010A&A...510A..61T}
{Torrej{\'o}n}, J.~M., {Negueruela}, I., {Smith}, D.~M., \& {Harrison}, T.~E.
  2010, \aap, 510, A61, \dodoi{10.1051/0004-6361/200912619}

\bibitem[{{Tsygankov} {et~al.}(2019{\natexlab{a}}){Tsygankov}, {Doroshenko},
  {Mushtukov}, {Lutovinov}, \& {Poutanen}}]{Tsygankov...2019A&A...621A.134T}
{Tsygankov}, S.~S., {Doroshenko}, V., {Mushtukov}, A.~A., {Lutovinov}, A.~A.,
  \& {Poutanen}, J. 2019{\natexlab{a}}, \aap, 621, A134,
  \dodoi{10.1051/0004-6361/201833786}

\bibitem[{{Tsygankov} {et~al.}(2012){Tsygankov}, {Krivonos}, \&
  {Lutovinov}}]{Tsygankov...2012MNRAS.421.2407T}
{Tsygankov}, S.~S., {Krivonos}, R.~A., \& {Lutovinov}, A.~A. 2012, \mnras, 421,
  2407, \dodoi{10.1111/j.1365-2966.2012.20475.x}

\bibitem[{{Tsygankov} {et~al.}(2016){Tsygankov}, {Lutovinov}, {Krivonos},
  {Molkov}, {Jenke}, {Finger}, \& {Poutanen}}]{Tsygankov...2016MNRAS.457..258T}
{Tsygankov}, S.~S., {Lutovinov}, A.~A., {Krivonos}, R.~A., {et~al.} 2016,
  \mnras, 457, 258, \dodoi{10.1093/mnras/stv2849}

\bibitem[{{Tsygankov} {et~al.}(2019{\natexlab{b}}){Tsygankov}, {Rouco
  Escorial}, {Suleimanov}, {Mushtukov}, {Doroshenko}, {Lutovinov}, {Wijnands},
  \& {Poutanen}}]{Tsygankov...2019MNRAS.483L.144T}
{Tsygankov}, S.~S., {Rouco Escorial}, A., {Suleimanov}, V.~F., {et~al.}
  2019{\natexlab{b}}, \mnras, 483, L144, \dodoi{10.1093/mnrasl/sly236}

\bibitem[{{Tsygankov} {et~al.}(2021){Tsygankov}, {Lutovinov}, {Molkov},
  {Djupvik}, {Karasev}, {Doroshenko}, {Mushtukov}, {Malacaria}, {Kretschmar},
  \& {Poutanen}}]{Tsygankov...2021ApJ...909..154T}
{Tsygankov}, S.~S., {Lutovinov}, A.~A., {Molkov}, S.~V., {et~al.} 2021, \apj,
  909, 154, \dodoi{10.3847/1538-4357/abddbd}

\bibitem[{{Tutukov} \& {Cherepashchuk}(2020)}]{Tutukov...2020PhyU...63..209T}
{Tutukov}, A.~V., \& {Cherepashchuk}, A.~M. 2020, Physics Uspekhi, 63, 209,
  \dodoi{10.3367/UFNe.2019.03.038547}

\bibitem[{{Uchida} {et~al.}(2021){Uchida}, {Takahashi}, {Fukazawa}, \&
  {Makishima}}]{Uchida...2021PASJ...73.1389U}
{Uchida}, N., {Takahashi}, H., {Fukazawa}, Y., \& {Makishima}, K. 2021, \pasj,
  73, 1389, \dodoi{10.1093/pasj/psab083}

\bibitem[{{UKIDSS Consortium}(2012)}]{UKIDSS...2012yCat.2316....0U}
{UKIDSS Consortium}. 2012, VizieR Online Data Catalog, II/316

\bibitem[{{van den Heuvel}(2017)}]{Heuvel...2017hsn..book.1527V}
{van den Heuvel}, E. P.~J. 2017, in Handbook of Supernovae, ed. A.~W. {Alsabti}
  \& P.~{Murdin} (Springer, Cham), 1527, \dodoi{10.1007/978-3-319-21846-5_75}

\bibitem[{{van Paradijs}(1995)}]{vanParadijs...1995xrbi...nasa...536V}
{van Paradijs}, J. 1995, in X-ray Binaries, 536--577

\bibitem[{{Wang}(2010)}]{Wang...2010A&A...516A..15W}
{Wang}, W. 2010, \aap, 516, A15, \dodoi{10.1051/0004-6361/200913196}

\bibitem[{{Wang}(2011)}]{Wang...2011MNRAS.413.1083W}
---. 2011, \mnras, 413, 1083, \dodoi{10.1111/j.1365-2966.2010.18192.x}

\bibitem[{{Wang}(2014)}]{Wang...2014RAA....14..565W}
---. 2014, Research in Astronomy and Astrophysics, 14, 565,
  \dodoi{10.1088/1674-4527/14/5/006}

\bibitem[{{Wen} {et~al.}(2000){Wen}, {Remillard}, \&
  {Bradt}}]{Wen...2000ApJ...532.1119W}
{Wen}, L., {Remillard}, R.~A., \& {Bradt}, H.~V. 2000, \apj, 532, 1119,
  \dodoi{10.1086/308604}

\bibitem[{{Wenger} {et~al.}(2000){Wenger}, {Ochsenbein}, {Egret}, {Dubois},
  {Bonnarel}, {Borde}, {Genova}, {Jasniewicz}, {Lalo{\"e}}, {Lesteven}, \&
  {Monier}}]{Wenger...2000AAS..143....9W}
{Wenger}, M., {Ochsenbein}, F., {Egret}, D., {et~al.} 2000, \aaps, 143, 9,
  \dodoi{10.1051/aas:2000332}

\bibitem[{{Wilson} {et~al.}(2008){Wilson}, {Finger}, \&
  {Camero-Arranz}}]{Wilson...2008ApJ...678.1263W}
{Wilson}, C.~A., {Finger}, M.~H., \& {Camero-Arranz}, A. 2008, \apj, 678, 1263,
  \dodoi{10.1086/587134}

\bibitem[{{Wilson} {et~al.}(2003){Wilson}, {Finger}, {Coe}, \&
  {Negueruela}}]{Wilson...2003ApJ...584..996W}
{Wilson}, C.~A., {Finger}, M.~H., {Coe}, M.~J., \& {Negueruela}, I. 2003, \apj,
  584, 996, \dodoi{10.1086/345791}

\bibitem[{{Wilson} {et~al.}(2002){Wilson}, {Finger},
  {G{\"o}{\v{g}}{\"u}{\c{s}}}, {Woods}, \&
  {Kouveliotou}}]{Wilson...2002ApJ...565.1150W}
{Wilson}, C.~A., {Finger}, M.~H., {G{\"o}{\v{g}}{\"u}{\c{s}}}, E., {Woods},
  P.~M., \& {Kouveliotou}, C. 2002, \apj, 565, 1150, \dodoi{10.1086/324707}

\bibitem[{{Wilson} {et~al.}(1998){Wilson}, {Finger}, {Harmon}, {Chakrabarty},
  \& {Strohmayer}}]{Wilson...1998ApJ...499..820W}
{Wilson}, C.~A., {Finger}, M.~H., {Harmon}, B.~A., {Chakrabarty}, D., \&
  {Strohmayer}, T. 1998, \apj, 499, 820, \dodoi{10.1086/305677}

\bibitem[{{Wilson} {et~al.}(1997){Wilson}, {Finger}, {Harmon}, {Scott},
  {Wilson}, {Bildsten}, {Chakrabarty}, \&
  {Prince}}]{Wilson...1997ApJ...479..388W}
{Wilson}, C.~A., {Finger}, M.~H., {Harmon}, B.~A., {et~al.} 1997, \apj, 479,
  388, \dodoi{10.1086/303841}

\bibitem[{{Wilson} {et~al.}(1999){Wilson}, {Finger}, \&
  {Scott}}]{Wilson...1999ApJ...511..367W}
{Wilson}, C.~A., {Finger}, M.~H., \& {Scott}, D.~M. 1999, \apj, 511, 367,
  \dodoi{10.1086/306660}

\bibitem[{{Wilson-Hodge}(1999)}]{Wilson...1999PhDT.........6W}
{Wilson-Hodge}, C.~A. 1999, PhD thesis, University of Alabama, Huntsville

\bibitem[{{Wilson-Hodge} {et~al.}(2018){Wilson-Hodge}, {Malacaria}, {Jenke},
  {Jaisawal}, {Kerr}, {Wolff}, {Arzoumanian}, {Chakrabarty}, {Doty},
  {Gendreau}, {Guillot}, {Ho}, {LaMarr}, {Markwardt}, {{\"O}zel}, {Prigozhin},
  {Ray}, {Ramos-Lerate}, {Remillard}, {Strohmayer}, {Vezie}, {Wood}, \& {NICER
  Science Team}}]{Wilson-Hodge...2018ApJ...863....9W}
{Wilson-Hodge}, C.~A., {Malacaria}, C., {Jenke}, P.~A., {et~al.} 2018, \apj,
  863, 9, \dodoi{10.3847/1538-4357/aace60}

\bibitem[{{Yamamoto} {et~al.}(2011){Yamamoto}, {Sugizaki}, {Mihara},
  {Nakajima}, {Yamaoka}, {Matsuoka}, {Morii}, \&
  {Makishima}}]{Yamamoto...2011PASJ...63S.751Y}
{Yamamoto}, T., {Sugizaki}, M., {Mihara}, T., {et~al.} 2011, \pasj, 63, S751,
  \dodoi{10.1093/pasj/63.sp3.S751}

\bibitem[{{Yan} {et~al.}(2012){Yan}, {Zurita Heras}, {Chaty}, {Li}, \&
  {Liu}}]{Yan...2012ApJ...753...73Y}
{Yan}, J., {Zurita Heras}, J.~A., {Chaty}, S., {Li}, H., \& {Liu}, Q. 2012,
  \apj, 753, 73, \dodoi{10.1088/0004-637X/753/1/73}

\bibitem[{{Yungelson} {et~al.}(2019){Yungelson}, {Kuranov}, \&
  {Postnov}}]{Yungelson...2019MNRAS.485..851Y}
{Yungelson}, L.~R., {Kuranov}, A.~G., \& {Postnov}, K.~A. 2019, \mnras, 485,
  851, \dodoi{10.1093/mnras/stz467}

\bibitem[{{Zacharias} {et~al.}(2012){Zacharias}, {Finch}, {Girard}, {Henden},
  {Bartlett}, {Monet}, \& {Zacharias}}]{Zacharias...2012yCat.1322....0Z}
{Zacharias}, N., {Finch}, C.~T., {Girard}, T.~M., {et~al.} 2012, VizieR Online
  Data Catalog, I/322A

\bibitem[{{Zhang} {et~al.}(2005){Zhang}, {Qu}, {Song}, \&
  {Torres}}]{Zhang...2005ApJ...630L..65Z}
{Zhang}, S., {Qu}, J.-L., {Song}, L.-M., \& {Torres}, D.~F. 2005, \apjl, 630,
  L65, \dodoi{10.1086/462415}

\bibitem[{{Zhao} {et~al.}(2019){Zhao}, {Heinke}, {Tsygankov}, {Ho}, {Potekhin},
  \& {Shaw}}]{Zhao...2019MNRAS.488.4427Z}
{Zhao}, Y., {Heinke}, C.~O., {Tsygankov}, S.~S., {et~al.} 2019, \mnras, 488,
  4427, \dodoi{10.1093/mnras/stz1946}

\bibitem[{{Zurita Heras} \& {Chaty}(2008)}]{Zurita...2008A&A...489..657Z}
{Zurita Heras}, J.~A., \& {Chaty}, S. 2008, \aap, 489, 657,
  \dodoi{10.1051/0004-6361:20079097}

\bibitem[{{Zurita Heras} {et~al.}(2006){Zurita Heras}, {De Cesare}, {Walter},
  {Bodaghee}, {B{\'e}langer}, {Courvoisier}, {Shaw}, \&
  {Stephen}}]{Zurita...2006A&A...448..261Z}
{Zurita Heras}, J.~A., {De Cesare}, G., {Walter}, R., {et~al.} 2006, \aap, 448,
  261, \dodoi{10.1051/0004-6361:20053876}

\end{thebibliography}
\bibliographystyle{aasjournal}

\end{document}